\documentclass[twocolumn]{aastex631}
\usepackage[utf8]{inputenc}
\usepackage[figure,figure*]{hypcap}
\usepackage{multirow}
\usepackage{natbib}
\usepackage{xcolor}
\usepackage{soul}
\usepackage{CJK}
\usepackage{enumitem}

\usepackage{etoolbox}
\makeatletter
\patchcmd{\NAT@citex}
  {\@citea\NAT@hyper@{%
     \NAT@nmfmt{\NAT@nm}%
     \hyper@natlinkbreak{\NAT@aysep\NAT@spacechar}{\@citeb\@extra@b@citeb}%
     \NAT@date}}
  {\@citea\NAT@nmfmt{\NAT@nm}%
   \NAT@aysep\NAT@spacechar\NAT@hyper@{\NAT@date}}{}{}
\patchcmd{\NAT@citex}
  {\@citea\NAT@hyper@{%
     \NAT@nmfmt{\NAT@nm}%
     \hyper@natlinkbreak{\NAT@spacechar\NAT@@open\if*#1*\else#1\NAT@spacechar\fi}%
       {\@citeb\@extra@b@citeb}%
     \NAT@date}}
  {\@citea\NAT@nmfmt{\NAT@nm}%
   \NAT@spacechar\NAT@@open\if*#1*\else#1\NAT@spacechar\fi\NAT@hyper@{\NAT@date}}
  {}{}
\makeatother

\newcolumntype{X}[1]{>{\raggedright\let\newline\\\arraybackslash\hspace{0pt}}p{#1}}

\setstcolor{red}
\hypersetup{linkcolor=red,citecolor=blue,filecolor=magenta,urlcolor=blue}

\newcommand{\uJy}{$\mu$Jy}

\newcommand{\eazy}{{\sc Eazy}}
\newcommand{\bpz}{{\sc BPZ}}
\newcommand{\lephare}{{\sc LePhare}}
\newcommand{\zphot}{{\sc zphot}}

\newcommand{\dbasis}{{\sc Dense Basis}}
\newcommand{\cigale}{{\sc Cigale}}

\newcommand\tblmark[1]{\tablenotemark{\scriptsize {#1}}}
\newcommand\tbltext[1]{\tablenotetext{{#1}}}

\shorttitle{UVCANDELS: photometric redshifts and galaxy physical properties}
\shortauthors{Mehta, Rafelski, et al.}

\graphicspath{{./}{figures/}}

\begin{document}

\title{UVCANDELS: Catalogs of photometric redshifts and galaxy physical properties}

\author[0000-0001-7166-6035]{Vihang Mehta}
\affiliation{IPAC, Mail Code 314-6, California Institute of Technology, 1200 E. California Blvd., Pasadena, CA, 91125, USA}
\email{vmehta@ipac.caltech.edu}

\author[0000-0002-9946-4731]{Marc Rafelski}
\affiliation{Space Telescope Science Institute, 3700 San Martin Drive, Baltimore, MD 21218, USA}
\affiliation{Department of Physics and Astronomy, Johns Hopkins University, Baltimore, MD 21218, USA}

\author[0000-0003-3759-8707]{Ben Sunnquist}
\affiliation{Space Telescope Science Institute, 3700 San Martin Drive, Baltimore, MD 21218, USA}


\author[0000-0002-7064-5424]{Harry I. Teplitz}
\affiliation{IPAC, Mail Code 314-6, California Institute of Technology, 1200 E. California Blvd., Pasadena, CA, 91125, USA}

\author[0000-0002-9136-8876]{Claudia Scarlata}
\affiliation{Minnesota Institute for Astrophysics, University of Minnesota, 116 Church St SE, Minneapolis, MN 55455, USA}

\author[0000-0002-9373-3865]{Xin Wang}
\affiliation{School of Astronomy and Space Science, University of Chinese Academy of Sciences (UCAS), Beijing 100049, China}
\affiliation{National Astronomical Observatories, Chinese Academy of Sciences, Beijing 100101, China}

\author[0000-0003-3820-2823]{Adriano Fontana}
\affiliation{INAF – Osservatorio Astronomico di Roma – via Frascati 33, Monte Porzio Catone, 00078 Rome, Italy}

\author[0000-0001-6145-5090]{Nimish P. Hathi}
\affiliation{Space Telescope Science Institute, 3700 San Martin Drive, Baltimore, MD 21218, USA}

\author[0000-0001-9298-3523]{Kartheik G. Iyer}
\affiliation{Columbia Astrophysics Laboratory, Columbia University, 550 West 120th Street, New York, NY 10027, USA}

\author[0000-0002-8630-6435]{Anahita Alavi}
\affiliation{IPAC, Mail Code 314-6, California Institute of Technology, 1200 E. California Blvd., Pasadena, CA, 91125, USA}

\author[0000-0001-6482-3020]{James Colbert}
\affiliation{IPAC, Mail Code 314-6, California Institute of Technology, 1200 E. California Blvd., Pasadena, CA, 91125, USA}


\author[0000-0001-9440-8872]{Norman Grogin}
\affiliation{Space Telescope Science Institute, 3700 San Martin Drive, Baltimore, MD 21218, USA}

\author[0000-0002-6610-2048]{Anton Koekemoer}
\affiliation{Space Telescope Science Institute, 3700 San Martin Drive, Baltimore, MD 21218, USA}

\author[0000-0001-5294-8002]{Kalina V. Nedkova}
\affiliation{Department of Physics and Astronomy, Johns Hopkins University, Baltimore, MD 21218, USA}
\affiliation{Space Telescope Science Institute, 3700 San Martin Drive, Baltimore, MD 21218, USA}

\author[0000-0001-8587-218X]{Matthew Hayes}
\affiliation{Stockholm University, Department of Astronomy and Oskar Klein Centre for Cosmoparticle Physics, SE-10691, Stockholm, Sweden}

\author[0000-0002-0604-654X]{Laura Prichard}
\affiliation{Space Telescope Science Institute, 3700 San Martin Drive, Baltimore, MD 21218, USA}

\author[0000-0002-4935-9511]{Brian Siana}
\affiliation{Department of Physics and Astronomy, University of California, Riverside, Riverside, CA 92521, USA}

\author[0000-0002-0648-1699]{Brent M. Smith}
\affil{School of Earth and Space Exploration, Arizona State University, Tempe, AZ 85287, USA}

\author[0000-0001-8156-6281]{Rogier Windhorst}
\affiliation{School of Earth and Space Exploration, Arizona State University, Tempe, AZ 85287, USA}

\author[0000-0003-4439-6003]{Teresa Ashcraft}
\affiliation{School of Earth and Space Exploration, Arizona State University, Tempe, AZ 85287, USA}

\author[0000-0002-9921-9218]{Micaela Bagley}
\affiliation{Department of Astronomy, The University of Texas at Austin, Austin, TX 78712, USA}

\author[0000-0003-0556-2929]{Ivano Baronchelli}
\affiliation{INAF -- Osservatorio Astronomico di Padova, Vicolo dell'Osservatorio 5, I-35122, Padova, Italy}

\author[0000-0001-6813-875X]{Guillermo Barro}
\affiliation{Department of Physics, University of the Pacific, Stockton, CA 95211, USA}

\author[0000-0003-2102-3933]{Alex Blanche}
\affiliation{School of Earth and Space Exploration, Arizona State University, Tempe, AZ 85287, USA}

\author[0000-0002-7767-5044]{Adam Broussard}
\affiliation{Department of Physics and Astronomy, Rutgers, The State University of New Jersey, Piscataway, NJ 08854, USA}

\author[0000-0001-6650-2853]{Timothy Carleton}
\affiliation{School of Earth and Space Exploration, Arizona State University, Tempe, AZ 85287, USA}

\author[0000-0003-3691-937X]{Nima Chartab}
\affiliation{Observatories of the Carnegie Institution of Washington, Pasadena, CA 91101, US}

\author[0000-0001-9560-9174]{Alex Codoreanu}
\affiliation{Centre for Astrophysics and Supercomputing, Swinburne University of Technology, Hawthorn VIC 3122, Australia}

\author[0000-0003-3329-1337]{Seth Cohen} 
\affiliation{School of Earth and Space Exploration, Arizona State University, Tempe, AZ 85287, USA}

\author[0000-0003-1949-7638]{Christopher Conselice}
\affiliation{School of Physics and Astronomy, The University of Nottingham, University Park, Nottingham NG7 2RD, UK}

\author[0000-0002-7928-416X]{Y. Sophia Dai}
\affiliation{National Astronomical Observatories, Chinese Academy of Sciences, Beijing 100101, China}

\author[0000-0003-4919-9017]{Behnam Darvish}
\affiliation{Department of Physics and Astronomy, University of California, Riverside, Riverside, CA 92521, USA}

\author[0000-0003-2842-9434]{Romeel Dav\'{e}}
\affiliation{Institute for Astronomy, University of Edinburgh, Edinburgh, EH9 3HJ, UK}

\author[0000-0001-9022-665X]{Laura DeGroot}
\affiliation{College of Wooster, Wooster, OH 44691, USA}

\author[0000-0003-1624-8425]{Duilia De Mello}
\affiliation{Department of Physics, The Catholic University of America, Washington, DC 20064, USA}

\author[0000-0001-5414-5131]{Mark Dickinson}
\affiliation{NSF's NOIRLab, Tucson, AZ 85719, USA}

\author[0000-0003-2047-1689]{Najmeh Emami}
\affiliation{Department of Physics and Astronomy, University of California, Riverside, Riverside, CA 92521, USA}

\author[0000-0001-7113-2738]{Henry Ferguson}
\affiliation{Space Telescope Science Institute, 3700 San Martin Drive, Baltimore, MD 21218, USA}

\author[0000-0002-8919-079X]{Leonardo Ferreira}
\affiliation{Centre for Astronomy and Particle Physics, School of Physics and Astronomy, University of Nottingham, NG7 2RD, UK}

\author[0000-0003-0792-5877]{Keely Finkelstein}
\affiliation{Department of Astronomy, The University of Texas at Austin, Austin, TX 78712, USA}

\author[0000-0001-8519-1130]{Steven Finkelstein}
\affiliation{Department of Astronomy, The University of Texas at Austin, Austin, TX 78712, USA}

\author[0000-0003-2098-9568]{Jonathan P. Gardner}
\affiliation{Astrophysics Science Division, NASA Goddard Space Flight Center, Greenbelt, MD 20771, USA}

\author[0000-0003-1530-8713]{Eric Gawiser}
\affiliation{Department of Physics and Astronomy, Rutgers, The State University of New Jersey, Piscataway, NJ 08854, USA}

\author[0000-0002-7732-9205]{Timothy Gburek}
\affiliation{Department of Physics and Astronomy, University of California, Riverside, Riverside, CA 92521, USA}

\author[0000-0002-7831-8751]{Mauro Giavalisco}
\affiliation{Department of Astronomy, University of Massachusetts, Amherst, MA 01003, USA}

\author[0000-0002-5688-0663]{Andrea Grazian}
\affiliation{INAF -- Osservatorio Astronomico di Padova, Vicolo dell'Osservatorio 5, I-35122, Padova, Italy}

\author[0000-0001-6842-2371]{Caryl Gronwall}
\affiliation{Department of Astronomy \& Astrophysics, The Pennsylvania State University, University Park, PA 16802, USA}

\author[0000-0003-2775-2002]{Yicheng Guo}
\affiliation{Department of Physics and Astronomy, University of Missouri, Columbia, MO 65211, USA}

\author[0000-0002-7959-8783]{Pablo Arrabal Haro}
\affiliation{NSF's NOIRLab, Tucson, AZ 85719, USA}

\author[0000-0003-2226-5395]{Shoubaneh Hemmati}
\affiliation{IPAC, Mail Code 314-6, California Institute of Technology, 1200 E. California Blvd., Pasadena, CA, 91125, USA}

\author[0000-0002-5924-0629]{Justin Howell}
\affiliation{IPAC, Mail Code 314-6, California Institute of Technology, 1200 E. California Blvd., Pasadena, CA, 91125, USA}

\author[0000-0003-1268-5230]{Rolf A. Jansen}
\affiliation{School of Earth and Space Exploration, Arizona State University, Tempe, AZ 85287, USA}

\author[0000-0001-7673-2257]{Zhiyuan Ji}
\affiliation{Department of Astronomy, University of Massachusetts, Amherst, MA 01003, USA}

\author[0000-0002-5601-575X]{Sugata Kaviraj}
\affiliation{Centre for Astrophysics Research, University of Hertfordshire, Hatfield, AL10 9AB, UK}

\author[0000-0001-6506-0293]{Keunho J. Kim}
\affiliation{IPAC, Mail Code 314-6, California Institute of Technology, 1200 E. California Blvd., Pasadena, CA, 91125, USA}

\author[0000-0002-8816-5146]{Peter Kurczynski}
\affiliation{Astrophysics Science Division, NASA Goddard Space Flight Center, Greenbelt, MD 20771, USA}

\author[0009-0000-1797-0300]{Ilin Lazar}
\affiliation{Department of Galaxies and Cosmology, Max Planck Institute for Astronomy, K\"{o}nigstuhl 17, 69117 Heidelberg}

\author[0000-0003-1581-7825]{Ray A. Lucas}
\affiliation{Space Telescope Science Institute, 3700 San Martin Drive, Baltimore, MD 21218, USA}

\author[0000-0001-6529-8416]{John MacKenty}
\affiliation{Space Telescope Science Institute, 3700 San Martin Drive, Baltimore, MD 21218, USA}

\author[0000-0002-6016-300X]{Kameswara Bharadwaj Mantha}
\affiliation{Minnesota Institute for Astrophysics, University of Minnesota, 116 Church St SE, Minneapolis, MN 55455, USA}

\author[0000-0002-6632-4046]{Alec Martin}
\affiliation{Department of Physics and Astronomy, University of Missouri, Columbia, MO 65211, USA}

\author[0000-0003-2939-8668]{Garreth Martin}
\affiliation{Korea Astronomy and Space Science Institute, Yuseong-gu, Daejeon 34055, Korea}
\affiliation{Steward Observatory, University of Arizona, Tucson, AZ 85719, USA}

\author[0000-0002-5506-3880]{Tyler McCabe}
\affiliation{School of Earth and Space Exploration, Arizona State University, Tempe, AZ 85287, USA}

\author[0000-0001-5846-4404]{Bahram Mobasher}
\affiliation{Department of Physics and Astronomy, University of California, Riverside, Riverside, CA 92521, USA}

\author[0000-0003-4965-0402]{Alexa M. Morales}
\affiliation{Department of Astronomy, The University of Texas at Austin, Austin, TX 78712, USA}

\author[0000-0002-8190-7573]{Robert O'Connell}
\affiliation{Department of Astronomy, University of Virginia, Charlottesville, VA 22904}

\author[0000-0002-8085-7578]{Charlotte Olsen}
\affiliation{Department of Physics and Astronomy, Rutgers, The State University of New Jersey, Piscataway, NJ 08854, USA}

\author[0000-0001-9665-3003]{Lillian Otteson}
\affiliation{School of Earth and Space Exploration, Arizona State University, Tempe, AZ 85287, USA}

\author[0000-0002-5269-6527]{Swara Ravindranath}
\affiliation{Space Telescope Science Institute, 3700 San Martin Drive, Baltimore, MD 21218, USA}

\author[0000-0002-9961-2984]{Caleb Redshaw}
\affiliation{School of Earth and Space Exploration, Arizona State University, Tempe, AZ 85287, USA}

\author[0000-0001-7016-5220]{Michael Rutkowski}
\affiliation{Department of Physics and Astronomy, Minnesota State University Mankato, Mankato, MN 56001, USA}

\author[0000-0002-4271-0364]{Brant Robertson}
\affiliation{Department of Astronomy and Astrophysics, University of California, Santa Cruz, Santa Cruz, CA 95064, USA}

\author[0000-0002-0364-1159]{Zahra Sattari}
\affiliation{Department of Physics and Astronomy, University of California, Riverside, Riverside, CA 92521, USA}

\author[0000-0002-2390-0584]{Emmaris Soto}
\affiliation{Computational Physics, Inc., Springfield, VA 22151, USA}

\author[0009-0004-6325-7839]{Lei Sun}
\affil{School of Astronomy and Space Science, University of Chinese Academy of Sciences (UCAS), Beijing 100049, China}

\author[0000-0003-0749-4667]{Sina Taamoli}
\affiliation{Department of Physics and Astronomy, University of California, Riverside, Riverside, CA 92521, USA}

\author[0000-0002-5057-135X]{Eros Vanzella}
\affiliation{INAF -- Osservatorio di Astrofisica e Scienza dello Spazio di Bologna, via Gobetti 93/3, I-40129 Bologna, Italy}

\author[0000-0003-3466-035X]{L. Y. Aaron Yung}
\affiliation{Space Telescope Science Institute, 3700 San Martin Drive, Baltimore, MD 21218, USA}

\author[0000-0002-7830-363X]{Bonnabelle Zabelle}
\affiliation{Minnesota Institute for Astrophysics, University of Minnesota, 116 Church St SE, Minneapolis, MN 55455, USA}

\author{the UVCANDELS team}

\begin{abstract}
The UltraViolet imaging of the Cosmic Assembly Near-infrared Deep Extragalactic Legacy Survey Fields (UVCANDELS) program provides deep \textit{HST} F275W and F435W imaging over four CANDELS fields (GOODS-N, GOODS-S, COSMOS, and EGS). We combine this newly acquired UV imaging with existing \textit{HST} imaging from CANDELS as well as existing ancillary data to obtain robust photometric redshifts and reliable estimates for galaxy physical properties for over 150,000 galaxies in the $\sim$430~arcmin$^2$ UVCANDELS area. Here, we leverage the power of the new UV photometry to not only improve the photometric redshift measurements in these fields, but also constrain the full redshift probability distribution combining multiple redshift fitting tools. Furthermore, using the full UV--to--IR photometric dataset, we measure the galaxy physical properties by fitting templates from population synthesis models with two different parameterizations (flexible and fixed-form) of the star-formation histories (SFHs). Compared to the flexible SFH parametrization, we find that the fixed-form SFHs systematically underestimate the galaxy stellar masses, both at the low- ($\lesssim10^9~M_\odot$) and high- ($\gtrsim10^{10}~M_\odot$) mass end, by as much as $\sim0.5$~dex. This underestimation is primarily due the limited ability of fixed-form SFH parameterization to simultaneously capture the chaotic nature of star-formation in these galaxies.
\end{abstract}

\keywords{catalogs --- galaxies: redshifts --- galaxies: fundamental parameters --- methods: observational --- techniques: photometric}

\section{Introduction}
\label{sec:intro}

Over the past few decades, our knowledge of galaxy formation and evolution has seen significant advancement and large multi-wavelength photometric surveys have been one of the cornerstones enabling this progress. Space-based facilities such as \textit{HST}, \textit{Spitzer}, and now \textit{JWST} have granted access not only to high-resolution and extremely sensitive imaging, but also to wavelengths that are otherwise inaccessible from the ground. The combination of space- and ground-based observations covering wavelengths from the far ultraviolet (UV) out to mid-infrared (IR) have facilitated detailed studies of galaxies and the physical processes that govern how they grow and evolve over time -- namely, tracing the evolution of the stellar mass function over time \citep[e.g., ][]{Marchesini:2009,Muzzin:2013,Davidzon:2017,Weaver:2023,Weibel:2024}, the cosmic star-formation rate and its evolution \citep[e.g., ][]{Madau:2014,Alavi:2014,Alavi:2016,Mehta:2017,Moutard:2020,Picouet:2023}, the star-forming main sequence \citep[e.g., ][]{Speagle:2014,Whitaker:2014,Kurczynski:2016,Boogaard:2018,Merida:2023}, and the mass-size relation \citep[e.g., ][]{Shen:2003,Baldry:2012,Morishita:2017,Nedkova:2021,Nedkova:2024}, among others. Particularly, the WFC3/UVIS instrument on-board \textit{HST} has been monumental by providing access to the rest-frame UV light from galaxies which is key for studying their star-formation properties and hence, tracking their growth as well as for constraining the amount of dust in them.

One such pivotal photometric survey has been the Cosmic Assembly Near-Infrared Deep Extragalactic Legacy Survey \citep[CANDELS; ][]{Grogin:2011,Koekemoer:2011}, which is a legacy \textit{HST} program that obtained WFC3/IR and ACS/WFC optical imaging in up to 10 broadband filters for five extragalactic fields (GOODS-N, GOODS-S, COSMOS, EGS, and UDS) over a combined area of $\sim$0.2~deg$^2$. Complemented with additional observations with \textit{Spitzer} and various other ground-based instruments, the CANDELS fields serve as one of the premier observational datasets for studying galaxy evolution and have already enabled a wide range of science. The CANDELS team has previously assembled catalogs that include multi-wavelength photometric coverage for the \textit{HST} as well as all ancillary data available on these fields \citep{Guo:2013,Nayyeri:2017,Stefanon:2017,Barro:2019}. 

Recently, another \textit{HST} Treasury campaign, UVCANDELS \citep[PI: H. Teplitz; PID: 15647; ][]{Wang:2024}, accomplished the task of obtaining UV imaging with F275W and F435W (in parallel) over four of the five CANDELS fields (GOODS-N, GOODS-S, COSMOS, EGS). The new F435W coverage in COSMOS and EGS fields overlaps with that of F275W which previously did not have F435W; while in GOODS-N and GOODS-S, the parallel F435W coverage is limited to the deeper regions (not overlapping with F275W) given the existing F435W data already available for the UVCANDELS regions. The additional UV coverage improves the photometric redshift estimates, particularly for objects that have degenerate solutions, typically characterized as multiple peaks in their photometric redshift probability distributions that are significant (e.g., $\gtrsim$10\%\ of the primary peak) and sufficiently separated in redshift (e.g., $\Delta z/(1+z)$$\gtrsim$0.06), which when not appropriately accounted for can lead to catastrophic errors in the measured redshifts and associated uncertainties \citep[e.g., ][]{Rafelski:2009, Rafelski:2015}. The UV imaging helps prevent this for example by sampling the Lyman break of $z\sim2$ galaxies removing the degeneracy with the Balmer break of $z\sim0.3$ galaxies. Similarly, the UV photometry is critical to accurately constrain the recent star-formation activity and dust content of galaxies \citep[e.g., ][]{Mehta:2017, Mehta:2023}. In this work, we leverage the existing multi-wavelength CANDELS photometry with the addition of the new F275W and F435W photometry from UVCANDELS to provide improved photometric redshifts as well as galaxy physical properties for over 150,000 galaxies in these four CANDELS fields.

Estimating both photometric redshifts and physical parameters via modeling the galaxy spectral energy distribution (SED) has been well established with a vast library of tools with varying levels of sophistication now available for both \citep[e.g., ][]{Rafelski:2015,Mehta:2018,Kodra:2023,Pacifici:2023}. However, variations in template choices and codes result in slightly different results \citep[e.g., ][]{Dahlen:2013}. In this work, we leverage the power of multiple photometric redshift fitting tools and template sets and combine their results to yield robust redshift estimates as well as to accurately quantify the redshift probability distributions. When estimating the galaxy physical properties, the star-formation history (SFH) plays an essential role in determining the galaxy SED \citep[e.g., ][]{Iyer:2019, Leja:2019} and traditionally, SFHs in typical SED fitting tools have been assumed to have a fixed, functional forms. However, there is increasing evidence suggesting that galaxy SFHs, particularly in the early universe as well as toward the low-mass end, are not smoothly evolving, rather they are chaotic and often bursty in nature, both from a theoretical \citep[see e.g.,][]{Hopkins:2014,Dominguez:2015,Sun:2023} as well as observational perspectives \citep[see e.g.][]{Weisz:2012,Emami:2019,Mehta:2017,Mehta:2023}. In this work, we explore the impact of the traditional fixed-form SFH assumption for even the most basic physical parameter -- i.e., galaxy stellar mass -- by comparing to results from fitting tools that accommodate more flexible forms for the galaxy SFH.

This manuscript is organized as follows: Section~\ref{sec:data} describes the UVCANDELS dataset as well as the ancillary photometric datasets used for this analysis; Section~\ref{sec:photoz} describes the techniques used for estimating photometric redshifts and discusses the improvement in the redshift estimates by the inclusion of the UV photometry; Section~\ref{sec:phys_pars} describes the methodologies used for the measurements of galaxy physical parameters and discusses the impact of the fixed vs. flexible-form for the galaxy SFHs on the estimated physical parameters; Section~\ref{sec:catalogs} describes the final output catalogs that present the various measurements from this work and presents the publicly released catalogs; and Section~\ref{sec:summary} summarizes our findings.

Throughout this paper, we adopt cosmological parameters from Table 3 of \citet{Planck:2016}: $\Omega_m=0.315$, $\Omega_\lambda=0.685$ and $H_0=67.31$~km~s$^{-1}$~Mpc$^{-1}$ and all magnitudes used are AB magnitudes \citep{Oke:1983}.

\section{Data}
\label{sec:data}
\subsection{UVCANDELS Imaging}

The Ultraviolet Imaging of the Cosmic Assembly Near-infrared Deep Extragalactic Legacy Survey fields \citep[UVCANDELS; PI: H. Teplitz; PID: 15647; ][]{Wang:2024} is a Cycle~26 \textit{Hubble} Treasury program that obtained WFC3/UVIS F275W and ACS/WFC F435W (in parallel) for four of the deep-wide survey fields defined by CANDELS \citep{Grogin:2011,Koekemoer:2011}, namely GOODS-N, GOODS-S, COSMOS, and EGS, covering a combined area of $\sim$430~arcmin$^2$. The UVCANDELS imaging reaches an depth of AB=27 for compact galaxies in the WFC3/UVIS F275W filter, and AB$\le$28 in ACS/F435W. The UVCANDELS F275W mosaics are available at the Mikulski Archive for Space Telescopes (MAST; \dataset[doi:10.17909/8s31-f778]{https://dx.doi.org/10.17909/8s31-f778})\footnote{\url{https://archive.stsci.edu/hlsp/uvcandels}} for all four fields along with F435W mosaics for COSMOS and EGS\footnote{The F435W mosaics for GOODS-N and GOODS-S covered by UVCANDELS are already available from CANDELS.}. These UVCANDELS image mosaics are aligned and registered to the CANDELS astrometry. With the newly acquired UVCANDELS imaging, we have measured the fluxes in the F275W filter for the four fields and F435W for COSMOS and EGS. See \citet{Wang:2024} and \citet{Sun:2024} for full details on the methodology to generate the mosaics and photometric catalogs, respectively.

\subsection{Photometric Datasets}
\label{sec:photometry}
The photometric dataset used for modeling the galaxy spectral energy distributions (SEDs) in this work are primarily based on the publicly available CANDELS catalogs\footnote{\url{https://archive.stsci.edu/hlsp/candels}}, with the addition of the new F275W and F435W UVCANDELS photometry. Specifically, the CANDELS multi-wavelength photometric catalogs that we use for GOODS-N, GOODS-S, COSMOS and EGS fields are described and presented in \citet{Barro:2019}, \citet{Guo:2013}, \citet{Nayyeri:2017}, and \citet{Stefanon:2017}, respectively. Similarly, the photometric techniques used for measuring the F275W and F435W fluxes as well as the catalogs from UVCANDELS are fully described in detail in \citet{Sun:2024}. These F275W and F435W fluxes from UVCANDELS are measured in a consistent fashion as the rest of the CANDELS photometry (see Section~\ref{sec:VtoH} for more details). For completeness, we briefly summarize the various surveys that make up the full photometric dataset for each field in the subsections below as well as in Table~\ref{tab:filters}.

\begin{deluxetable*}{c l X{0.2\linewidth} X{0.48\linewidth}}
\centering
\tablecaption{The set of photometric filters included for SED modeling, both when measuring the photometric redshifts as well as the galaxy physical parameters\label{tab:filters}}
\tablewidth{\linewidth}
\tabletypesize{\footnotesize}
\tablehead{\colhead{Field} & \colhead{Instrument} & \colhead{Filter} & \colhead{Survey/Reference}}
\startdata
  \multirow{9}{*}{GOODS-N\tblmark{a}}  & \textit{HST}/UVIS      & \textbf{F275W}\tblmark{b}    & \textbf{UVCANDELS} \citep{Wang:2024,Sun:2024} \\
  \quad                                & \textit{HST}/ACS       & F435W, F606W, F775W, F814W, F850LP             & GOODS \citep{Giavalisco:2004}, CANDELS \citep{Grogin:2011,Koekemoer:2011}, \citet{Riess:2007} \\
  \quad                                & \textit{HST}/WFC3      & F105W, F125W, F140W, F160W                     & CANDELS \citep{Grogin:2011,Koekemoer:2011}, AGHAST \citep{Weiner:2009} \\
  \quad                                & \textit{LBT}/LBC       & $U'$                                           & \citet{Grazian:2017} \\
  \quad                                & \textit{Subaru}/MOIRCS & $K_s$                                          & \citet{Kajisawa:2011} \\
  \quad                                & \textit{CFHT}/WIRCam   & $K_s$                                          & \citet{Hsu:2019} \\
  \quad                                & \textit{Spitzer}/IRAC  & ch. 1, 2, 3, 4                                 & SEDS \citep{Ashby:2013}, S-CANDELS \citep{Ashby:2015}, Spitzer-GOODS \citep{Dickinson:2003} \\
\tableline
  \multirow{10}{*}{GOODS-S\tblmark{a}} & \textit{HST}/UVIS         & \textbf{F275W}\tblmark{b}    & \textbf{UVCANDELS} \citep{Wang:2024,Sun:2024} \\
  \quad                                & \textit{HST}/ACS          & F435W, F606W, F775W, F814W, F850LP             & GOODS \citep{Giavalisco:2004}, CANDELS \citep{Grogin:2011,Koekemoer:2011}, \citet{Riess:2007} \\
  \quad                                & \textit{HST}/WFC3         & F098M, F105W, F125W, F160W                     & CANDELS \citep{Grogin:2011,Koekemoer:2011}, HUDF \citep{Bouwens:2010}, ERS \citep{Windhorst:2011} \\
  \quad                                & \textit{Blanco}/MOSAIC~II & $U$                                            & \citet{Guo:2013} \\
  \quad                                & \textit{VLT}/VIMOS        & $U$                                            & \citet{Nonino:2009} \\
  \quad                                & \textit{VLT}/ISAAC        & $K_s$                                          & \citet{Retzlaff:2010} \\
  \quad                                & \textit{VLT}/HAWK-I       & $K_s$                                          & \citet{Fontana:2014} \\
  \quad                                & \textit{Spitzer}/IRAC     & ch. 1, 2, 3, 4                                 & SEDS \citep{Ashby:2013}, Spitzer-GOODS \citep{Dickinson:2003} \\
\tableline
  \multirow{8}{*}{COSMOS\tblmark{a}}   & \textit{HST}/UVIS           & \textbf{F275W}\tblmark{b}    & \textbf{UVCANDELS} \citep{Wang:2024,Sun:2024} \\
  \quad                                & \textit{HST}/ACS            & \textbf{F435W}\tblmark{b}, F606W, F814W  & \textbf{UVCANDELS} \citep{Wang:2024,Sun:2024}, CANDELS \citep{Grogin:2011,Koekemoer:2011} \\
  \quad                                & \textit{HST}/WFC3           & F125W, F160W                                   & CANDELS \citep{Grogin:2011,Koekemoer:2011} \\
  \quad                                & \textit{CFHT}/MegaPrime     & $ugriz$                                        & CHFT-LS \citep{Gwyn:2012} \\
  \quad                                & \textit{Subaru}/Suprime-Cam & $Bg^+Vr^+i^+z^+$                               & \citet{Taniguchi:2007, Taniguchi:2015} \\
  \quad                                & \textit{VISTA}/VIRCAM       & $YJHK_s$                                       & UltraVISTA \citep{McCracken:2012} \\
  \quad                                & \textit{Mayall}/NEWFIRM     & $J1,J2,J3,H1,H2,K$                             & NMBS \citep{Whitaker:2011} \\
  \quad                                & \textit{Spitzer}/IRAC       & ch. 1, 2, 3, 4                                 & SEDS \citep{Ashby:2013}, S-COSMOS \citep{Sanders:2007} \\
\tableline
  \multirow{9}{*}{EGS\tblmark{a}}      & \textit{HST}/UVIS       & \textbf{F275W}\tblmark{b}    & \textbf{UVCANDELS} \citep{Wang:2024,Sun:2024} \\
  \quad                                & \textit{HST}/ACS        & \textbf{F435W}\tblmark{b}, F606W, F814W  & \textbf{UVCANDELS} \citep{Wang:2024,Sun:2024}, CANDELS \citep{Grogin:2011,Koekemoer:2011} \\
  \quad                                & \textit{HST}/WFC3       & F125W, F140W, F160W                            & CANDELS \citep{Grogin:2011,Koekemoer:2011}, 3D-HST \citep{Brammer:2012,Skelton:2014} \\
  \quad                                & \textit{CFHT}/MegaCam   & $ugriz$                                        & CFHT-LS \citep{Gwyn:2012} \\
  \quad                                & \textit{CFHT}/WIRCam    & $JHK_s$                                        & WIRDS \citep{Bielby:2012} \\
  \quad                                & \textit{Mayall}/NEWFIRM & $J1~J2~J3~H1~H2~K$                             & NMBS \citep{Whitaker:2011} \\
  \quad                                & \textit{Spitzer}/IRAC   & ch. 1, 2, 3, 4                                 & SEDS \citep{Ashby:2013}, S-CANDELS \citep{Ashby:2015}, AEGIS \citep{Barmby:2008}
\enddata
\tbltext{a}{All photometry except for that provided by UVCANDELS is from the publicly available CANDELS multi-wavelength photometric catalogs -- i.e., \citet{Barro:2019} for GOODS-N; \citet{Guo:2013} for GOODS-S; \citet{Nayyeri:2017} for COSMOS; \citet{Stefanon:2017} for EGS.}
\tbltext{b}{Photometry for the F275W (all four fields) and F435W (COSMOS and EGS) bands is the new addition from UVCANDELS \citep{Wang:2024,Sun:2024}.}
\end{deluxetable*}

\subsubsection{GOODS-N}
We use the photometric catalog from \citet{Barro:2019} for the GOODS-N field \citep{Giavalisco:2004}. For the SED modeling, we consider the photometry for \textit{HST}/ACS optical imaging in F435W, F606W, F775W, F814W, and F850LP from the GOODS \textit{HST}/ACS Treasury Program \citep{Giavalisco:2004}, the CANDELS survey \citep{Grogin:2011,Koekemoer:2011}, as well as the search for high-redshift Type Ia supernovae \cite[e.g., ][]{Riess:2007}. We include photometry for \textit{HST}/WFC3 imaging in the NIR filters F105W, F125W, and F160W from the CANDELS survey as well as additional F140W coverage from the G141 AGHAST survey \citep{Weiner:2009}. We further include the new \textit{HST}/WFC3 F275W photometry available from the UVCANDELS imaging \citep{Wang:2024,Sun:2024}.

In addition to the \textit{HST} imaging, we also use the ground-based \textit{LBT}/LBC $U'$-band photometry from \citet{Grazian:2017} as well as the $K_s$-band photometry from \textit{Subaru}/MOIRCS \citep{Kajisawa:2011} and \textit{CFHT}/WIRCam \citep{Hsu:2019}. Lastly, we include \textit{Spitzer}/IRAC ch. 1 (3.6~\micron), 2 (4.5~\micron), 3 (5.8~\micron), 4 (8.0~\micron) photometry available from a combination of the \textit{Spitzer} Extended Deep Survey \citep[SEDS, ][]{Ashby:2013}, the \textit{Spitzer}-CANDELS \citep[S-CANDELS, ][]{Ashby:2015}, and \textit{Spitzer}-GOODS \citep{Dickinson:2003} surveys.

\subsubsection{GOODS-S}
\citet{Guo:2013} present in full detail the photometric catalog that we use for the GOODS-S field. Similar to GOODS-N, the photometry for \textit{HST}/ACS F435W, F606W, F775W, F814W, and F850LP is provided from the GOODS \textit{HST}/ACS Treasury Program, the CANDELS survey, as well as the search for high-redshift Type Ia supernovae. The \textit{HST}/WFC3 photometry for F098M, F105W, F125W, and F160W is provided from the CANDELS, HUDF \citep{Bouwens:2010}, and ERS \citep{Windhorst:2011} programs. Critically, we add the new \textit{HST}/WFC3 F275W photometry from UVCANDELS \citep{Wang:2024,Sun:2024}.

Furthermore, photometry from ground-based imaging in the U-band from \textit{Blanco}/MOSAIC~II and \textit{VLT}/VIMOS \citep{Nonino:2009} is included as well as the $K_s$-band from \textit{VLT}/ISAAC \citep{Retzlaff:2010} and \textit{VLT}/HAWK~I \citep{Fontana:2014}. Lastly, \textit{Spitzer}/IRAC ch. 1, 2, 3, 4 photometry available from a combination of SEDS \citep{Ashby:2013}, and the \textit{Spitzer}-GOODS \citep{Dickinson:2003} survey is also included.

\subsubsection{COSMOS}
For the COSMOS field \citep{Scoville:2007}, we utilize the catalog from \citet{Nayyeri:2017} for all photometry except for \textit{HST}/WFC3 F275W and \textit{HST}/ACS F435W which are provided from UVCANDELS \citep{Wang:2024,Sun:2024}. We include the photometry in \textit{HST}/ACS imaging filters F606W, F814W as well as in \textit{HST}/WFC3 NIR filters F125W, F140W from the CANDELS survey.

There exists a wealth of ancillary ground-based data for the COSMOS field. Here, we include the photometry from CFHT-LS $ugriz$-bands \citep{Gwyn:2012}, \textit{Subaru}/Suprime-Cam $Bg^+Vr^+i^+z^+$ from \citet{Taniguchi:2015}, and \textit{VISTA}/VIRCAM $YJHK_s$ from the UltraVISTA survey \citep{McCracken:2012}. We also include the medium-band photometry from the NEWFIRM Medium Band Survey \citep[NMBS, ][]{Whitaker:2011} for \textit{Mayall}/NEWFIRM $J1,J2,J3,H1,H2,K$-bands. Lastly, the \textit{Spitzer}/IRAC ch. 1, 2, 3, 4 photometry from the S-COSMOS survey \citep{Sanders:2007} is also used.

\subsubsection{EGS}
The catalog from \citet{Stefanon:2017} is used for all photometry in the EGS field except for \textit{HST}/WFC3 F275W and \textit{HST}/ACS F435W filters which are provided from the UVCANDELS imaging \citep{Wang:2024,Sun:2024}. We include the photometry for \textit{HST}/ACS F606W and F814W imaging taken as part of the AEGIS project and the CANDELS survey. The \textit{HST}/WFC3 photometry available in the NIR filters F125W, F140W, and F160W is contributed from a combination of the CANDELS and the 3D-HST \cite{Brammer:2012} surveys.

Additionally, photometry from ground-based imaging in $ugriz$ from CFHT-LS \citep{Gwyn:2012} is also included along side \textit{CFHT}/WIRCam $JHK_s$ imaging from the WIRCam Deep Survey \citep[WIRDS, ][]{Bielby:2012}. Similar to COSMOS, we include the medium-band imaging in the \textit{Mayall}/NEWFIRM $J1,J2,J3,H1,H2,K$-bands from NMBS \citep{Whitaker:2011}. Lastly, we include the \textit{Spitzer}/IRAC ch. 1, 2, 3, 4 photometry from the SEDS \citep{Ashby:2013}, S-CANDELS \citep{Ashby:2015}, and AEGIS \citep{Barmby:2008}.

\subsection{Photometric Measurements}
\label{sec:VtoH}
While we refer the reader to the respective publications for the exact details of the photometric techniques used to assemble the multi-wavelength catalogs, we briefly summarize the pertinent details here. For all the CANDELS catalogs considered here, the object detection is performed on the F160W-band image. The isophotes defined on the F160W image are then used to measure the photometry in the remaining bands. For the high-resolution data (i.e., all HST bands), the point-spread function (PSF) for each image is convolved to match the F160W resolution, which has the broadest PSF amongst the HST filters. For the low-resolution images (i.e., ground-based and \textit{Spitzer}/IRAC), the photometry was performed using the TFIT software \citep{Laidler:2007}, which uses a morphological template-fitting technique.

For the F275W/F435W photometry added from the UVCANDELS imaging \citep{Wang:2024,Sun:2024}, the F275W and F435W magnitudes are computed using UV-optimized isophotes following the methodology from the Hubble Ultra-Deep Field \citep[UVUDF, ][]{Teplitz:2013} UV analysis from \citet{Rafelski:2015}. Specifically, instead of following the typical approach of performing PSF-matched photometry using the isophotes defined on images with the broadest PSF (typically F160W for HST imaging data), we use isophotes defined on the F606W image instead to measure the fluxes  from the unconvolved F275W/F435W images (without matching PSF), followed by applying an adjustment factor to correct for the aperture and PSF differences to match back to the F160W isophote-based photometry. This has the advantage of having isophotes/apertures that are more appropriately matched to the detected UV sizes, and thus return significantly better signal-to-noise ratio (SNR) for the measured fluxes. In order to keep the photometry consistent with the rest from CANDELS, these F606W-defined isophotes are matched and reconciled back to the F160W-isophotes used by the CANDELS photometric catalogs. This is described as the photometry from the "VtoH" segmentation maps in \citet{Sun:2024}, which also provides the full detailed description of this photometric technique and methodology.

For the SED modeling presented in this work, we use the photometry from a combination of the above-described CANDELS and UVCANDELS catalogs. Before proceeding with the fitting, we correct the photometry for galactic extinction using the \citet{Schlegel:1998} dust maps\footnote{Specifically, we use the Python implementation from \url{https://github.com/adrn/SFD} to query the \citet{Schlegel:1998} maps.}. The reddening $E(B-V)$ is queried at the position of each galaxy and converted to an extinction assuming a \citet{Cardelli:1989} extinction law, for the respective filters.

\section{Photometric Redshifts}
\label{sec:photoz}

There are many different photometric redshift codes available, and they all yield slightly different results when run on the same data \citep[e.g.][]{Dahlen:2013, Kodra:2023}. In this work, we opt to utilize multiple codes and combine the results, which yields more robust redshifts than any individual code due to different systematic uncertainties in the codes and choice of template spectra. Specifically, we calculate photometric redshifts by combining the results from four different codes: \eazy\ \citep{Brammer:2008}, \bpz\ \citep{Coe:2006}, \lephare\ \citep{Arnouts:1999,Ilbert:2006}, and \zphot\ \citep{Giallongo:1998,Fontana:2000}. These codes were chosen as they were consistently among the top performers in photometric redshift review papers of CANDELS fields \citep{Hildebrandt:2010,Dahlen:2013}. 

We run two separate iterations of \eazy\ \citep{Brammer:2008} to give us a total of 5 independent code run results. The first \eazy\ iteration uses the default template spectra (\texttt{eazy\_v1.2\_dusty}) which includes the 6 original templates from \citet{Brammer:2008}, emission lines from \citet{Ilbert:2009}, an old and red SED from \citet{Maraston:2005}, and a dusty SED from \citet{Bruzual:2003}. The second \eazy\ iteration uses the five templates from \citet{Blanton:2007} (\texttt{br07\_default}). Both template sets use a non-negative matrix factorization to optimally select a reduced set of spectral templates that reproduce observed data \citep{Blanton:2007}. The first template set is optimized for high redshift galaxies from theoretical models, while the second is optimized for lower redshifts based on empirical SDSS data. For both iterations, \eazy\ is restricted to a redshift range 0 to 12.

For \bpz\ \citep{Benitez:2000,Benitez:2004,Coe:2006}, we use the improved software and procedure described in \citet{Rafelski:2015}, which includes 11 template SEDs based on those from PEGASE \citep{Fioc:1997} but re-calibrated based on photometric and spectroscopic redshifts from FIREWORKS \citep{Wuyts:2008}, and a prior based on luminosity functions observed in COSMOS \citep{Ilbert:2009}, GOODS-MUSIC \citep{Grazian:2006,Santini:2009}, and the UDF \citep{Coe:2006}. 

The other two codes were used with their default template sets. 
For \lephare\ \citep{Arnouts:1999,Ilbert:2006}, we use the 32 COSMOS SEDs described in \citet{Ilbert:2009}, and correct for emission lines and extinction using the reddening laws from \citet{Calzetti:2000} and \citet{Prevot:1984}. For \zphot\ \citep{Giallongo:1998,Fontana:2000}, we use SED templates from \citet{Bruzual:2003} with the addition of \citet{Calzetti:2000} extinction and \citet{Fan:2006} absorption due to the inter-galactic medium (IGM).

We note that sometimes photometric redshifts are calculated iteratively, implementing zero-point offsets to the photometry or modifying the templates based on the spectroscopic redshifts \citep[e.g.][]{Barro:2019}. This can often improve the photometric redshift accuracy, although a number of the templates used in this study are already optimized based on empirical data. We found that, on average, the dispersion of the redshifts were improved by modifying the zero-points, but that it worsened the results at $z\gtrsim3$. Also, each code would need a different set of zero-point offsets, which would add a level of inconsistency in the input photometry. We therefore opted not to modify the zeropoints or modify the templates based on the spectroscopic redshifts. 

\subsection{Spectroscopic Redshifts}
In order to test the accuracy of our photometric redshifts as well as to combine the results from the individual codes together, we require a highly vetted spectroscopic redshift catalog. We use a compilation of publicly available spectroscopic redshifts obtained from various facilities/sources in the UVCANDELS fields. The final spectroscopic reference sample used in this analysis is assembled from a combination of the 3D-HST \citep{Brammer:2012,Momcheva:2016}, zCOSMOS \citep{Lilly:2007,Lilly:2009}, hCOSMOS \citep{Damjanov:2018}, PRIMUS \citep{Coil:2011}, MOSDEF \citep{Kriek:2015}, DEIMOS 10K \citep{Hasinger:2018}, DEEP2 \citep{Newman:2013}, DEEP3 \citep{Cooper:2011,Cooper:2012}, MUSE-Wide \citep{Herenz:2017}, MUSE-HUDF \citep{Inami:2017}, K20 \citep{Mignoli:2005}, C3R2 \citep{Masters:2019,Stanford:2021}, LEGA-C \citep{vanderWel:2016}, VVDS \citep{LeFevre:2013}, VUDS \citep{LeFevre:2015}, VANDELS \citep{McLure:2018,Pentericci:2018}, GMASS \citep{Kurk:2013}, FMOS-COSMOS \citep{Silverman:2015}, FIREWORKS \citep{Wuyts:2008}, GOODS-MUSIC \citep{Grazian:2006}, PEARS \citep{Straughn:2009, Ferreras:2009}, GRAPES-HUDF \citep{Hathi:2009, Pasquali:2006}, LCIRS \citep{Doherty:2005} spectroscopic surveys as well as compilations from the following publications: \citet{Balestra:2010, Barger:2008, Cristiani:2000, Croom:2001, Daddi:2004, Huang:2009, Krogager:2014, Ravikumar:2007, Roche:2006, Strolger:2004, Treister:2009, Trump:2009, Trump:2011, Trump:2013, Trump:2015, Vanzella:2008, Vanzella:2009, vanderWel:2005, Wirth:2015, Wolf:2004, Wuyts:2009, Yoshikawa:2010}, including some spectroscopic redshifts provided to our team via private communication.

For our spectroscopic reference sample, we only include those redshifts with the highest data quality flags from the individual papers, and those 3D-HST grism redshifts with good grism data quality flags (\texttt{use\_zgrism=1}). We omit sources with X-ray detections to avoid AGN contaminants, \citep[][D. Kocevski 2023, private communication]{Xue:2016}, which amounts to a total of 858 objects excluded. To avoid potential confusion from neighbors in ground-based spectroscopy, we also exclude any sources that have a relatively bright neighbor, defined as a source within 3\arcsec\ and within 2 mag or brighter in F160W. As a final measure to ensure only the highest quality redshifts remain, we visually inspect all sources with photometric redshift outliers (defined as $|z_{\rm spec}-z_{\rm phot}| / (1+z_{\rm spec}) > 0.15$) and remove those with anomalous behavior affecting the input photometry. These are typically sources contaminated by very bright, extended sources not caught by the automated neighbor cut, but also include sources with other contaminants, including stars or diffraction spikes. The final visual vetting removes ~0.5\% of all available redshifts. After the vetting procedure, we compile a spectroscopic redshift catalog including 8,081 redshifts in total over the four CANDELS fields. 

The spectroscopic redshifts from this compilation (limited to those available publicly) are included in the photometric redshift catalog released as part of this work (see Section~\ref{sec:catalogs} and Table~\ref{tab:photz_columns}).

\begin{figure}[!b]
\centering
\includegraphics[width=\columnwidth]{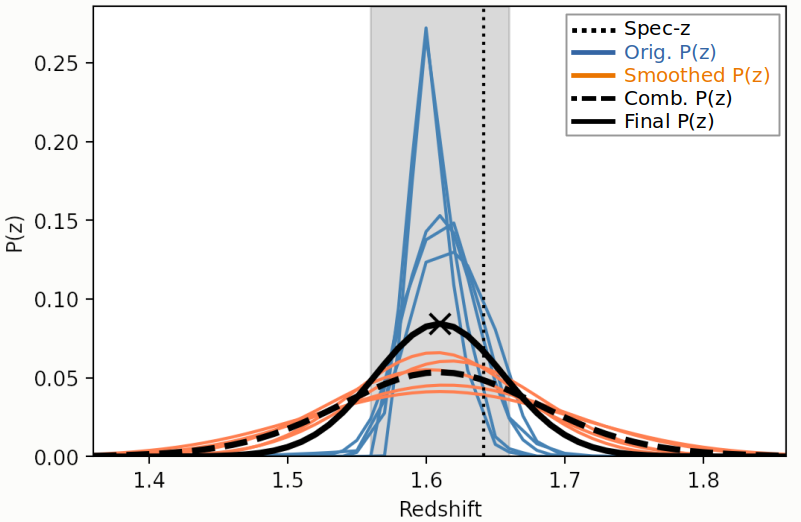}
\caption{An example of the photometric redshift probability distribution combination process for an example source (COSMOS-697: $z_{spec}=1.6418$; $z_{phot}=1.61\pm0.05$). \textit{Blue curves} show the $P(z)$’s from the individual codes, \textit{orange curves} show these $P(z)$’s after smoothing, and the \textit{black curves} show the final combined $P(z)$ before (\textit{dashed}) and after (\textit{solid}) the final sharpening procedure. The final photometric redshift (\textit{cross symbol}) and confidence intervals (\textit{shaded gray region}) of the final $P(z)$ are shown, along with the spectroscopic redshift (\textit{dotted line}).}
\label{fig:probfunction}
\end{figure}

\subsection{Combining Photometric Redshifts}
In order to combine the results from the 5 individual runs into a single estimate for the photometric redshift, we add together the individual probability distribution functions, $P(z)$, following the procedure outlined in \cite{Dahlen:2013} (see their Section~{5.2}). 

Before combining the individual $P(z)$, it is imperative to ensure that the photometric redshift accuracy estimated from the individual codes is consistent and representative of the actual sample of galaxies for which the photometric redshifts are derived. One common check is to compare the $1\sigma$ uncertainty from the $P(z)$ with that derived directly from the offsets with respect to the spectroscopic reference sample \citep[e.g., ][]{Ilbert:2009}. In the cases where the $P(z)$ underestimates (overestimates) the statistical errors from the spectroscopic sample comparison, a smoothing (sharpening) can be applied to the $P(z)$ to ensure that the error estimates on the photometric redshifts are accurate and consistent \citep[e.g., ][]{Dahlen:2013}.

Similar to \citet{Dahlen:2013}, we find that the individual codes used in our analysis underestimate their confidence intervals, i.e. fewer than 68.3\%\ (1$\sigma$) of galaxies with known spectroscopic redshifts fall within their 68.3\% confidence intervals from the photometric redshift probability distributions $P(z)$. Hence, to alleviate this, we iteratively smooth each $P(z)_i$ for code $i$ and redshift bin $j$ by
\begin{equation}
    P(z_j)_i=0.25P(z_j-1)_i+0.5P(z_j)_i+0.25P(z_j+1)_i
\end{equation}
until 68.3\%\ of the known spectroscopic redshifts fall within the smoothed $P(z)$ 68.3\%\ confidence intervals. We perform this smoothing procedure for each code separately, as each code requires a different number of smoothing iterations. Next, we add the smoothed probability distributions from each code together and renormalize. Similar to \citet{Dahlen:2013}, we find that these resulting probability distributions after combining tend to overestimate the confidence intervals, and hence we apply a sharpening according to ${P(z_j)_i}^{1/\alpha}$, where the exponent $\alpha$ is chosen to ensure 68.3\% of the spectroscopic redshifts fall within the 68.3\% confidence intervals. This results in a single $P(z)$ that combines the knowledge for all individual fitting codes with a corresponding error (or confidence interval) that is statistically consistent with the spectroscopic reference sample used for calibration. The final probability distributions are on a redshift grid of $z=0-12$ in steps of 0.01. Figure~\ref{fig:probfunction} shows an example of this combination procedure for an individual source.

\begin{figure*}[!ht]
\centering
\includegraphics[width=0.4\textwidth]{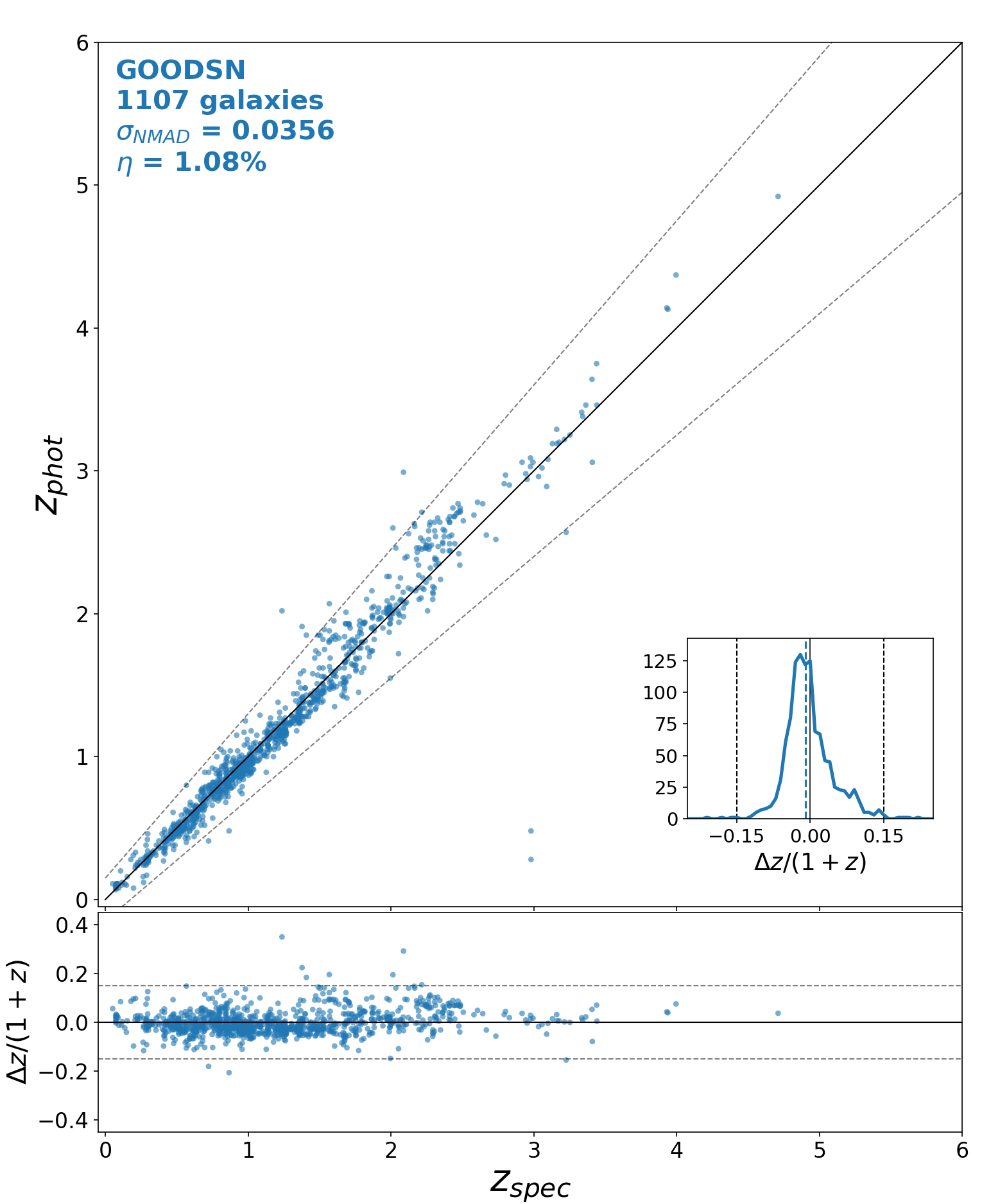} \hspace{0.3in}
\includegraphics[width=0.4\textwidth]{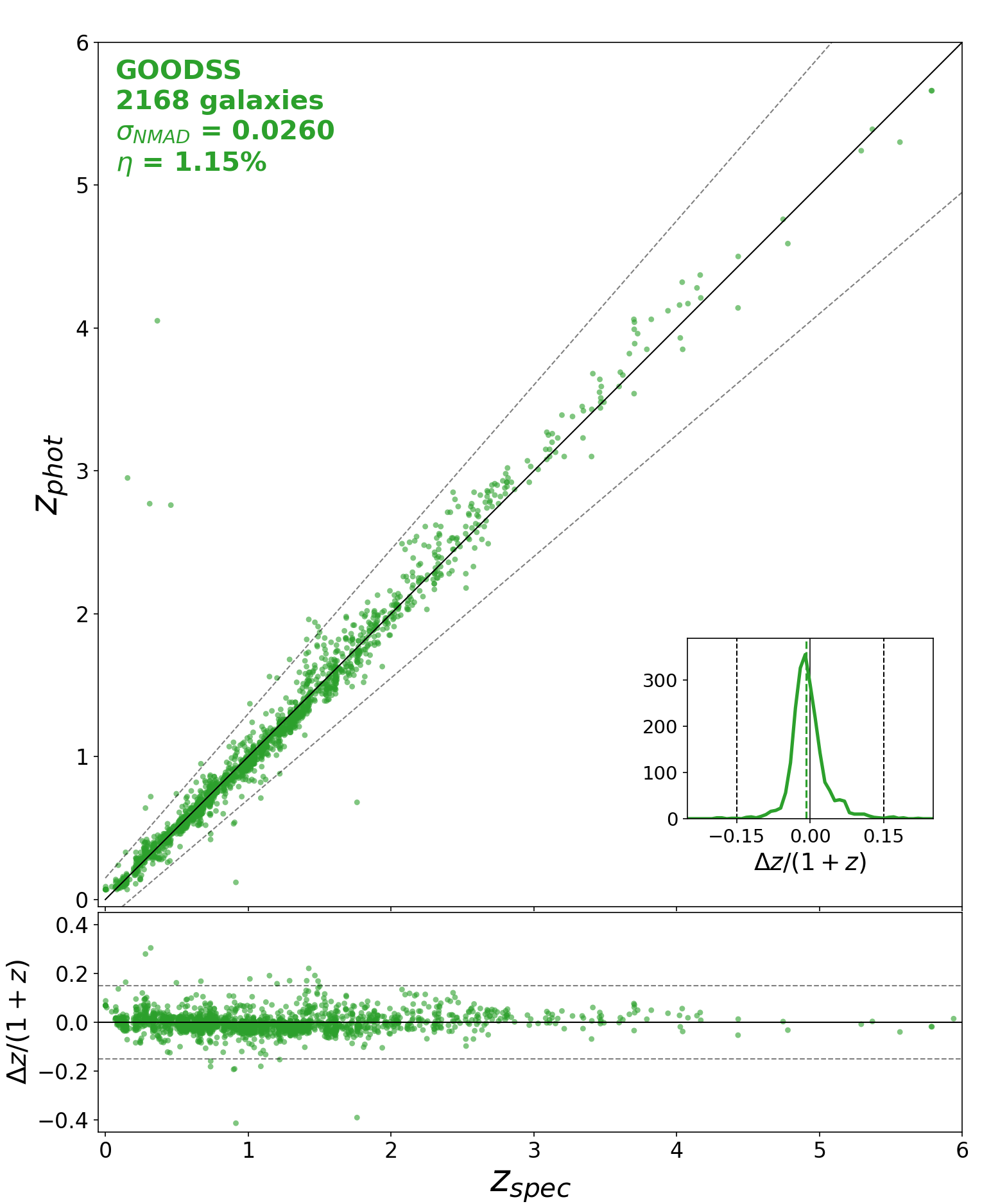} \\
\includegraphics[width=0.4\textwidth]{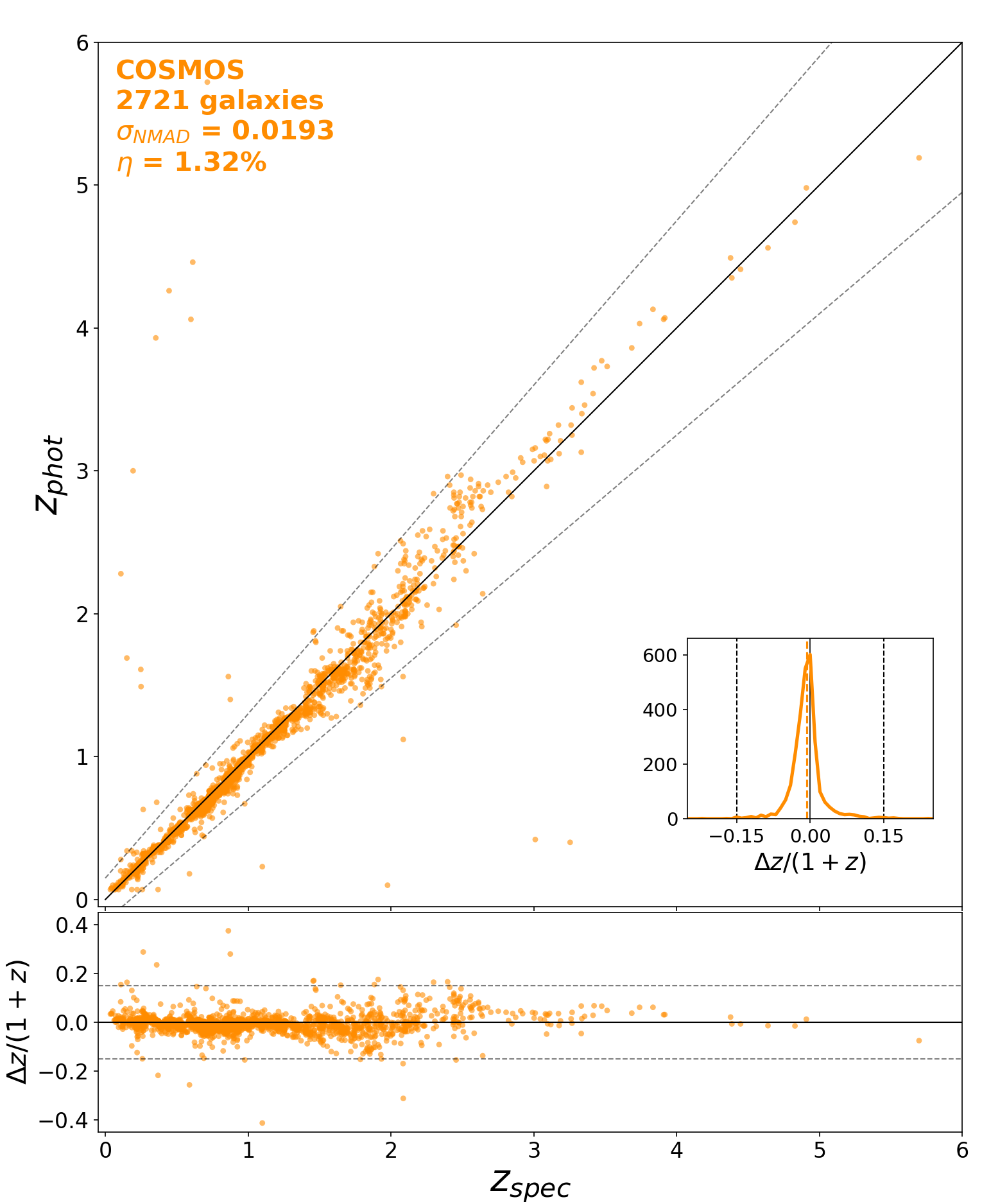} \hspace{0.3in}
\includegraphics[width=0.4\textwidth]{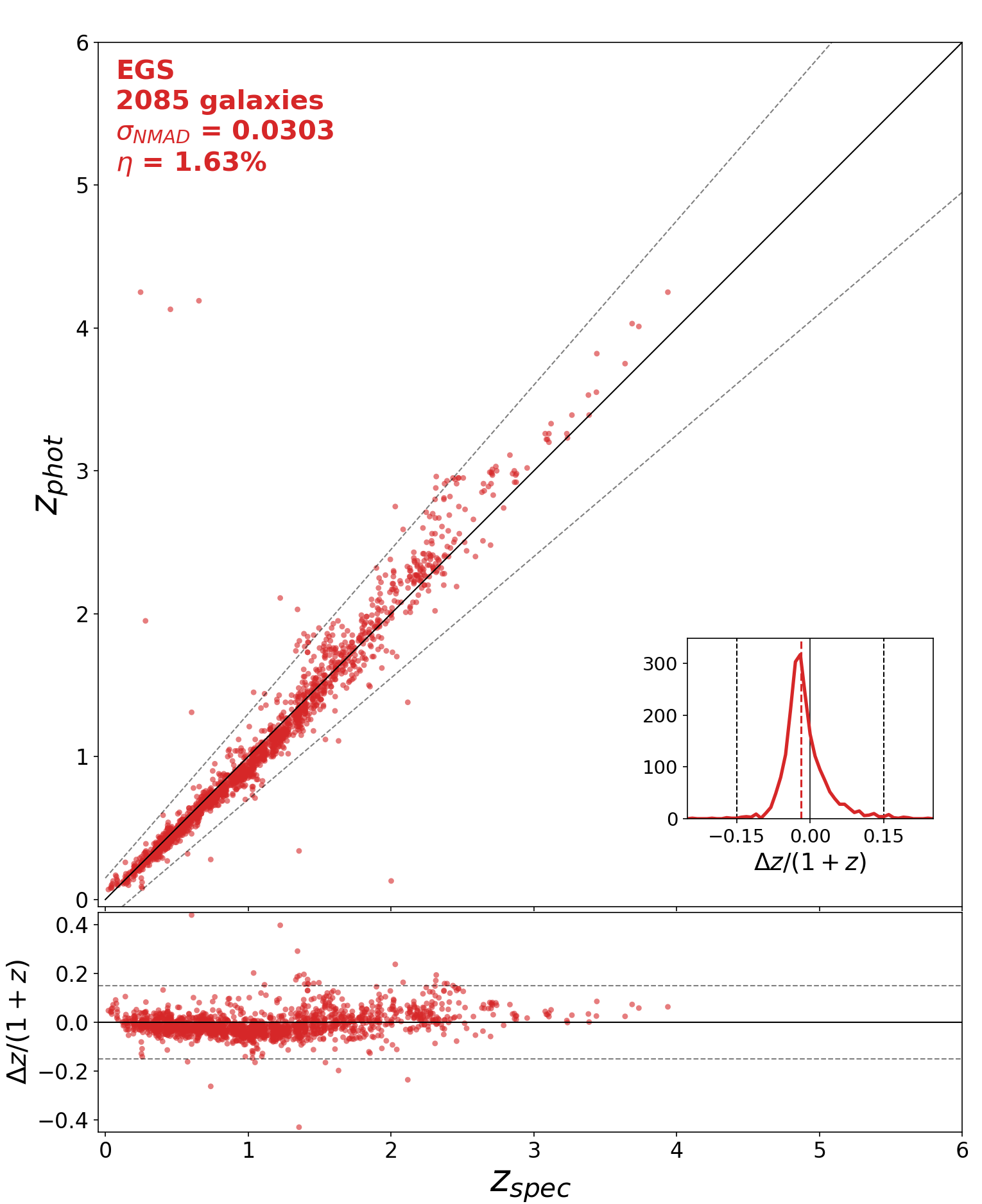}
\caption{Comparison of the performance of the photometric redshifts with the spectroscopic sample for each of the individual fields. The key statistics, i.e., the scatter ($\sigma_{NMAD}$) and the outlier fraction ($\eta=|\Delta z|/(1+z)>0.15$) are reported in the \textit{upper left} corner. The inset shows the distribution of the fractional differences between the photometric and spectroscopic redshifts. The \textit{dotted} lines show the outlier criterion: $z_{phot} = z_{spec} \pm 0.15(1+z_{spec})$ and the \textit{dashed} lines show the median of the distribution.}
\label{fig:photoz}
\end{figure*}

We note that when the photometric redshift codes fail, they output a redshift of $\sim0$ at high probability, which can affect the combined redshifts. For example, a failure mode of EAZY returns a redshift of $\sim$0.01 with 100\% confidence in the P(z). This throws off the combined P(z) distribution, resulting in a redshift at $z<0.06$ due to the high confidence over a small redshift range. Since the number of expected galaxies at $z<0.06$ in the field of view is extremely small, we truncated the redshift catalog at $z=0.06$ to avoid cases with code failures.  

We identify multiple peaks in each $P(z)$ as those with peak and prominence at least 10\%\ that of the main peaks, and with no higher peak within $0.06(1+z)$. The confidence intervals for each peak are calculated from the minimums between two neighboring peaks. We record up to 3 separate redshift peaks in the final photometric redshift catalogs, sorted by height (\texttt{zpeak}, \texttt{zpeak2}, \texttt{zpeak3}). Overall, we find that 20.5\%\ of sources have 2 peaks, and 3.4\%\ have 3 peaks. Typically, the final photometric redshift and its associated error is recorded as the median and 68.3\%\ confidence intervals of the $P(z)$. However, for the $\sim$1\% of sources that have multiple peaks with the main peak at $z<0.1$, we instead record the second peak as the final photometric redshift, as the main peak could be due to some codes failing to find a good fit. 

The final photometric redshifts along with their errors are provided in the photometric redshift catalog (see Table~\ref{tab:photz_columns}) and the full redshift probability distributions are also provided separately (see Table~\ref{tab:photz_pdfz_columns}).

\begin{figure*}[!t]
\centering
\includegraphics[width=0.4\textwidth]{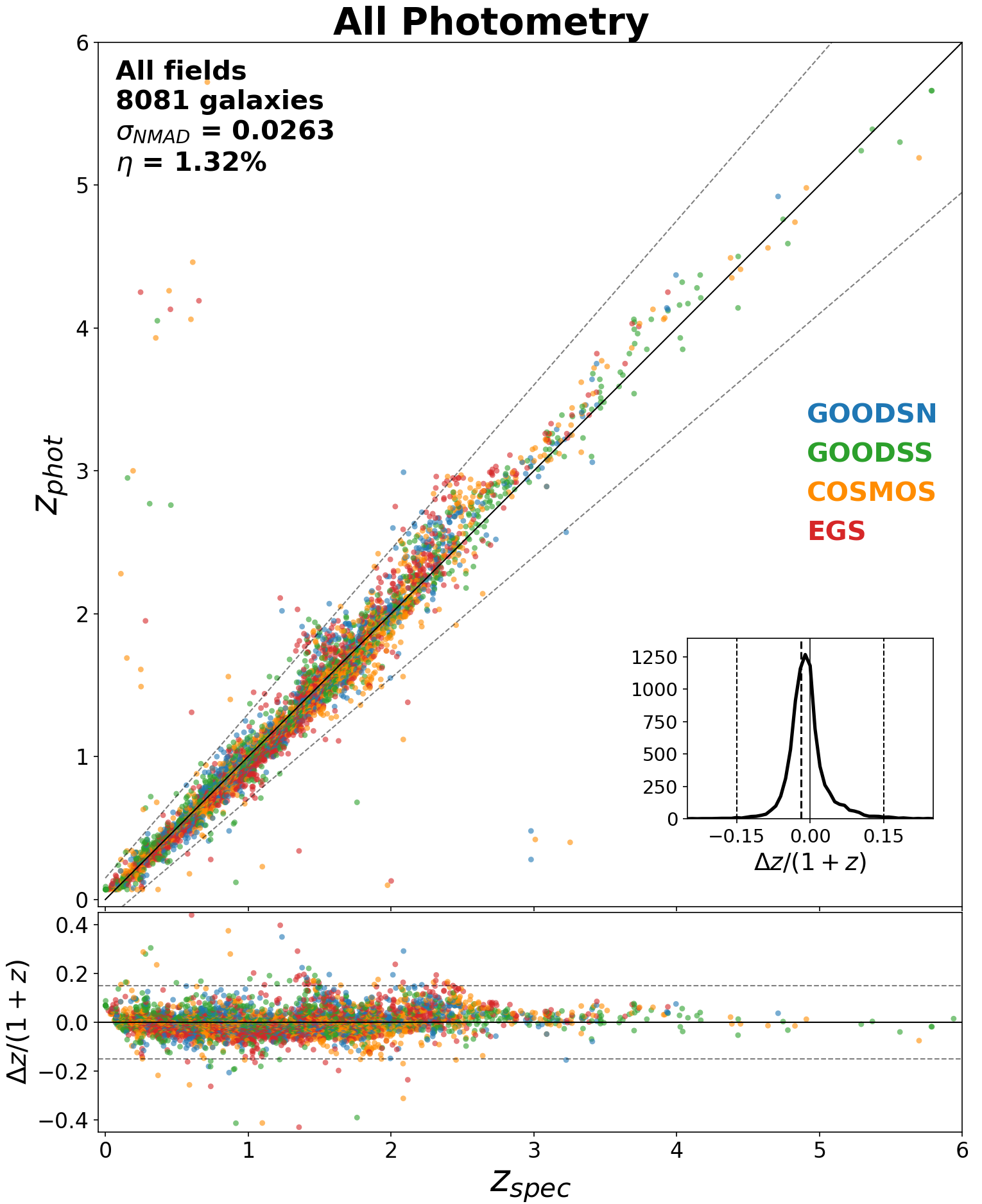} 
\hspace{0.3in}
\includegraphics[width=0.4\textwidth]{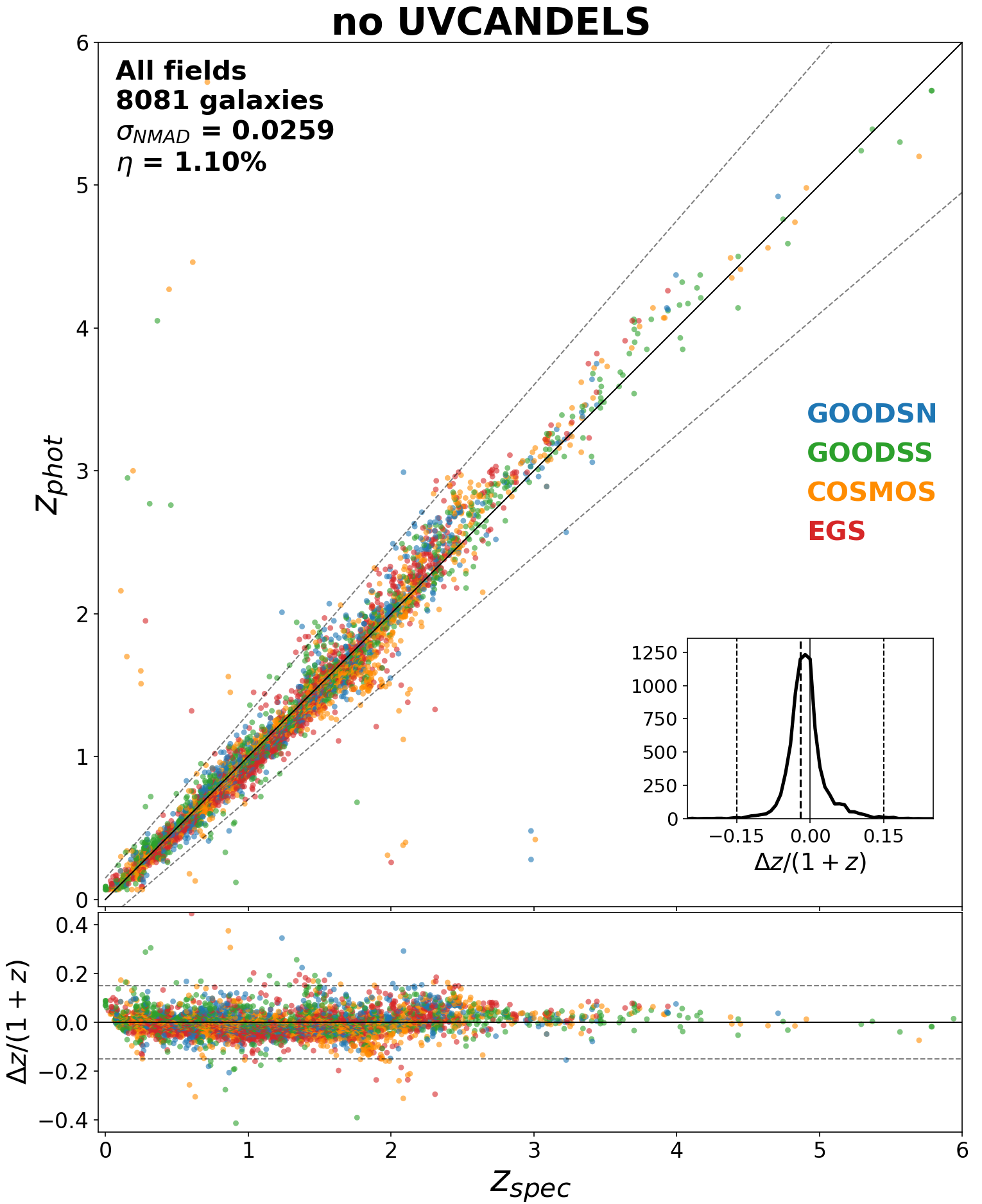} \\
\includegraphics[width=0.4\textwidth,trim={0 0 0 1cm},clip]{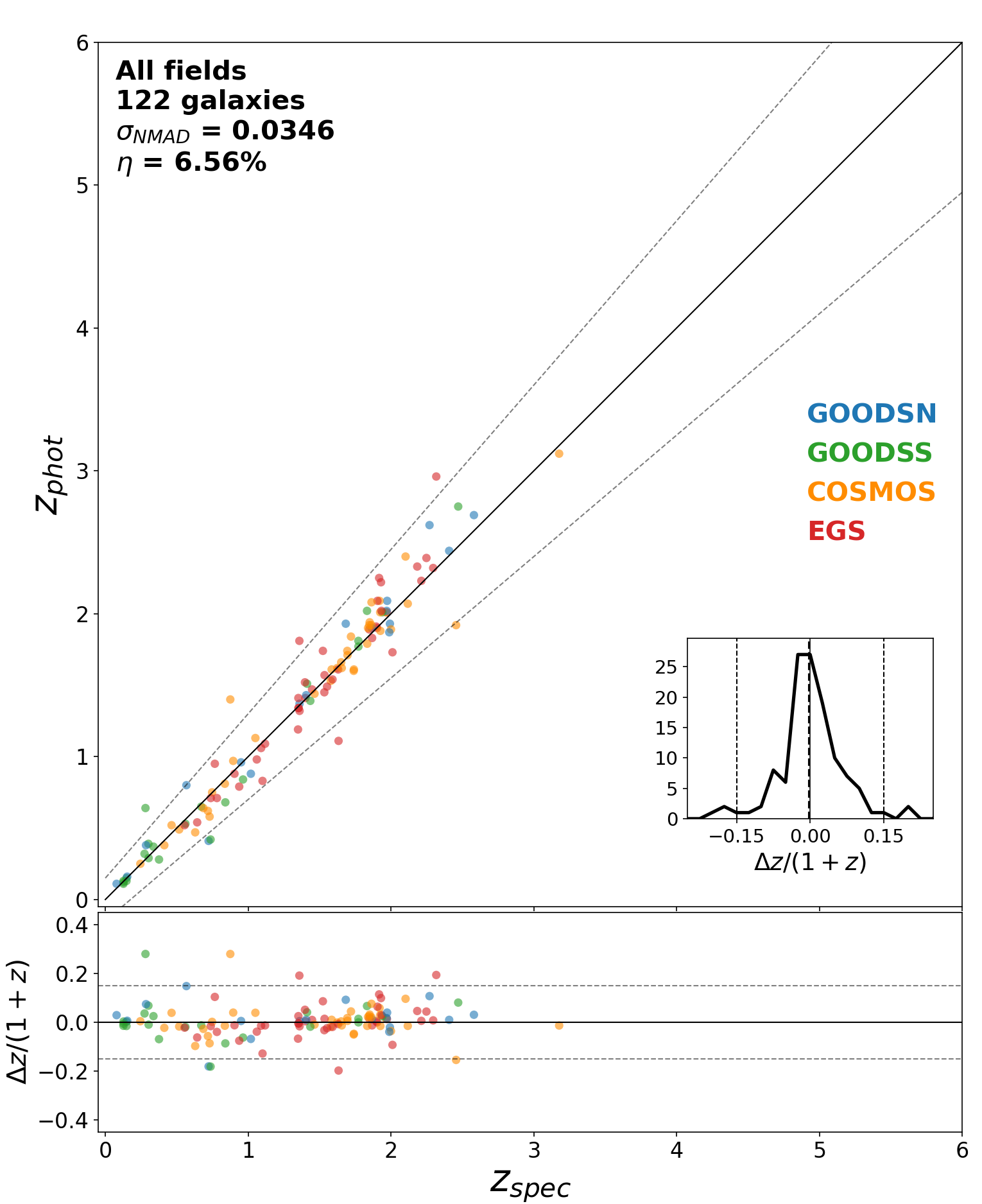} 
\hspace{0.3in}
\includegraphics[width=0.4\textwidth,trim={0 0 0 1cm},clip]{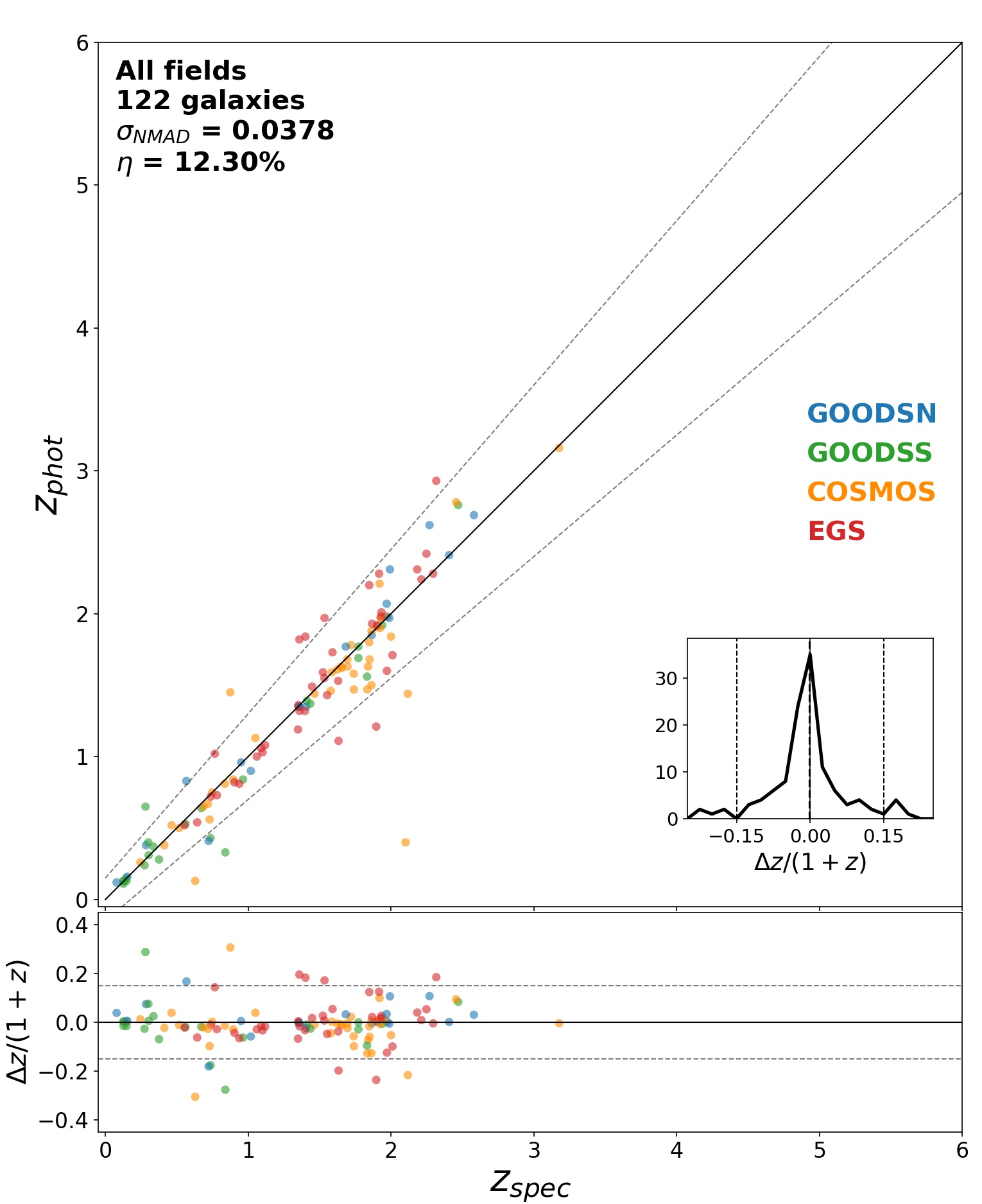}
\caption{An aggregate comparison between the photometric and spectroscopic redshifts for all four CANDELS fields combined. The \textit{left column} plots show the comparison where all available filters were included in the photometric redshift fitting, where as the \textit{right column} plots show the same without using filters from the UVCANDELS program (i.e. F275W/F435W for COSMOS and EGS; F275W for GOODSN and GOODSS). The \textit{top row} plots include all sources, where as the \textit{bottom row} plots only shows sources with multiple peaks in their probability distributions $P(z)$ and with SNR$>3$ in F275W. The improvement of the photometric redshift by including the UV photometry is evident in this case, reducing both $\sigma_{\rm NMAD}$ and $\eta$ for sources detected in the UV or that previously had multiple peaks in $P(z)$.}
\label{fig:uvphotoz}
\end{figure*}

\subsection{Quality of Resultant Redshifts}
Figure~\ref{fig:photoz} shows the comparison between the photometric redshifts and spectroscopic redshifts for each CANDELS field, as well as the overall scatter (NMAD; normalized median absolute deviation) defined by 
\begin{equation}
\sigma_{\rm NMAD}=1.48 \times {\rm median} (|\Delta z| / (1+z_{spec}))
\end{equation}

and outlier fraction $\eta$, where

\begin{equation}
\eta=|\Delta z| / (1+z_{spec}) > 0.15.
\end{equation}
Figure~\ref{fig:photoz} shows the performance in each individual field. We find that $\eta$ is  lower for fields with more filters (e.g. COSMOS has the lowest $\eta$). On the other hand, we see the best $\sigma_{\rm NMAD}$ for GOODS-N and GOODS-S, which have the most filters and deepest imaging. Overall, EGS performs the worst with the fewest filters, but still have relatively low $\eta$ and $\sigma_{\rm NMAD}$ given the excellent coverage.

The top left panel of Figure~\ref{fig:uvphotoz} shows the performance on the combined sample across all fields. We obtain an average $\sigma_{\rm NMAD}$ of 0.0263 and outlier fraction $\eta$ of 1.32\%. This is within family of the other photometric redshifts determined for the CANDELS fields \citep[e.g.][]{Dahlen:2013, Skelton:2014, Bezanson:2016, Kodra:2023}. We note that the quality of $\sigma_{\rm NMAD}$ and $\eta$ depends on the quality of the spectroscopic sample being compared to. Therefore, it is not useful to carefully compare the values of $\sigma_{\rm NMAD}$ and $\eta$ between studies of the same fields determined from different spectroscopic samples. In fact, we could probably improve our $\eta$ further by careful inspection of all the spectroscopic data, but that is out of the scope of this paper. On the other hand, we can investigate the improvements in photometric redshifts with the addition of the UVCANDELS data with the same spectroscopic sample. 

\subsection{Improvement of Photometric Redshifts with UVCANDELS data}

Aside from the fiducial case (all photometry), in order to quantify the impact of including the new UV photometry from UVCANDELS, we repeat the analysis of estimating photometric redshifts for an additional scenario without including the UVCANDELS photometry, i.e., no F275W for all fields and no F435W for COSMOS and EGS. The results from this test are shown in Figure~\ref{fig:uvphotoz}.

As evident from Figure~\ref{fig:uvphotoz}, high-fidelity photometric redshift estimates can be obtained with and without the inclusion of UVCANDELS photometry (i.e. F275W/F435W); the scatter and outlier fraction are within 0.004 and 0.22\%, respectively. This is because the vast majority of galaxies in the catalogs do not have detections in the UV, either because they are too faint, or because they are at high redshift. Unlike in the UVUDF where the UV limits significantly improved the photometric redshifts \citep{Rafelski:2015}, the depth of the UVCANDELS data only provide modest limits for non-detections, which do not significantly help the redshift determinations which already are based on many filters. Additionally, these fields have ground-based $u$-band data as well, already providing useful limits on non-detections at similar wavelengths, and further reducing the benefits of the F275W and F435W from UVCANDELS for photometric redshifts. 

For a more useful measure of the improvement provided by the UV data to the photometric redshifts, we limit ourselves to sources with significant F275W fluxes. The \textit{bottom row} in Figure~\ref{fig:uvphotoz} shows only those sources with multiple $P(z)$ peaks and SNR$>$3 in F275W. In this case, the improvement in the photometric redshift quality when including UV data is significantly more evident; the scatter and outlier fraction falls by 0.003 and 5.7\%, respectively. The percentage of sources with multiple peaks also decreases from 23.4\%\ to 20.5\%\ after including the UV data.

\section{Galaxy Physical Properties}
\label{sec:phys_pars}
The star-formation history (SFH) represents one of the core components for defining a stellar population for a galaxy and consequently, its SED. Traditionally, galaxy SED modeling tools have typically implemented fixed, functional forms to parameterize galaxy SFHs, such as exponentially rising/declining, delayed exponential, constant, short bursts, or some linear combination of thereof. However, modern SED modeling tools have evolved beyond the basic functional forms to now allow for more flexible, and even non-parametric forms for the galaxy SFH. Several recent studies have investigated the differences in the stellar physical parameters estimated from parametric and non-parametric approaches \citep[e.g., ][]{Lower:2020,Pacifici:2023,Kaushal:2024,Jain:2024} and the general consensus from comparisons to simulations prefers the non-parametric or flexible SFH approach, which returns more accurate physical parameters \citep[e.g., ][]{Iyer:2017,Iyer:2019,Leja:2019}.

With the rich UVCANDELS+CANDELS multi-wavelength photometric dataset and redshifts described in Section~\ref{sec:photometry} and \ref{sec:photoz} respectively, here we investigate the impact of the SFH parameterization on the inferred galaxy physical properties from the observational perspective. In this work, we implement two parallel approaches for modeling galaxy SEDs and estimating galaxy physical parameters: \textit{(i)} a modern approach that facilitates flexible SFHs, and \textit{(ii)} a fiducial approach that utilizes traditional, fixed functional SFHs. In the following sections, we describe our methodology for each of these approaches. For both, we use the photometric dataset described in Section~\ref{sec:photometry} for the SED modeling. 

\begin{figure*}[!ht]
\centering
\includegraphics[width=\textwidth]{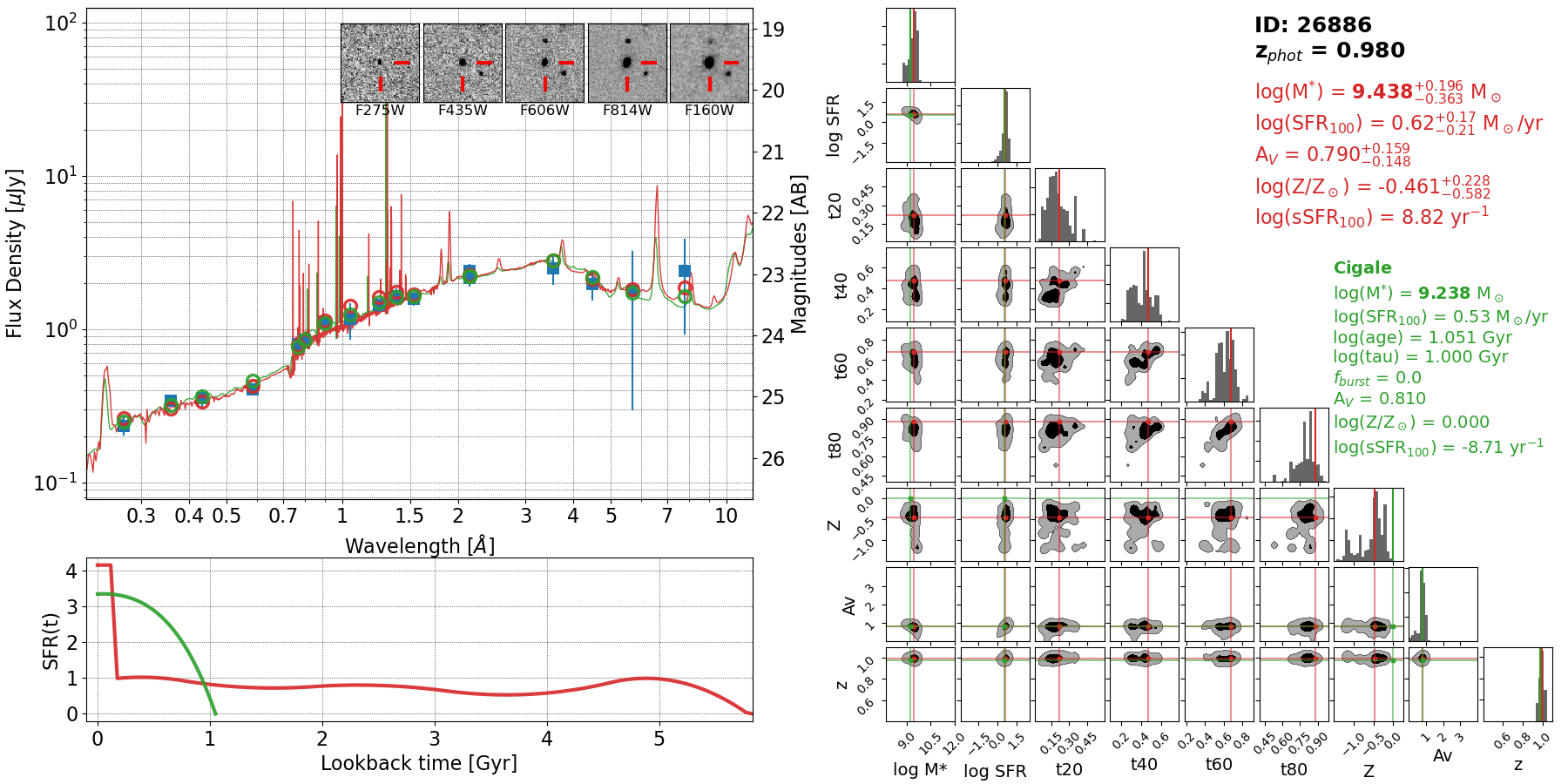}
\caption{The SED modeling shown for an example galaxy (GOODSN-26886). The \textit{top left} panel shows the observed photometry (shown in \textit{blue}) along with the best-fit model template from the flexible (\dbasis, in \textit{red}) and fixed (\cigale, in \textit{green}) SFH assumptions. The insets show 5\arcsec$\times$5\arcsec\ postage stamps of the galaxy in the respective filters. The \textit{bottom left} panel shows the best-fit SFH for both cases demonstrating how the flexible-form SFH captures SF at earlier times that is missed by the fixed-form approach, while both return excellent overall fit to the observed photometry. For the \dbasis\ result, a corner plot with the posterior distributions of the free parameters is shown on the \textit{right} along with the best-fit parameter values for both cases.}
\label{fig:example_sedfit}
\end{figure*}

\subsection{Fitting flexible SFHs with \dbasis}
\label{sec:flexible_SFH}
For the flexible SFH approach, we use the SED modeling tool \dbasis\footnote{\url{https://dense-basis.readthedocs.io/en/latest/}\\\url{https://dfm.io/python-fsps/}} \citep{Iyer:2019}, which uses Gaussian Process Regression to model highly flexible SFHs. The key advantage of the Gaussian Process (GP) based formalism is its ability to encode maximum amount of information in the complex galaxy SFHs using a minimal number of parameters. As demonstrated in \citet{Iyer:2019}, this formalism minimizes the bias in estimating galaxy SFHs at all lookback times, compared to the parametric (functional) approaches.

\subsubsection{Defining the SFHs}
The galaxy SFHs in \dbasis\ are described using a fixed number of ``shape'' parameters ($N$). These shape parameters ($t_X$) are $N$ lookback times at which the galaxy formed equally spaced quantiles of its total mass (see \citealt{Iyer:2019} for a more detailed description). For this work, we adopt $N=4$ shape parameters for describing the galaxy SFHs with 4~$t_X$ parameters: $t_{20}$, $t_{40}$, $t_{60}$, $t_{80}$ representing the lookback times\footnote{These lookback times are defined as the fraction of the age of the universe (at the given redshift for a galaxy).} at which the galaxy assembled 20\%, 40\%, 60\%, 80\%\ of its total stellar mass, respectively. We choose $N=4$ for the GP-based SFH parameterization as it allows for sufficient flexibility \citep{Iyer:2019} while minimizing the computation time for the SED modeling.

Furthermore, we allow the GP-based SFHs to have a "de-coupled" star-formation episode over the most recent 100~Myr, which still retains continuity with the rest of the SFH but allows for further flexibility in fitting any recent star-formation traced by the newly added rest-frame UV data from UVCANDELS.

\subsubsection{Galaxy model template library}
The galaxy model templates in \dbasis\ are defined using the FSPS\footnote{\url{https://github.com/cconroy20/fsps}} \citep[Flexible Stellar Population Synthesis; ][]{Conroy:2009,Conroy:2010} library, which includes a self-consistent prescription for both the stellar continuum as well as nebular emission. The nebular emission lines and continuum implementation in FSPS is based on the CLOUDY models from \citet{Byler:2017}. We also use the built-in \citet{Draine:2007} model for dust emission.

For this analysis, we assume a \citet{Chabrier:2003} IMF, which is a well-established choice when fitting the SEDs of a general galaxy population, particularly at higher redshifts where galaxies tend to be relatively more star-forming and have lower metallicities \citep[see e.g.,][]{Chabrier:2014,Hopkins:2018}. We adopt the \citet{Madau:1995} model for IGM absorption, and the \citet{Calzetti:2000} dust attenuation law, which couples the attenuation for young and older stars (i.e., assumes the same dust attenuation for the nebular and stellar components) with one free parameter: $A_V$, the $V$-band dust attenuation. We also assume the gas-phase metallicity to be equal to the stellar metallicity (left as a free parameter), and the ionization parameter for the nebular gas is left as default ($\log{U}=-2$).

The template redshifts are set to the galaxy's photometric redshifts defined in Section~\ref{sec:photoz}. During fitting, a nominal error of $\delta(z)=0.025 \cdot z/(1+z)$ is assumed to ensure a robust estimation for the galaxy physical parameters.

\begin{figure*}[!ht]
\centering
\includegraphics[width=0.49\textwidth]{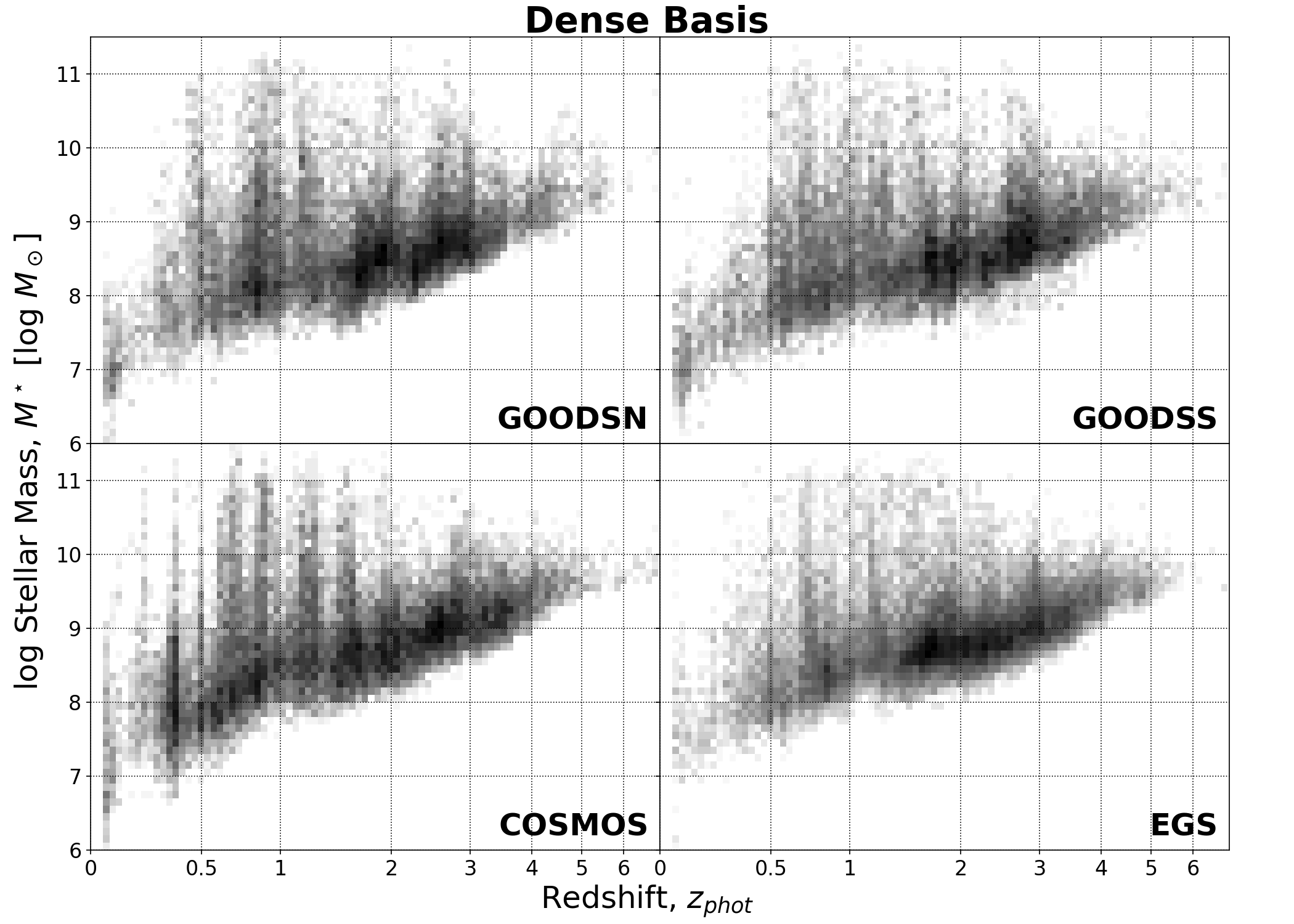}
\includegraphics[width=0.49\textwidth]{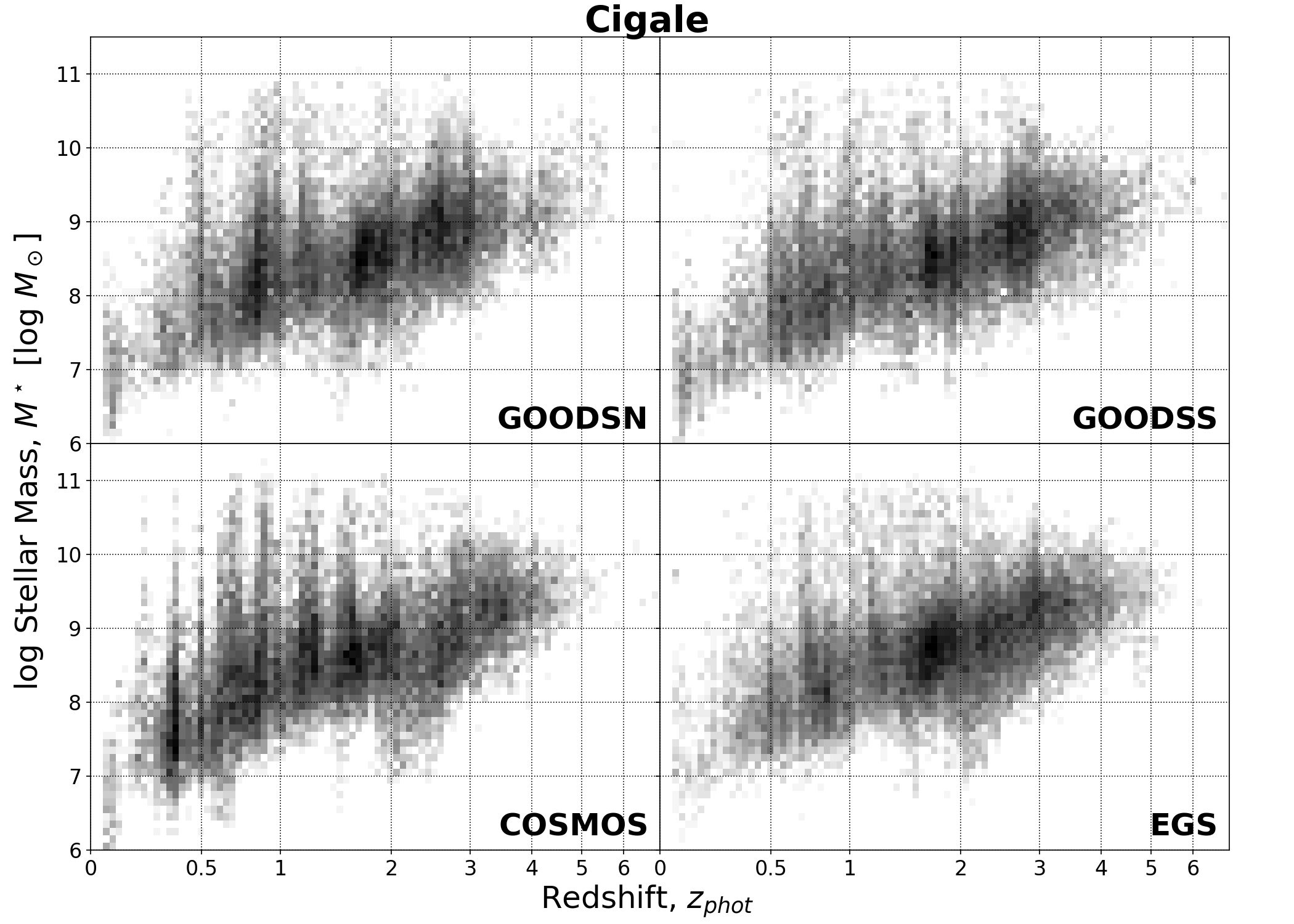}
\caption{The mass distribution as a function of redshift for each field, with the masses estimated using the flexible (\dbasis, \textit{left panel}) and fixed (\cigale, \textit{right panel}) SFH assumptions.}
\label{fig:mass_redshift}
\end{figure*}

\subsubsection{Priors and Parameter estimation}
Ultimately, we perform the fitting with 8 free parameters: the galaxy stellar mass ($M^\star$), star-formation rate (SFR; over the recent 100~Myr), the stellar dust attenuation ($A_V$), the stellar metallicity ($Z^\star$), and the 4~$t_X$ parameters ($t_{20}$, $t_{40}$, $t_{60}$, $t_{80}$) that define the galaxy SFH.

We jointly define the prior for $M^\star$ and SFR as the specific star-formation rate (sSFR$=$SFR/$M^\star$), which is assumed as a uniform (flat) prior spanning a range of log(sSFR)$=-12$ to $-7~yr^{-1}$. We adopt uniform (flat) prior for $A_V$ spanning from $A_V=0$ to $4$, and for $Z^\star$ over $log(Z/Z_\odot)=-1.5$ to $0.25$. For the GP-based SFH parameters, i.e., the $t_X$ lookback times, we adopt a Dirichlet prior with $\alpha=5$ as per the recommendation from \citet{Iyer:2019}, which leads to a distribution of parameters that is well-matched to galaxies in simulations.

We perform parameter estimation for \dbasis\ in two ways:
  \textit{(i)} For each galaxy, a likelihood is computed from the sufficiently large number of model SEDs realized in the pre-generated grid library with parameters drawn according to the priors described above. This results in a posterior distribution for each of the free parameters. The posterior medians and 1$\sigma$ (68\%) percentiles are then used to define the best-fit parameters and the associated errors.
  \textit{(ii)} A simpler, frequentist approach is also implemented that uses the (single) SED from the pre-generated library with the minimum $\chi^2$ value to define the best-fit parameters, with no uncertainties reported.

The estimates from the full posterior analysis are generally more robust and preferred over the simple minimum $\chi^2$ result, but we provide both for completeness and as a consistency check (see Table~\ref{tab:physpars_dbasis_columns}). Figure~\ref{fig:example_sedfit} demonstrates the modeling for an example galaxy along with the full posterior distributions for the free parameters. The full mass-redshift distribution for each of the fields is shown in Figure~\ref{fig:mass_redshift}.

\subsection{Fitting fixed SFHs with \cigale}
\label{sec:fixed_SFH}
For our fiducial approach, we use the traditional approach of using fixed, functional forms for the galaxy SFHs. In this case, we implement the \cigale\footnote{\url{https://cigale.lam.fr/}} \citep[Code Investigating GALaxy Emission; ][]{Boquien:2019,Burgarella:2005,Noll:2009} tool for the estimation of the galaxy physical properties.

\subsubsection{Defining the SFHs}
For the galaxy SFH parameterization within \cigale, we adopt a delayed-exponential functional form\footnote{SFR$(t) \propto t~e^{-t/\tau}$} with an additional possibility for an episode of recent star-burst (similar as with \dbasis) with an exponentially declining functional form. The SFH is parameterized with an overall age ($t_{age}$) and an $e$-folding timescale ($\tau$) for the main stellar population, along with an age ($t_{burst}$) and an $e$-folding timescale ($\tau_{burst}$) for the burst episode as well as the fraction of mass formed in the burst phase ($f_{burst}$).

In order to keep the computational times manageable, we fix the burst population's $t_{burst}$ and $\tau_{burst}$ parameters to 10~Myr and 50~Myr, respectively, while the remaining parameters ($t_{age}$, $\tau_{age}$, and $f_{burst}$) are varied freely.

\begin{figure*}[!ht]
\centering
\includegraphics[width=0.7\textwidth]{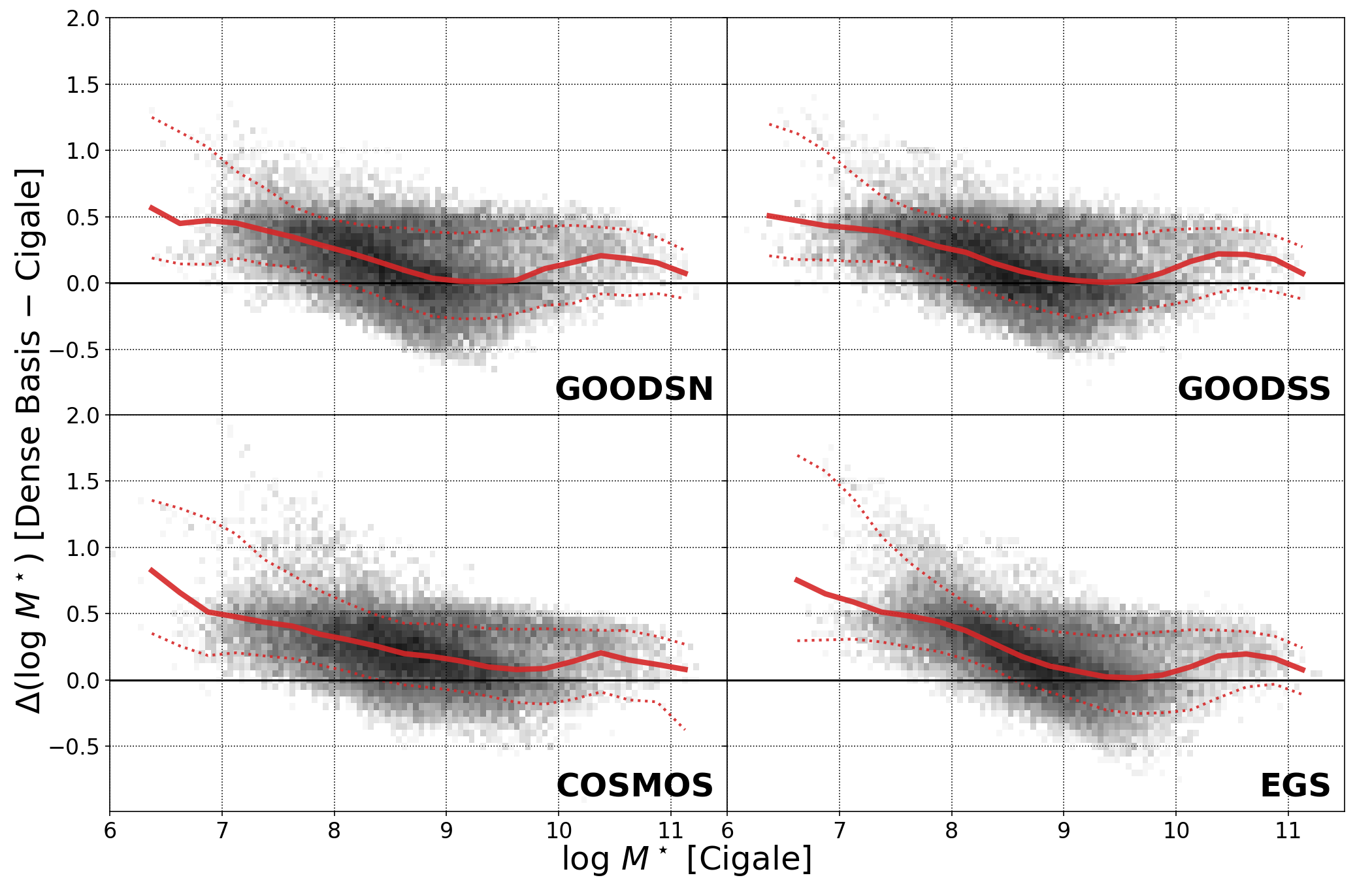}
\caption{The difference between the galaxy stellar masses estimated from flexible (\dbasis) and fixed (\cigale) SFH approaches. The \textit{red} curves show the running median (\textit{solid}) and the 68\%\ (1$\sigma$) percentile for the population. Even when fitting the same photometric dataset with consistent assumptions, the inferred stellar masses are significantly impacted by the parameterization of the galaxy SFH during SED modelling.}
\label{fig:mass_compare}
\end{figure*}

\subsubsection{Galaxy model template library}
Here, we use the \citet{Bruzual:2003} stellar population models, along with the prescription for nebular emission (both emission line and continuum) included within \cigale, which is based on \citet{Inoue:2011}. We note that the IGM prescriptions between \cigale\ and \dbasis\ (Section~\ref{sec:flexible_SFH}) are not identically similar, which is a technical limitation as these are the only prescriptions available within the respective codes. We use the dust emission model from \citet{Dale:2014} included in \cigale.

For this analysis, we assume a \citet{Chabrier:2003} IMF and the prescription from \citet{Meiksin:2006} is used for the IGM absorption. Same as the flexible SFH case, we adopt the \citet{Calzetti:2000} dust law (which assumes same attenuation for young and older stars (and hence the nebular and stellar dust components) parameterized by the free parameter $A_V$ or $E(B-V)=A_V/4.05$. Here, the gas-phase metallicity is assumed to be solar and the gas ionization parameter is left at its default value ($\log{U}=-2$).

The template redshifts are set to the galaxy's photometric redshifts defined in Section~\ref{sec:photoz}. In the case of \cigale, since all models are generated at the input redshifts, the galaxy redshifts are treated as exact, albeit rounded to 2 decimals to manage computation time.

\subsubsection{Parameter estimation}
For this \cigale\ implementation, we have 7 free parameters: the galaxy stellar mass ($M^\star$), star-formation rate (SFR), stellar dust attenuation ($A_V$), stellar metallicity ($Z^\star$) and the SFH-related parameters, $t_{age}$, $\tau_{age}$, and $f_{burst}$.

For the main stellar population in the SFH, the age, $t_{age}$, is allowed to vary over a grid of 21 points from 10~Myr to the age of the universe and similarly, $\tau_{age}$ is varied freely over a grid spanning over 30~Myr to 30~Gyr. For the burst component, $f_{burst}$ is allowed to vary (over gridded values from 0 to 1). Lastly, the dust attenuation parameter $E(B-V)=A_V/4.05$ is varied over $[0, 1.2]$ and the stellar metallicity over $log(Z/Z_\odot)=[-2.3, 0.4]$.

The best-fit parameters are estimated by performing a minimum $\chi^2$ search over a template library generated from the full grid of free parameters. As part of the output, an instantaneous as well as a 100~Myr averaged SFR is reported (see Table~\ref{tab:physpars_cigale_columns}). The resulting best-fit solutions from both the flexible (\dbasis) and fixed (\cigale) approaches for an example galaxy is shown in Figure~\ref{fig:example_sedfit} while the full mass-redshift distribution for each field is shown in Figure~\ref{fig:mass_redshift}.

\subsection{The impact of the SFH assumption on parameter estimates}
From a statistical perspective, the SED modeling with either of the approaches (flexible/\dbasis\ or fixed/\cigale) are relatively robust with no significant systematic biases in the fitting residuals. As illustrated in Figure~\ref{fig:example_sedfit} with an example galaxy, the best-fit model template from both methodologies represents a qualitatively good fit to the observed photometry.

However, a population comparison of the best-fit parameters reveals significant systematic biases in even the most basic physical property, the galaxy stellar mass. Figure~\ref{fig:mass_compare} shows the difference in the stellar mass inferred from the two approaches and it is evident that the masses estimated from the flexible SFH assumption (\dbasis) are systematically higher, particularly at the low-mass end. For typical galaxies ($\sim10^{9-10}~M_\odot$), the stellar masses are in good agreement regardless of the SFH parameterization. However, below $\sim10^9~M_\odot$, the flexible-SFH stellar masses get increasingly biased by as much as $0.5$~dex at $\sim10^7~M_\odot$. Similarly, at the high-mass end ($\gtrsim10^{10}~M_\odot$), the masses from \dbasis\ are also systematically higher, albeit only by $\sim0.2$~dex.

This discrepancy in the estimated stellar masses due to different SFH assumptions is not new and has already been reported in the literature \citep[e.g., ][]{Leja:2019,Lower:2020,Jain:2024}. Our results here are broadly consistent with the previous findings. From comparisons to simulated galaxies, it is clear that the stellar masses estimated using flexible SFHs are closer to the true value and it is the fixed-form SFHs that underestimate the stellar masses \citep[e.g., ][]{Lower:2020}. This is due to their inability to simultaneously capture the multiple episodes of star-formation and other gradual changes in the SFHs. This effect is also evident in the example shown in Figure~\ref{fig:example_sedfit}, where the fixed SFH (from \cigale; in \textit{green}) is missing the early episode of star-formation captured by the flexible SFH (from \dbasis; in \textit{black}). This effect is amplified particularly at the lower mass end ($\lesssim10^9~M_\odot$) where galaxies tend to have more chaotic, bursty SFHs with multiple episodes of prominent star-formation activity, which can contribute relatively large differences in their galaxies' stellar masses \citep[e.g., ][]{Emami:2019}. On a similar note, at higher masses, it is likely that the functional form of the fixed SFHs is missing the extended periods of star-formation activity due to the choice of parameterization, again resulting in underestimated masses.

\section{Catalog descriptions}
\label{sec:catalogs}

From the analysis presented here, we produce catalogs listing new photometric redshifts as well as galaxy physical properties for the four CANDELS fields covered by UVCANDELS (GOODS-N, GOODS-S, COSMOS, EGS). These catalogs will be publicly released via MAST (\dataset[doi:10.17909/8s31-f778]{https://dx.doi.org/10.17909/8s31-f778}\footnote{\url{https://archive.stsci.edu/hlsp/uvcandels}}). The full output is delivered with the following catalogs:
\begin{itemize}
  \item[\textit{(i)}] A catalog presenting the photometric redshifts from Section~\ref{sec:photoz} (see column description in Table~\ref{tab:photz_columns});
  \item[\textit{(ii)}] The full probability distribution, $P(z)$, for the photometric redshifts from Section~\ref{sec:photoz} (see column description in Table~\ref{tab:photz_pdfz_columns});
  \item[\textit{(iii)}] Catalogs for each field listing the best-fit galaxy physical properties estimated using \dbasis\ with the flexible SFH assumption from Section~\ref{sec:flexible_SFH} (see column description in Table~\ref{tab:physpars_dbasis_columns}); and
  \item[\textit{(iv)}] Per-field galaxy physical properties catalogs from \cigale\ (fixed SFH assumption) from Section~\ref{sec:fixed_SFH} (see column description in Table~\ref{tab:physpars_cigale_columns}).
\end{itemize}

\begin{deluxetable}{r l X{0.5\linewidth}}
\centering
\tablecaption{Column descriptions for the photometric redshift catalog \label{tab:photz_columns}}
\tablewidth{\linewidth}
\tabletypesize{\footnotesize}
\tablehead{\colhead{\#} & \colhead{Parameter} & \colhead{Description}}
\startdata
     1 & \texttt{field} & The CANDELS field \\
     2 & \texttt{ID} & The CANDELS ID\\
     3 & \texttt{RA}    & Right ascension\\
     4 & \texttt{DEC}   & Declination\\
     5 & \texttt{photz} & Best photometric redshift \\
     6 & \texttt{l68} & Lower 68.3\% confidence interval of photz \\
     7 & \texttt{u68} & Upper 68.3\% confidence interval of photz\\
     8 & \texttt{zpeak} & Highest $P(z)$ peak\\
     9 & \texttt{zpeak\_l68} & Lower 68.3\% confidence interval of zpeak\\
    10 & \texttt{zpeak\_u68} & Upper 68.3\% confidence interval of zpeak\\
    11 & \texttt{zpeak2} & Second highest $P(z)$ peak\\
    12 & \texttt{zpeak2\_l68} & Lower 68.3\% confidence interval of zpeak2\\
    13 & \texttt{zpeak2\_u68} & Upper 68.3\% confidence interval of zpeak2\\
    14 & \texttt{zpeak3} & Third highest $P(z)$ peak\\
    15 & \texttt{zpeak3\_l68} & Lower 68.3\% confidence interval of zpeak3\\
    16 & \texttt{zpeak3\_u68} & Upper 68.3\% confidence interval of zpeak3\\
    17 & \texttt{photz\_med} & Median of the $P(z)$\\
    18 & \texttt{photz\_med\_l68} & Lower 68.3\% confidence interval of photz\_med\\
    19 & \texttt{photz\_med\_u68} & Upper 68.3\% confidence inverval of photz\_med\\
    20 & \texttt{specz} & The spectroscopic redshift\\
    21 & \texttt{specz\_ref} & The source of the spectroscopic redshift
\enddata
\tablecomments{The \texttt{photz} column is the best redshift to use, and generally consists of the median of the $P(z)$ (i.e. \texttt{photz\_med}). However, sometimes if the first peak is at low redshift ($z<0.1$), it is instead the second $P(z)$ peak (i.e. \texttt{zpeak2}) to avoid issues with the photometric redshift software as described in the text. Missing or irrelevant values are set to -99. }
\end{deluxetable}

\begin{deluxetable}{r l X{0.45\linewidth}}
\centering
\tablecaption{Column descriptions for the catalog presenting the photometric redshift probability distribution $P(z)$ \label{tab:photz_pdfz_columns}}
\tablewidth{\linewidth}
\tabletypesize{\footnotesize}
\tablehead{\colhead{\#} & \colhead{Parameter} & \colhead{Description}}
\startdata
    1 & \texttt{field} & The CANDELS field \\
    2 & \texttt{ID}    & The CANDELS ID\\
    3 & \texttt{RA}    & Right ascension\\
    4 & \texttt{DEC}   & Declination\\
    5 & \texttt{pz}    & $P(z)$
\enddata
\tablecomments{The third column consists of the photometric redshift probability distribution $P(z)$ on a redshift grid from $z=0-12$ in steps of 0.01. This probability function is useful when considering how robust a photometric redshift is.}
\end{deluxetable}

\begin{deluxetable*}{r l l}
\centering
\tablecaption{Column descriptions for the catalog presenting the galaxy physical properties computed from \dbasis \label{tab:physpars_dbasis_columns}}
\tablewidth{\linewidth}
\tabletypesize{\footnotesize}
\tablehead{\colhead{\#} & \colhead{Parameter} & \colhead{Description}}
\startdata
       1  &  \texttt{ID}\tblmark{a}            & Identification number\\
       2  &  \texttt{RA}\tblmark{a}            & Right ascension\\
       3  &  \texttt{DEC}\tblmark{a}           & Declination\\
       4  &  \texttt{photz}                    & Photometric redshift used (from Table~\ref{tab:photz_columns})\\
       5  &  \texttt{delz}                     & Error on photometric redshift (assumed)\\
       6  &  \texttt{nbands}                   & Number of photometric bands used for fitting\\
\tableline
\multicolumn{3}{c}{Parameter estimates from the posterior distribution} \\
\tableline
  7$-$9   &  \texttt{logM}\tblmark{b}          & Galaxy stellar mass estimate [log $M_\odot$]\\
 10$-$12  &  \texttt{logSFR}\tblmark{b}        & Star-formation rate [log $M_\odot/yr$]\\
 13$-$15  &  \texttt{Av}\tblmark{b}            & Dust attenuation (V-band) [mag]\\
 16$-$18  &  \texttt{logZsol}\tblmark{b}       & Stellar metallicity [log $Z_\odot$]\\
 18$-$21  &  \texttt{t20}\tblmark{b}           & Time\tblmark{c} when 20\%\ of galaxy mass was formed\\
 22$-$24  &  \texttt{t40}\tblmark{b}           & Time\tblmark{c} when 40\%\ of galaxy mass was formed\\
 25$-$27  &  \texttt{t60}\tblmark{b}           & Time\tblmark{c} when 60\%\ of galaxy mass was formed\\
 28$-$30  &  \texttt{t80}\tblmark{b}           & Time\tblmark{c} when 80\%\ of galaxy mass was formed\\
 31$-$33  &  \texttt{color\_nuvu}\tblmark{b}   & Rest-frame NUV-U color\\
 34$-$36  &  \texttt{color\_nuvr}\tblmark{b}   & Rest-frame NUV-R color\\
 37$-$39  &  \texttt{color\_uv}\tblmark{b}     & Rest-frame U-V color\\
 40$-$42  &  \texttt{color\_vj}\tblmark{b}     & Rest-frame V-J color\\
 43$-$45  &  \texttt{color\_rj}\tblmark{b}     & Rest-frame R-J color\\
      46  &  \texttt{flags}                    & Quality flags\tblmark{d}\\
\tableline
\multicolumn{3}{c}{Parameter estimates from the minimum-$\chi^2$ model} \\
\tableline
      47  &  \texttt{logM\_chi2}               & Galaxy stellar mass estimate [log $M_\odot$] \\
      48  &  \texttt{logSFR\_chi2}             & Star-formation rate [log $M_\odot/yr$] \\
      49  &  \texttt{Av\_chi2}                 & Dust attenuation (V-band) [mag] \\
      50  &  \texttt{logZsol\_chi2}            & Stellar metallicity [log $Z_\odot$] \\
      51  &  \texttt{t20\_chi2}                & Time\tblmark{c} when 20\%\ of galaxy mass was formed\\
      52  &  \texttt{t40\_chi2}                & Time\tblmark{c} when 40\%\ of galaxy mass was formed\\
      53  &  \texttt{t60\_chi2}                & Time\tblmark{c} when 60\%\ of galaxy mass was formed\\
      54  &  \texttt{t80\_chi2}                & Time\tblmark{c} when 80\%\ of galaxy mass was formed\\
      55  &  \texttt{color\_nuvu\_chi2}        & Rest-frame NUV-U color \\
      56  &  \texttt{color\_nuvr\_chi2}        & Rest-frame NUV-R color \\
      57  &  \texttt{color\_uv\_chi2}          & Rest-frame U-V color \\
      58  &  \texttt{color\_vj\_chi2}          & Rest-frame V-J color \\
      59  &  \texttt{color\_rj\_chi2}          & Rest-frame R-J color \\
      60  &  \texttt{chi2}                     & $\chi^2$ value for the minimum-$\chi^2$ solution\\
\tableline
\multicolumn{3}{c}{Model magnitudes from the best-fit template\tblmark{e}} \\
\tableline
  61$-$69 &  \texttt{model\_Lnu\_rest*}        & Model template fluxes for rest-frame Johnson filters (FUV, NUV, $UBVRIJK$) [\uJy]\\
  70$-$   &  \texttt{model\_Fnu\_*}\tblmark{f} & Model template fluxes for observed photometric bands [\uJy]
\enddata
\tbltext{a}{The galaxy ID number, RA and DEC are consistent with that from the CANDELS photometric catalogs and the catalogs for each field are provided separately.}
\tbltext{b}{For these columns, each has two additional columns named \texttt{*\_16} and \texttt{*\_84} denoting the 1$\sigma$ (68\%) confidence interval for the best estimate.}
\tbltext{c}{The time is given as the fraction of the age of the universe (at the redshift of the galaxy).}
\tbltext{d}{Quality flags: 1$=$invalid fit (\texttt{nan} for mass estimate); 2$=$large SFR uncertainties ($\Delta(\log{SFR})>2$); 3$=$ objects flagged as stars in photometry catalog (\texttt{CLASS\_STAR}$>$0.5); 4$=$poor fit/large $\chi^2$ ($>$1000).}
\tbltext{e}{These model magnitudes are computed for the best-fit parameters estimated from the full Bayesian posterior distribution (\textit{not} the minimum-$\chi^2$) solution.}
\tbltext{f}{The model magnitudes are available for the specific set of photometric bands included in the SED modeling, which differs from field to field.}
\end{deluxetable*}

\begin{deluxetable}{r l X{0.5\linewidth}}
\centering
\tablecaption{Column descriptions for the catalog presenting the galaxy physical properties computed from \cigale \label{tab:physpars_cigale_columns}}
\tablewidth{\linewidth}
\tabletypesize{\footnotesize}
\tablehead{\colhead{\#} & \colhead{Parameter} & \colhead{Description}}
\startdata
     1  &  \texttt{ID}             & Identification number\\
     2  &  \texttt{RA}             & Right ascension\\
     3  &  \texttt{DEC}            & Declination\\
     4  &  \texttt{photz}          & Photometric redshift used (from Table~\ref{tab:photz_columns})\\
     5  &  \texttt{logM}           & Galaxy stellar mass [log $M_\odot$]\\
     6  &  \texttt{logSFR}         & Instantaneous star-formation rate [log $M_\odot/yr$]\\
     7  &  \texttt{logSFR100}      & 100~Myr-averaged star-formation rate [log $M_\odot/yr$]\\
     8  &  \texttt{Av}             & Dust attenuation (V-band) [mag]\\
     9  &  \texttt{logZsol}        & Stellar metallicity [log $Z_\odot$]\\\\
    10  &  \texttt{logage}         & Age of the main stellar population [log $yr$]\\
    11  &  \texttt{logtau}         & $e$-folding timescale of the main stellar population [log $yr$]\\
    12  &  \texttt{f\_burst}       & Fraction of mass formed in the late burst\tblmark{a}\\
    13  &  \texttt{color\_fuvnuv}  & Rest-frame FUV-NUV color\\
    14  &  \texttt{color\_nuvr}    & Rest-frame NUV-R color\\
    15  &  \texttt{chi2}           & Reduced $\chi^2$ value \\
  16$-$ &  \texttt{model\_Fnu\_*}  & Model template fluxes for observed photometric bands [\uJy]
\enddata
\tbltext{a}{The late burst is characterized as a (fixed) 10~Myr old exponentially declining episode of star-formation with an $e$-folding timescale of 50~Myr.}
\end{deluxetable}

\section{Summary}
\label{sec:summary}
We present a set of new measurements for photometric redshifts and galaxy physical parameters for over 150,000 galaxies in four of the five CANDELS fields (GOODS-N, GOODS-S, COSMOS, EGS; combined area of $\sim$430~arcmin$^2$) covered by the recent \textit{Hubble} Treasury program, UVCANDELS. In this work, we utilize the UV-to-IR photometric dataset available for these fields from a combination of the existing \textit{HST} optical/NIR, \textit{Spitzer} mid-IR, and other ancillary ground-based imaging as well as the newly acquired WFC3/UVIS F275W and ACS/WFC F435W (for COSMOS and EGS) imaging from UVCANDELS, to model the galaxy SEDs using state-of-the-art fitting tools.

Our key findings are as follows:
\begin{itemize}
    \item The inclusion of UV photometry significantly improves the photometric redshift for sources with significant UV detections and degenerate redshift solutions. After adding UV photometry when measuring the redshifts, we find significantly lower outlier fractions for sources with SNR$>3$ in F275W and multiple peaks in their probability distributions, P(z).
    \item When modeling galaxy SEDs, the typical assumption of a fixed, functional form for the galaxy SFH systematically biases the measurement of even the most basic physical parameter -- the galaxy stellar mass -- at both the low- ($\lesssim10^9~M_\odot$) as well as the high-mass ($\gtrsim10^{10}~M_\odot$) end. We find the fixed-form SFHs to underestimate the stellar mass by as much as $\sim0.5$~dex at 10$^7~M_\odot$ and $\sim0.2$ at 10$^{10.5}~M_\odot$.
    \item This underestimation for the stellar mass is mainly driven by the inability of the fixed-form SFHs to simultaneously capture the major star-formation episodes as well as the prolonged lower-level star-formation and/or other gradual changes in the galaxy SFHs.
\end{itemize}

The final catalogs presenting the photometric redshifts and physical parameters will be publicly available via the MAST archive\footnote{\url{https://archive.stsci.edu/hlsp/uvcandels}}.

\begin{acknowledgments}
We thank the anonymous referee for their constructive feedback which helped improve the quality of the manuscript. The analysis presented in this paper is based on observations with the NASA/ESA Hubble Space Telescope obtained at the Space Telescope Science Institute, which is operated by the Association of Universities for Research in Astronomy, Incorporated, under NASA contract NAS5-26555. Support for Program Number HST-GO-15647 was provided through a grant from the STScI under NASA contract NAS5-26555. V.M., H.T., and X.W. acknowledge work carried out, in part, by IPAC at the California Institute of Technology, as was sponsored by the National Aeronautics and Space Administration.
\end{acknowledgments} 

\software{NumPy \citep{Numpy:2020}; SciPy \citep{Scipy:2020}; AstroPy \citep{Astropy:2013,Astropy:2018,Astropy:2022}; Matplotlib \citep{Matplotlib:2007}; FSPS \citep{FSPS}; \dbasis\ \citep{Iyer:2019}; \cigale\ \citep{Boquien:2019}; \eazy\ \citep{Brammer:2008}; \bpz\ \citep{Benitez:2000,Benitez:2004,Coe:2006}; \lephare\ \citep{Arnouts:1999,Ilbert:2006}; \zphot\ \citep{Giallongo:1998,Fontana:2000}}


\bibliographystyle{aasjournal}
\bibliography{UVC_sedfitting, UVC_specz_refs}

\begin{thebibliography}{}
\expandafter\ifx\csname natexlab\endcsname\relax\def\natexlab#1{#1}\fi
\providecommand{\url}[1]{\href{#1}{#1}}
\providecommand{\dodoi}[1]{doi:~\href{http://doi.org/#1}{\nolinkurl{#1}}}
\providecommand{\doeprint}[1]{\href{http://ascl.net/#1}{\nolinkurl{http://ascl.net/#1}}}
\providecommand{\doarXiv}[1]{\href{https://arxiv.org/abs/#1}{\nolinkurl{https://arxiv.org/abs/#1}}}

\bibitem[{{Alavi} {et~al.}(2014){Alavi}, {Siana}, {Richard}, {Stark},
  {Scarlata}, {Teplitz}, {Freeman}, {Dominguez}, {Rafelski}, {Robertson}, \&
  {Kewley}}]{Alavi:2014}
{Alavi}, A., {Siana}, B., {Richard}, J., {et~al.} 2014, \apj, 780, 143,
  \dodoi{10.1088/0004-637X/780/2/143}

\bibitem[{{Alavi} {et~al.}(2016){Alavi}, {Siana}, {Richard}, {Rafelski},
  {Jauzac}, {Limousin}, {Freeman}, {Scarlata}, {Robertson}, {Stark}, {Teplitz},
  \& {Desai}}]{Alavi:2016}
---. 2016, \apj, 832, 56, \dodoi{10.3847/0004-637X/832/1/56}

\bibitem[{{Arnouts} {et~al.}(1999){Arnouts}, {Cristiani}, {Moscardini},
  {Matarrese}, {Lucchin}, {Fontana}, \& {Giallongo}}]{Arnouts:1999}
{Arnouts}, S., {Cristiani}, S., {Moscardini}, L., {et~al.} 1999, \mnras, 310,
  540, \dodoi{10.1046/j.1365-8711.1999.02978.x}

\bibitem[{{Ashby} {et~al.}(2013){Ashby}, {Willner}, {Fazio}, {Huang}, {Arendt},
  {Barmby}, {Barro}, {Bell}, {Bouwens}, {Cattaneo}, {Croton}, {Dav{\'e}},
  {Dunlop}, {Egami}, {Faber}, {Finlator}, {Grogin}, {Guhathakurta},
  {Hernquist}, {Hora}, {Illingworth}, {Kashlinsky}, {Koekemoer}, {Koo},
  {Labb{\'e}}, {Li}, {Lin}, {Moseley}, {Nandra}, {Newman}, {Noeske}, {Ouchi},
  {Peth}, {Rigopoulou}, {Robertson}, {Sarajedini}, {Simard}, {Smith}, {Wang},
  {Wechsler}, {Weiner}, {Wilson}, {Wuyts}, {Yamada}, \& {Yan}}]{Ashby:2013}
{Ashby}, M.~L.~N., {Willner}, S.~P., {Fazio}, G.~G., {et~al.} 2013, \apj, 769,
  80, \dodoi{10.1088/0004-637X/769/1/80}

\bibitem[{{Ashby} {et~al.}(2015){Ashby}, {Willner}, {Fazio}, {Dunlop}, {Egami},
  {Faber}, {Ferguson}, {Grogin}, {Hora}, {Huang}, {Koekemoer}, {Labb{\'e}}, \&
  {Wang}}]{Ashby:2015}
---. 2015, \apjs, 218, 33, \dodoi{10.1088/0067-0049/218/2/33}

\bibitem[{{Astropy Collaboration} {et~al.}(2013){Astropy Collaboration},
  {Robitaille}, {Tollerud}, {Greenfield}, {Droettboom}, {Bray}, {Aldcroft},
  {Davis}, {Ginsburg}, {Price-Whelan}, {Kerzendorf}, {Conley}, {Crighton},
  {Barbary}, {Muna}, {Ferguson}, {Grollier}, {Parikh}, {Nair}, {Unther},
  {Deil}, {Woillez}, {Conseil}, {Kramer}, {Turner}, {Singer}, {Fox}, {Weaver},
  {Zabalza}, {Edwards}, {Azalee Bostroem}, {Burke}, {Casey}, {Crawford},
  {Dencheva}, {Ely}, {Jenness}, {Labrie}, {Lim}, {Pierfederici}, {Pontzen},
  {Ptak}, {Refsdal}, {Servillat}, \& {Streicher}}]{Astropy:2013}
{Astropy Collaboration}, {Robitaille}, T.~P., {Tollerud}, E.~J., {et~al.} 2013,
  \aap, 558, A33, \dodoi{10.1051/0004-6361/201322068}

\bibitem[{{Astropy Collaboration} {et~al.}(2018){Astropy Collaboration},
  {Price-Whelan}, {Sip{\H{o}}cz}, {G{\"u}nther}, {Lim}, {Crawford}, {Conseil},
  {Shupe}, {Craig}, {Dencheva}, {Ginsburg}, {VanderPlas}, {Bradley},
  {P{\'e}rez-Su{\'a}rez}, {de Val-Borro}, {Aldcroft}, {Cruz}, {Robitaille},
  {Tollerud}, {Ardelean}, {Babej}, {Bach}, {Bachetti}, {Bakanov}, {Bamford},
  {Barentsen}, {Barmby}, {Baumbach}, {Berry}, {Biscani}, {Boquien}, {Bostroem},
  {Bouma}, {Brammer}, {Bray}, {Breytenbach}, {Buddelmeijer}, {Burke},
  {Calderone}, {Cano Rodr{\'\i}guez}, {Cara}, {Cardoso}, {Cheedella}, {Copin},
  {Corrales}, {Crichton}, {D'Avella}, {Deil}, {Depagne}, {Dietrich}, {Donath},
  {Droettboom}, {Earl}, {Erben}, {Fabbro}, {Ferreira}, {Finethy}, {Fox},
  {Garrison}, {Gibbons}, {Goldstein}, {Gommers}, {Greco}, {Greenfield},
  {Groener}, {Grollier}, {Hagen}, {Hirst}, {Homeier}, {Horton}, {Hosseinzadeh},
  {Hu}, {Hunkeler}, {Ivezi{\'c}}, {Jain}, {Jenness}, {Kanarek}, {Kendrew},
  {Kern}, {Kerzendorf}, {Khvalko}, {King}, {Kirkby}, {Kulkarni}, {Kumar},
  {Lee}, {Lenz}, {Littlefair}, {Ma}, {Macleod}, {Mastropietro}, {McCully},
  {Montagnac}, {Morris}, {Mueller}, {Mumford}, {Muna}, {Murphy}, {Nelson},
  {Nguyen}, {Ninan}, {N{\"o}the}, {Ogaz}, {Oh}, {Parejko}, {Parley}, {Pascual},
  {Patil}, {Patil}, {Plunkett}, {Prochaska}, {Rastogi}, {Reddy Janga},
  {Sabater}, {Sakurikar}, {Seifert}, {Sherbert}, {Sherwood-Taylor}, {Shih},
  {Sick}, {Silbiger}, {Singanamalla}, {Singer}, {Sladen}, {Sooley},
  {Sornarajah}, {Streicher}, {Teuben}, {Thomas}, {Tremblay}, {Turner},
  {Terr{\'o}n}, {van Kerkwijk}, {de la Vega}, {Watkins}, {Weaver}, {Whitmore},
  {Woillez}, {Zabalza}, \& {Astropy Contributors}}]{Astropy:2018}
{Astropy Collaboration}, {Price-Whelan}, A.~M., {Sip{\H{o}}cz}, B.~M., {et~al.}
  2018, \aj, 156, 123, \dodoi{10.3847/1538-3881/aabc4f}

\bibitem[{{Astropy Collaboration} {et~al.}(2022){Astropy Collaboration},
  {Price-Whelan}, {Lim}, {Earl}, {Starkman}, {Bradley}, {Shupe}, {Patil},
  {Corrales}, {Brasseur}, {N{"o}the}, {Donath}, {Tollerud}, {Morris},
  {Ginsburg}, {Vaher}, {Weaver}, {Tocknell}, {Jamieson}, {van Kerkwijk},
  {Robitaille}, {Merry}, {Bachetti}, {G{"u}nther}, {Aldcroft},
  {Alvarado-Montes}, {Archibald}, {B{'o}di}, {Bapat}, {Barentsen}, {Baz{'a}n},
  {Biswas}, {Boquien}, {Burke}, {Cara}, {Cara}, {Conroy}, {Conseil}, {Craig},
  {Cross}, {Cruz}, {D'Eugenio}, {Dencheva}, {Devillepoix}, {Dietrich},
  {Eigenbrot}, {Erben}, {Ferreira}, {Foreman-Mackey}, {Fox}, {Freij}, {Garg},
  {Geda}, {Glattly}, {Gondhalekar}, {Gordon}, {Grant}, {Greenfield}, {Groener},
  {Guest}, {Gurovich}, {Handberg}, {Hart}, {Hatfield-Dodds}, {Homeier},
  {Hosseinzadeh}, {Jenness}, {Jones}, {Joseph}, {Kalmbach}, {Karamehmetoglu},
  {Ka{l}uszy{'n}ski}, {Kelley}, {Kern}, {Kerzendorf}, {Koch}, {Kulumani},
  {Lee}, {Ly}, {Ma}, {MacBride}, {Maljaars}, {Muna}, {Murphy}, {Norman},
  {O'Steen}, {Oman}, {Pacifici}, {Pascual}, {Pascual-Granado}, {Patil},
  {Perren}, {Pickering}, {Rastogi}, {Roulston}, {Ryan}, {Rykoff}, {Sabater},
  {Sakurikar}, {Salgado}, {Sanghi}, {Saunders}, {Savchenko}, {Schwardt},
  {Seifert-Eckert}, {Shih}, {Jain}, {Shukla}, {Sick}, {Simpson},
  {Singanamalla}, {Singer}, {Singhal}, {Sinha}, {Sip{H{o}}cz}, {Spitler},
  {Stansby}, {Streicher}, {{{S}}umak}, {Swinbank}, {Taranu}, {Tewary},
  {Tremblay}, {Val-Borro}, {Van Kooten}, {Vasovi{'c}}, {Verma}, {de Miranda
  Cardoso}, {Williams}, {Wilson}, {Winkel}, {Wood-Vasey}, {Xue}, {Yoachim},
  {Zhang}, {Zonca}, \& {Astropy Project Contributors}}]{Astropy:2022}
{Astropy Collaboration}, {Price-Whelan}, A.~M., {Lim}, P.~L., {et~al.} 2022,
  \apj, 935, 167, \dodoi{10.3847/1538-4357/ac7c74}

\bibitem[{{Baldry} {et~al.}(2012){Baldry}, {Driver}, {Loveday}, {Taylor},
  {Kelvin}, {Liske}, {Norberg}, {Robotham}, {Brough}, {Hopkins}, {Bamford},
  {Peacock}, {Bland-Hawthorn}, {Conselice}, {Croom}, {Jones}, {Parkinson},
  {Popescu}, {Prescott}, {Sharp}, \& {Tuffs}}]{Baldry:2012}
{Baldry}, I.~K., {Driver}, S.~P., {Loveday}, J., {et~al.} 2012, \mnras, 421,
  621, \dodoi{10.1111/j.1365-2966.2012.20340.x}

\bibitem[{{Balestra} {et~al.}(2010){Balestra}, {Mainieri}, {Popesso},
  {Dickinson}, {Nonino}, {Rosati}, {Teimoorinia}, {Vanzella}, {Cristiani},
  {Cesarsky}, {Fosbury}, {Kuntschner}, \& {Rettura}}]{Balestra:2010}
{Balestra}, I., {Mainieri}, V., {Popesso}, P., {et~al.} 2010, \aap, 512, A12,
  \dodoi{10.1051/0004-6361/200913626}

\bibitem[{{Barger} {et~al.}(2008){Barger}, {Cowie}, \& {Wang}}]{Barger:2008}
{Barger}, A.~J., {Cowie}, L.~L., \& {Wang}, W.~H. 2008, \apj, 689, 687,
  \dodoi{10.1086/592735}

\bibitem[{{Barmby} {et~al.}(2008){Barmby}, {Huang}, {Ashby}, {Eisenhardt},
  {Fazio}, {Willner}, \& {Wright}}]{Barmby:2008}
{Barmby}, P., {Huang}, J.~S., {Ashby}, M.~L.~N., {et~al.} 2008, \apjs, 177,
  431, \dodoi{10.1086/588583}

\bibitem[{{Barro} {et~al.}(2019){Barro}, {P{\'e}rez-Gonz{\'a}lez}, {Cava},
  {Brammer}, {Pandya}, {Eliche Moral}, {Esquej}, {Dom{\'\i}nguez-S{\'a}nchez},
  {Alcalde Pampliega}, {Guo}, {Koekemoer}, {Trump}, {Ashby}, {Cardiel},
  {Castellano}, {Conselice}, {Dickinson}, {Dolch}, {Donley}, {Espino Briones},
  {Faber}, {Fazio}, {Ferguson}, {Finkelstein}, {Fontana}, {Galametz},
  {Gardner}, {Gawiser}, {Giavalisco}, {Grazian}, {Grogin}, {Hathi}, {Hemmati},
  {Hern{\'a}n-Caballero}, {Kocevski}, {Koo}, {Kodra}, {Lee}, {Lin}, {Lucas},
  {Mobasher}, {McGrath}, {Nandra}, {Nayyeri}, {Newman}, {Pforr}, {Peth},
  {Rafelski}, {Rodr{\'\i}guez-Munoz}, {Salvato}, {Stefanon}, {van der Wel},
  {Willner}, {Wiklind}, \& {Wuyts}}]{Barro:2019}
{Barro}, G., {P{\'e}rez-Gonz{\'a}lez}, P.~G., {Cava}, A., {et~al.} 2019, \apjs,
  243, 22, \dodoi{10.3847/1538-4365/ab23f2}

\bibitem[{{Ben{\'\i}tez}(2000)}]{Benitez:2000}
{Ben{\'\i}tez}, N. 2000, \apj, 536, 571, \dodoi{10.1086/308947}

\bibitem[{{Ben{\'\i}tez} {et~al.}(2004){Ben{\'\i}tez}, {Ford}, {Bouwens},
  {Menanteau}, {Blakeslee}, {Gronwall}, {Illingworth}, {Meurer}, {Broadhurst},
  {Clampin}, {Franx}, {Hartig}, {Magee}, {Sirianni}, {Ardila}, {Bartko},
  {Brown}, {Burrows}, {Cheng}, {Cross}, {Feldman}, {Golimowski}, {Infante},
  {Kimble}, {Krist}, {Lesser}, {Levay}, {Martel}, {Miley}, {Postman}, {Rosati},
  {Sparks}, {Tran}, {Tsvetanov}, {White}, \& {Zheng}}]{Benitez:2004}
{Ben{\'\i}tez}, N., {Ford}, H., {Bouwens}, R., {et~al.} 2004, \apjs, 150, 1,
  \dodoi{10.1086/380120}

\bibitem[{{Bezanson} {et~al.}(2016){Bezanson}, {Wake}, {Brammer}, {van Dokkum},
  {Franx}, {Labb{\'e}}, {Leja}, {Momcheva}, {Nelson}, {Quadri}, {Skelton},
  {Weiner}, \& {Whitaker}}]{Bezanson:2016}
{Bezanson}, R., {Wake}, D.~A., {Brammer}, G.~B., {et~al.} 2016, \apj, 822, 30,
  \dodoi{10.3847/0004-637X/822/1/30}

\bibitem[{{Bielby} {et~al.}(2012){Bielby}, {Hudelot}, {McCracken}, {Ilbert},
  {Daddi}, {Le F{\`e}vre}, {Gonzalez-Perez}, {Kneib}, {Marmo}, {Mellier},
  {Salvato}, {Sanders}, \& {Willott}}]{Bielby:2012}
{Bielby}, R., {Hudelot}, P., {McCracken}, H.~J., {et~al.} 2012, \aap, 545, A23,
  \dodoi{10.1051/0004-6361/201118547}

\bibitem[{{Blanton} \& {Roweis}(2007)}]{Blanton:2007}
{Blanton}, M.~R., \& {Roweis}, S. 2007, \aj, 133, 734, \dodoi{10.1086/510127}

\bibitem[{{Boogaard} {et~al.}(2018){Boogaard}, {Brinchmann}, {Bouch{\'e}},
  {Paalvast}, {Bacon}, {Bouwens}, {Contini}, {Gunawardhana}, {Inami}, {Marino},
  {Maseda}, {Mitchell}, {Nanayakkara}, {Richard}, {Schaye}, {Schreiber},
  {Tacchella}, {Wisotzki}, \& {Zabl}}]{Boogaard:2018}
{Boogaard}, L.~A., {Brinchmann}, J., {Bouch{\'e}}, N., {et~al.} 2018, \aap,
  619, A27, \dodoi{10.1051/0004-6361/201833136}

\bibitem[{{Boquien} {et~al.}(2019){Boquien}, {Burgarella}, {Roehlly}, {Buat},
  {Ciesla}, {Corre}, {Inoue}, \& {Salas}}]{Boquien:2019}
{Boquien}, M., {Burgarella}, D., {Roehlly}, Y., {et~al.} 2019, \aap, 622, A103,
  \dodoi{10.1051/0004-6361/201834156}

\bibitem[{{Bouwens} {et~al.}(2010){Bouwens}, {Illingworth}, {Oesch},
  {Stiavelli}, {van Dokkum}, {Trenti}, {Magee}, {Labb{\'e}}, {Franx},
  {Carollo}, \& {Gonzalez}}]{Bouwens:2010}
{Bouwens}, R.~J., {Illingworth}, G.~D., {Oesch}, P.~A., {et~al.} 2010, \apjl,
  709, L133, \dodoi{10.1088/2041-8205/709/2/L133}

\bibitem[{{Brammer} {et~al.}(2008){Brammer}, {van Dokkum}, \&
  {Coppi}}]{Brammer:2008}
{Brammer}, G.~B., {van Dokkum}, P.~G., \& {Coppi}, P. 2008, \apj, 686, 1503,
  \dodoi{10.1086/591786}

\bibitem[{{Brammer} {et~al.}(2012){Brammer}, {van Dokkum}, {Franx},
  {Fumagalli}, {Patel}, {Rix}, {Skelton}, {Kriek}, {Nelson}, {Schmidt},
  {Bezanson}, {da Cunha}, {Erb}, {Fan}, {F{\"o}rster Schreiber}, {Illingworth},
  {Labb{\'e}}, {Leja}, {Lundgren}, {Magee}, {Marchesini}, {McCarthy},
  {Momcheva}, {Muzzin}, {Quadri}, {Steidel}, {Tal}, {Wake}, {Whitaker}, \&
  {Williams}}]{Brammer:2012}
{Brammer}, G.~B., {van Dokkum}, P.~G., {Franx}, M., {et~al.} 2012, \apjs, 200,
  13, \dodoi{10.1088/0067-0049/200/2/13}

\bibitem[{{Bruzual} \& {Charlot}(2003)}]{Bruzual:2003}
{Bruzual}, G., \& {Charlot}, S. 2003, \mnras, 344, 1000,
  \dodoi{10.1046/j.1365-8711.2003.06897.x}

\bibitem[{{Burgarella} {et~al.}(2005){Burgarella}, {Buat}, \&
  {Iglesias-P{\'a}ramo}}]{Burgarella:2005}
{Burgarella}, D., {Buat}, V., \& {Iglesias-P{\'a}ramo}, J. 2005, \mnras, 360,
  1413, \dodoi{10.1111/j.1365-2966.2005.09131.x}

\bibitem[{{Byler} {et~al.}(2017){Byler}, {Dalcanton}, {Conroy}, \&
  {Johnson}}]{Byler:2017}
{Byler}, N., {Dalcanton}, J.~J., {Conroy}, C., \& {Johnson}, B.~D. 2017, \apj,
  840, 44, \dodoi{10.3847/1538-4357/aa6c66}

\bibitem[{{Calzetti} {et~al.}(2000){Calzetti}, {Armus}, {Bohlin}, {Kinney},
  {Koornneef}, \& {Storchi-Bergmann}}]{Calzetti:2000}
{Calzetti}, D., {Armus}, L., {Bohlin}, R.~C., {et~al.} 2000, \apj, 533, 682,
  \dodoi{10.1086/308692}

\bibitem[{{Cardelli} {et~al.}(1989){Cardelli}, {Clayton}, \&
  {Mathis}}]{Cardelli:1989}
{Cardelli}, J.~A., {Clayton}, G.~C., \& {Mathis}, J.~S. 1989, \apj, 345, 245,
  \dodoi{10.1086/167900}

\bibitem[{{Chabrier}(2003)}]{Chabrier:2003}
{Chabrier}, G. 2003, \pasp, 115, 763, \dodoi{10.1086/376392}

\bibitem[{{Chabrier} {et~al.}(2014){Chabrier}, {Hennebelle}, \&
  {Charlot}}]{Chabrier:2014}
{Chabrier}, G., {Hennebelle}, P., \& {Charlot}, S. 2014, \apj, 796, 75,
  \dodoi{10.1088/0004-637X/796/2/75}

\bibitem[{{Coe} {et~al.}(2006){Coe}, {Ben{\'\i}tez}, {S{\'a}nchez}, {Jee},
  {Bouwens}, \& {Ford}}]{Coe:2006}
{Coe}, D., {Ben{\'\i}tez}, N., {S{\'a}nchez}, S.~F., {et~al.} 2006, \aj, 132,
  926, \dodoi{10.1086/505530}

\bibitem[{{Coil} {et~al.}(2011){Coil}, {Blanton}, {Burles}, {Cool},
  {Eisenstein}, {Moustakas}, {Wong}, {Zhu}, {Aird}, {Bernstein}, {Bolton}, \&
  {Hogg}}]{Coil:2011}
{Coil}, A.~L., {Blanton}, M.~R., {Burles}, S.~M., {et~al.} 2011, \apj, 741, 8,
  \dodoi{10.1088/0004-637X/741/1/8}

\bibitem[{{Conroy} \& {Gunn}(2010{\natexlab{a}})}]{Conroy:2010}
{Conroy}, C., \& {Gunn}, J.~E. 2010{\natexlab{a}}, \apj, 712, 833,
  \dodoi{10.1088/0004-637X/712/2/833}

\bibitem[{{Conroy} \& {Gunn}(2010{\natexlab{b}})}]{FSPS}
---. 2010{\natexlab{b}}, {FSPS: Flexible Stellar Population Synthesis},
  Astrophysics Source Code Library, record ascl:1010.043.
\newblock \doeprint{1010.043}

\bibitem[{{Conroy} {et~al.}(2009){Conroy}, {Gunn}, \& {White}}]{Conroy:2009}
{Conroy}, C., {Gunn}, J.~E., \& {White}, M. 2009, \apj, 699, 486,
  \dodoi{10.1088/0004-637X/699/1/486}

\bibitem[{{Cooper} {et~al.}(2011){Cooper}, {Aird}, {Coil}, {Davis}, {Faber},
  {Juneau}, {Lotz}, {Nandra}, {Newman}, {Willmer}, \& {Yan}}]{Cooper:2011}
{Cooper}, M.~C., {Aird}, J.~A., {Coil}, A.~L., {et~al.} 2011, \apjs, 193, 14,
  \dodoi{10.1088/0067-0049/193/1/14}

\bibitem[{{Cooper} {et~al.}(2012){Cooper}, {Griffith}, {Newman}, {Coil},
  {Davis}, {Dutton}, {Faber}, {Guhathakurta}, {Koo}, {Lotz}, {Weiner},
  {Willmer}, \& {Yan}}]{Cooper:2012}
{Cooper}, M.~C., {Griffith}, R.~L., {Newman}, J.~A., {et~al.} 2012, \mnras,
  419, 3018, \dodoi{10.1111/j.1365-2966.2011.19938.x}

\bibitem[{{Cristiani} {et~al.}(2000){Cristiani}, {Appenzeller}, {Arnouts},
  {Nonino}, {Arag{\'o}n-Salamanca}, {Benoist}, {da Costa}, {Dennefeld},
  {Rengelink}, {Renzini}, {Szeifert}, \& {White}}]{Cristiani:2000}
{Cristiani}, S., {Appenzeller}, I., {Arnouts}, S., {et~al.} 2000, \aap, 359,
  489, \dodoi{10.48550/arXiv.astro-ph/0004213}

\bibitem[{{Croom} {et~al.}(2001){Croom}, {Smith}, {Boyle}, {Shanks}, {Loaring},
  {Miller}, \& {Lewis}}]{Croom:2001}
{Croom}, S.~M., {Smith}, R.~J., {Boyle}, B.~J., {et~al.} 2001, \mnras, 322,
  L29, \dodoi{10.1046/j.1365-8711.2001.04474.x}

\bibitem[{{Daddi} {et~al.}(2004){Daddi}, {Cimatti}, {Renzini}, {Fontana},
  {Mignoli}, {Pozzetti}, {Tozzi}, \& {Zamorani}}]{Daddi:2004}
{Daddi}, E., {Cimatti}, A., {Renzini}, A., {et~al.} 2004, \apj, 617, 746,
  \dodoi{10.1086/425569}

\bibitem[{{Dahlen} {et~al.}(2013){Dahlen}, {Mobasher}, {Faber}, {Ferguson},
  {Barro}, {Finkelstein}, {Finlator}, {Fontana}, {Gruetzbauch}, {Johnson},
  {Pforr}, {Salvato}, {Wiklind}, {Wuyts}, {Acquaviva}, {Dickinson}, {Guo},
  {Huang}, {Huang}, {Newman}, {Bell}, {Conselice}, {Galametz}, {Gawiser},
  {Giavalisco}, {Grogin}, {Hathi}, {Kocevski}, {Koekemoer}, {Koo}, {Lee},
  {McGrath}, {Papovich}, {Peth}, {Ryan}, {Somerville}, {Weiner}, \&
  {Wilson}}]{Dahlen:2013}
{Dahlen}, T., {Mobasher}, B., {Faber}, S.~M., {et~al.} 2013, \apj, 775, 93,
  \dodoi{10.1088/0004-637X/775/2/93}

\bibitem[{{Dale} {et~al.}(2014){Dale}, {Helou}, {Magdis}, {Armus},
  {D{\'\i}az-Santos}, \& {Shi}}]{Dale:2014}
{Dale}, D.~A., {Helou}, G., {Magdis}, G.~E., {et~al.} 2014, \apj, 784, 83,
  \dodoi{10.1088/0004-637X/784/1/83}

\bibitem[{{Damjanov} {et~al.}(2018){Damjanov}, {Zahid}, {Geller}, {Fabricant},
  \& {Hwang}}]{Damjanov:2018}
{Damjanov}, I., {Zahid}, H.~J., {Geller}, M.~J., {Fabricant}, D.~G., \&
  {Hwang}, H.~S. 2018, \apjs, 234, 21, \dodoi{10.3847/1538-4365/aaa01c}

\bibitem[{{Davidzon} {et~al.}(2017){Davidzon}, {Ilbert}, {Laigle}, {Coupon},
  {McCracken}, {Delvecchio}, {Masters}, {Capak}, {Hsieh}, {Le F{\`e}vre},
  {Tresse}, {Bethermin}, {Chang}, {Faisst}, {Le Floc'h}, {Steinhardt}, {Toft},
  {Aussel}, {Dubois}, {Hasinger}, {Salvato}, {Sanders}, {Scoville}, \&
  {Silverman}}]{Davidzon:2017}
{Davidzon}, I., {Ilbert}, O., {Laigle}, C., {et~al.} 2017, \aap, 605, A70,
  \dodoi{10.1051/0004-6361/201730419}

\bibitem[{{Dickinson} {et~al.}(2003){Dickinson}, {Bergeron}, {Casertano},
  {Cesarsky}, {Chary}, {Cristiani}, {Eisenhardt}, {Elbaz}, {Elston}, {Fall},
  {Ferguson}, {Fosbury}, {Giacconi}, {Giavalisco}, {Grogin}, {Hauser},
  {Hanisch}, {Hook}, {J{\~a}{\c{R}}gensen}, {Koekemoer}, {Ledlow}, {Livio},
  {Mobasher}, {Padovani}, {Papovich}, {Primack}, {Rauscher}, {Reach},
  {Renzini}, {Rieke}, {Rosati}, {Roth}, {Roy}, {Schreier}, {Stern},
  {Stiavelli}, {Takamiya}, {Tollestrup}, {Urry}, {Williams}, {Winge}, \&
  {Wright}}]{Dickinson:2003}
{Dickinson}, M., {Bergeron}, J., {Casertano}, S., {et~al.} 2003, {Great
  Observatories Origins Deep Survey (GOODS) Validation Observations}, Spitzer
  Proposal ID 196

\bibitem[{{Doherty} {et~al.}(2005){Doherty}, {Bunker}, {Ellis}, \&
  {McCarthy}}]{Doherty:2005}
{Doherty}, M., {Bunker}, A.~J., {Ellis}, R.~S., \& {McCarthy}, P.~J. 2005,
  \mnras, 361, 525, \dodoi{10.1111/j.1365-2966.2005.09191.x}

\bibitem[{{Dom{\'\i}nguez} {et~al.}(2015){Dom{\'\i}nguez}, {Siana}, {Brooks},
  {Christensen}, {Bruzual}, {Stark}, \& {Alavi}}]{Dominguez:2015}
{Dom{\'\i}nguez}, A., {Siana}, B., {Brooks}, A.~M., {et~al.} 2015, \mnras, 451,
  839, \dodoi{10.1093/mnras/stv1001}

\bibitem[{{Draine} \& {Li}(2007)}]{Draine:2007}
{Draine}, B.~T., \& {Li}, A. 2007, \apj, 657, 810, \dodoi{10.1086/511055}

\bibitem[{{Emami} {et~al.}(2019){Emami}, {Siana}, {Weisz}, {Johnson}, {Ma}, \&
  {El-Badry}}]{Emami:2019}
{Emami}, N., {Siana}, B., {Weisz}, D.~R., {et~al.} 2019, \apj, 881, 71,
  \dodoi{10.3847/1538-4357/ab211a}

\bibitem[{{Fan} {et~al.}(2006){Fan}, {Strauss}, {Becker}, {White}, {Gunn},
  {Knapp}, {Richards}, {Schneider}, {Brinkmann}, \& {Fukugita}}]{Fan:2006}
{Fan}, X., {Strauss}, M.~A., {Becker}, R.~H., {et~al.} 2006, \aj, 132, 117,
  \dodoi{10.1086/504836}

\bibitem[{{Ferreras} {et~al.}(2009){Ferreras}, {Pasquali}, {Malhotra},
  {Rhoads}, {Cohen}, {Windhorst}, {Pirzkal}, {Grogin}, {Koekemoer}, {Lisker},
  {Panagia}, {Daddi}, \& {Hathi}}]{Ferreras:2009}
{Ferreras}, I., {Pasquali}, A., {Malhotra}, S., {et~al.} 2009, \apj, 706, 158,
  \dodoi{10.1088/0004-637X/706/1/158}

\bibitem[{{Fioc} \& {Rocca-Volmerange}(1997)}]{Fioc:1997}
{Fioc}, M., \& {Rocca-Volmerange}, B. 1997, \aap, 326, 950,
  \dodoi{10.48550/arXiv.astro-ph/9707017}

\bibitem[{{Fontana} {et~al.}(2000){Fontana}, {D'Odorico}, {Poli}, {Giallongo},
  {Arnouts}, {Cristiani}, {Moorwood}, \& {Saracco}}]{Fontana:2000}
{Fontana}, A., {D'Odorico}, S., {Poli}, F., {et~al.} 2000, \aj, 120, 2206,
  \dodoi{10.1086/316803}

\bibitem[{{Fontana} {et~al.}(2014){Fontana}, {Dunlop}, {Paris}, {Targett},
  {Boutsia}, {Castellano}, {Galametz}, {Grazian}, {McLure}, {Merlin},
  {Pentericci}, {Wuyts}, {Almaini}, {Caputi}, {Chary}, {Cirasuolo},
  {Conselice}, {Cooray}, {Daddi}, {Dickinson}, {Faber}, {Fazio}, {Ferguson},
  {Giallongo}, {Giavalisco}, {Grogin}, {Hathi}, {Koekemoer}, {Koo}, {Lucas},
  {Nonino}, {Rix}, {Renzini}, {Rosario}, {Santini}, {Scarlata}, {Sommariva},
  {Stark}, {van der Wel}, {Vanzella}, {Wild}, {Yan}, \&
  {Zibetti}}]{Fontana:2014}
{Fontana}, A., {Dunlop}, J.~S., {Paris}, D., {et~al.} 2014, \aap, 570, A11,
  \dodoi{10.1051/0004-6361/201423543}

\bibitem[{{Giallongo} {et~al.}(1998){Giallongo}, {D'Odorico}, {Fontana},
  {Cristiani}, {Egami}, {Hu}, \& {McMahon}}]{Giallongo:1998}
{Giallongo}, E., {D'Odorico}, S., {Fontana}, A., {et~al.} 1998, \aj, 115, 2169,
  \dodoi{10.1086/300361}

\bibitem[{{Giavalisco} {et~al.}(2004){Giavalisco}, {Ferguson}, {Koekemoer},
  {Dickinson}, {Alexander}, {Bauer}, {Bergeron}, {Biagetti}, {Brandt},
  {Casertano}, {Cesarsky}, {Chatzichristou}, {Conselice}, {Cristiani}, {Da
  Costa}, {Dahlen}, {de Mello}, {Eisenhardt}, {Erben}, {Fall}, {Fassnacht},
  {Fosbury}, {Fruchter}, {Gardner}, {Grogin}, {Hook}, {Hornschemeier}, {Idzi},
  {Jogee}, {Kretchmer}, {Laidler}, {Lee}, {Livio}, {Lucas}, {Madau},
  {Mobasher}, {Moustakas}, {Nonino}, {Padovani}, {Papovich}, {Park},
  {Ravindranath}, {Renzini}, {Richardson}, {Riess}, {Rosati}, {Schirmer},
  {Schreier}, {Somerville}, {Spinrad}, {Stern}, {Stiavelli}, {Strolger},
  {Urry}, {Vandame}, {Williams}, \& {Wolf}}]{Giavalisco:2004}
{Giavalisco}, M., {Ferguson}, H.~C., {Koekemoer}, A.~M., {et~al.} 2004, \apjl,
  600, L93, \dodoi{10.1086/379232}

\bibitem[{{Grazian} {et~al.}(2006){Grazian}, {Fontana}, {de Santis}, {Nonino},
  {Salimbeni}, {Giallongo}, {Cristiani}, {Gallozzi}, \&
  {Vanzella}}]{Grazian:2006}
{Grazian}, A., {Fontana}, A., {de Santis}, C., {et~al.} 2006, \aap, 449, 951,
  \dodoi{10.1051/0004-6361:20053979}

\bibitem[{{Grazian} {et~al.}(2017){Grazian}, {Giallongo}, {Paris}, {Boutsia},
  {Dickinson}, {Santini}, {Windhorst}, {Jansen}, {Cohen}, {Ashcraft},
  {Scarlata}, {Rutkowski}, {Vanzella}, {Cusano}, {Cristiani}, {Giavalisco},
  {Ferguson}, {Koekemoer}, {Grogin}, {Castellano}, {Fiore}, {Fontana},
  {Marchi}, {Pedichini}, {Pentericci}, {Amor{\'\i}n}, {Barro}, {Bonchi},
  {Bongiorno}, {Faber}, {Fumana}, {Galametz}, {Guaita}, {Kocevski}, {Merlin},
  {Nonino}, {O'Connell}, {Pilo}, {Ryan}, {Sani}, {Speziali}, {Testa}, {Weiner},
  \& {Yan}}]{Grazian:2017}
{Grazian}, A., {Giallongo}, E., {Paris}, D., {et~al.} 2017, \aap, 602, A18,
  \dodoi{10.1051/0004-6361/201730447}

\bibitem[{{Grogin} {et~al.}(2011){Grogin}, {Kocevski}, {Faber}, {Ferguson},
  {Koekemoer}, {Riess}, {Acquaviva}, {Alexander}, {Almaini}, {Ashby}, {Barden},
  {Bell}, {Bournaud}, {Brown}, {Caputi}, {Casertano}, {Cassata}, {Castellano},
  {Challis}, {Chary}, {Cheung}, {Cirasuolo}, {Conselice}, {Roshan Cooray},
  {Croton}, {Daddi}, {Dahlen}, {Dav{\'e}}, {de Mello}, {Dekel}, {Dickinson},
  {Dolch}, {Donley}, {Dunlop}, {Dutton}, {Elbaz}, {Fazio}, {Filippenko},
  {Finkelstein}, {Fontana}, {Gardner}, {Garnavich}, {Gawiser}, {Giavalisco},
  {Grazian}, {Guo}, {Hathi}, {H{\"a}ussler}, {Hopkins}, {Huang}, {Huang},
  {Jha}, {Kartaltepe}, {Kirshner}, {Koo}, {Lai}, {Lee}, {Li}, {Lotz}, {Lucas},
  {Madau}, {McCarthy}, {McGrath}, {McIntosh}, {McLure}, {Mobasher},
  {Moustakas}, {Mozena}, {Nandra}, {Newman}, {Niemi}, {Noeske}, {Papovich},
  {Pentericci}, {Pope}, {Primack}, {Rajan}, {Ravindranath}, {Reddy}, {Renzini},
  {Rix}, {Robaina}, {Rodney}, {Rosario}, {Rosati}, {Salimbeni}, {Scarlata},
  {Siana}, {Simard}, {Smidt}, {Somerville}, {Spinrad}, {Straughn}, {Strolger},
  {Telford}, {Teplitz}, {Trump}, {van der Wel}, {Villforth}, {Wechsler},
  {Weiner}, {Wiklind}, {Wild}, {Wilson}, {Wuyts}, {Yan}, \&
  {Yun}}]{Grogin:2011}
{Grogin}, N.~A., {Kocevski}, D.~D., {Faber}, S.~M., {et~al.} 2011, \apjs, 197,
  35, \dodoi{10.1088/0067-0049/197/2/35}

\bibitem[{{Guo} {et~al.}(2013){Guo}, {Ferguson}, {Giavalisco}, {Barro},
  {Willner}, {Ashby}, {Dahlen}, {Donley}, {Faber}, {Fontana}, {Galametz},
  {Grazian}, {Huang}, {Kocevski}, {Koekemoer}, {Koo}, {McGrath}, {Peth},
  {Salvato}, {Wuyts}, {Castellano}, {Cooray}, {Dickinson}, {Dunlop}, {Fazio},
  {Gardner}, {Gawiser}, {Grogin}, {Hathi}, {Hsu}, {Lee}, {Lucas}, {Mobasher},
  {Nandra}, {Newman}, \& {van der Wel}}]{Guo:2013}
{Guo}, Y., {Ferguson}, H.~C., {Giavalisco}, M., {et~al.} 2013, \apjs, 207, 24,
  \dodoi{10.1088/0067-0049/207/2/24}

\bibitem[{{Gwyn}(2012)}]{Gwyn:2012}
{Gwyn}, S. D.~J. 2012, \aj, 143, 38, \dodoi{10.1088/0004-6256/143/2/38}

\bibitem[{Harris {et~al.}(2020)Harris, Millman, van~der Walt, Gommers,
  Virtanen, Cournapeau, Wieser, Taylor, Berg, Smith, Kern, Picus, Hoyer, van
  Kerkwijk, Brett, Haldane, del R{\'{i}}o, Wiebe, Peterson,
  G{\'{e}}rard-Marchant, Sheppard, Reddy, Weckesser, Abbasi, Gohlke, \&
  Oliphant}]{Numpy:2020}
Harris, C.~R., Millman, K.~J., van~der Walt, S.~J., {et~al.} 2020, Nature, 585,
  357, \dodoi{10.1038/s41586-020-2649-2}

\bibitem[{{Hasinger} {et~al.}(2018){Hasinger}, {Capak}, {Salvato}, {Barger},
  {Cowie}, {Faisst}, {Hemmati}, {Kakazu}, {Kartaltepe}, {Masters}, {Mobasher},
  {Nayyeri}, {Sanders}, {Scoville}, {Suh}, {Steinhardt}, \&
  {Yang}}]{Hasinger:2018}
{Hasinger}, G., {Capak}, P., {Salvato}, M., {et~al.} 2018, \apj, 858, 77,
  \dodoi{10.3847/1538-4357/aabacf}

\bibitem[{{Hathi} {et~al.}(2009){Hathi}, {Ferreras}, {Pasquali}, {Malhotra},
  {Rhoads}, {Pirzkal}, {Windhorst}, \& {Xu}}]{Hathi:2009}
{Hathi}, N.~P., {Ferreras}, I., {Pasquali}, A., {et~al.} 2009, \apj, 690, 1866,
  \dodoi{10.1088/0004-637X/690/2/1866}

\bibitem[{{Herenz} {et~al.}(2017){Herenz}, {Urrutia}, {Wisotzki}, {Kerutt},
  {Saust}, {Werhahn}, {Schmidt}, {Caruana}, {Diener}, {Bacon}, {Brinchmann},
  {Schaye}, {Maseda}, \& {Weilbacher}}]{Herenz:2017}
{Herenz}, E.~C., {Urrutia}, T., {Wisotzki}, L., {et~al.} 2017, \aap, 606, A12,
  \dodoi{10.1051/0004-6361/201731055}

\bibitem[{{Hildebrandt} {et~al.}(2010){Hildebrandt}, {Arnouts}, {Capak},
  {Moustakas}, {Wolf}, {Abdalla}, {Assef}, {Banerji}, {Ben{\'\i}tez},
  {Brammer}, {Budav{\'a}ri}, {Carliles}, {Coe}, {Dahlen}, {Feldmann}, {Gerdes},
  {Gillis}, {Ilbert}, {Kotulla}, {Lahav}, {Li}, {Miralles}, {Purger},
  {Schmidt}, \& {Singal}}]{Hildebrandt:2010}
{Hildebrandt}, H., {Arnouts}, S., {Capak}, P., {et~al.} 2010, \aap, 523, A31,
  \dodoi{10.1051/0004-6361/201014885}

\bibitem[{{Hopkins}(2018)}]{Hopkins:2018}
{Hopkins}, A.~M. 2018, \pasa, 35, e039, \dodoi{10.1017/pasa.2018.29}

\bibitem[{{Hopkins} {et~al.}(2014){Hopkins}, {Kere{\v{s}}}, {O{\~n}orbe},
  {Faucher-Gigu{\`e}re}, {Quataert}, {Murray}, \& {Bullock}}]{Hopkins:2014}
{Hopkins}, P.~F., {Kere{\v{s}}}, D., {O{\~n}orbe}, J., {et~al.} 2014, \mnras,
  445, 581, \dodoi{10.1093/mnras/stu1738}

\bibitem[{{Hsu} {et~al.}(2019){Hsu}, {Lin}, {Dickinson}, {Yan}, {Bau-Ching},
  {Wang}, {Lee}, {Yan}, {Scott}, {Willner}, {Ouchi}, {Ashby}, {Chen}, {Daddi},
  {Elbaz}, {Fazio}, {Foucaud}, {Huang}, {Koo}, {Morrison}, {Owen}, {Pannella},
  {Pope}, {Simard}, \& {Wang}}]{Hsu:2019}
{Hsu}, L.-T., {Lin}, L., {Dickinson}, M., {et~al.} 2019, \apj, 871, 233,
  \dodoi{10.3847/1538-4357/aaf9a7}

\bibitem[{{Huang} {et~al.}(2009){Huang}, {Faber}, {Daddi}, {Laird}, {Lai},
  {Omont}, {Wu}, {Younger}, {Bundy}, {Cattaneo}, {Chapman}, {Conselice},
  {Dickinson}, {Egami}, {Fazio}, {Im}, {Koo}, {Le Floc'h}, {Papovich},
  {Rigopoulou}, {Smail}, {Song}, {Van de Werf}, {Webb}, {Willmer}, {Willner},
  \& {Yan}}]{Huang:2009}
{Huang}, J.~S., {Faber}, S.~M., {Daddi}, E., {et~al.} 2009, \apj, 700, 183,
  \dodoi{10.1088/0004-637X/700/1/183}

\bibitem[{Hunter(2007)}]{Matplotlib:2007}
Hunter, J.~D. 2007, Computing in Science \& Engineering, 9, 90,
  \dodoi{10.1109/MCSE.2007.55}

\bibitem[{{Ilbert} {et~al.}(2006){Ilbert}, {Arnouts}, {McCracken},
  {Bolzonella}, {Bertin}, {Le F{\`e}vre}, {Mellier}, {Zamorani}, {Pell{\`o}},
  {Iovino}, {Tresse}, {Le Brun}, {Bottini}, {Garilli}, {Maccagni}, {Picat},
  {Scaramella}, {Scodeggio}, {Vettolani}, {Zanichelli}, {Adami}, {Bardelli},
  {Cappi}, {Charlot}, {Ciliegi}, {Contini}, {Cucciati}, {Foucaud}, {Franzetti},
  {Gavignaud}, {Guzzo}, {Marano}, {Marinoni}, {Mazure}, {Meneux}, {Merighi},
  {Paltani}, {Pollo}, {Pozzetti}, {Radovich}, {Zucca}, {Bondi}, {Bongiorno},
  {Busarello}, {de La Torre}, {Gregorini}, {Lamareille}, {Mathez}, {Merluzzi},
  {Ripepi}, {Rizzo}, \& {Vergani}}]{Ilbert:2006}
{Ilbert}, O., {Arnouts}, S., {McCracken}, H.~J., {et~al.} 2006, \aap, 457, 841,
  \dodoi{10.1051/0004-6361:20065138}

\bibitem[{{Ilbert} {et~al.}(2009){Ilbert}, {Capak}, {Salvato}, {Aussel},
  {McCracken}, {Sanders}, {Scoville}, {Kartaltepe}, {Arnouts}, {Le Floc'h},
  {Mobasher}, {Taniguchi}, {Lamareille}, {Leauthaud}, {Sasaki}, {Thompson},
  {Zamojski}, {Zamorani}, {Bardelli}, {Bolzonella}, {Bongiorno}, {Brusa},
  {Caputi}, {Carollo}, {Contini}, {Cook}, {Coppa}, {Cucciati}, {de la Torre},
  {de Ravel}, {Franzetti}, {Garilli}, {Hasinger}, {Iovino}, {Kampczyk},
  {Kneib}, {Knobel}, {Kovac}, {Le Borgne}, {Le Brun}, {Le F{\`e}vre}, {Lilly},
  {Looper}, {Maier}, {Mainieri}, {Mellier}, {Mignoli}, {Murayama}, {Pell{\`o}},
  {Peng}, {P{\'e}rez-Montero}, {Renzini}, {Ricciardelli}, {Schiminovich},
  {Scodeggio}, {Shioya}, {Silverman}, {Surace}, {Tanaka}, {Tasca}, {Tresse},
  {Vergani}, \& {Zucca}}]{Ilbert:2009}
{Ilbert}, O., {Capak}, P., {Salvato}, M., {et~al.} 2009, \apj, 690, 1236,
  \dodoi{10.1088/0004-637X/690/2/1236}

\bibitem[{{Inami} {et~al.}(2017){Inami}, {Bacon}, {Brinchmann}, {Richard},
  {Contini}, {Conseil}, {Hamer}, {Akhlaghi}, {Bouch{\'e}}, {Cl{\'e}ment},
  {Desprez}, {Drake}, {Hashimoto}, {Leclercq}, {Maseda}, {Michel-Dansac},
  {Paalvast}, {Tresse}, {Ventou}, {Kollatschny}, {Boogaard}, {Finley},
  {Marino}, {Schaye}, \& {Wisotzki}}]{Inami:2017}
{Inami}, H., {Bacon}, R., {Brinchmann}, J., {et~al.} 2017, \aap, 608, A2,
  \dodoi{10.1051/0004-6361/201731195}

\bibitem[{{Inoue}(2011)}]{Inoue:2011}
{Inoue}, A.~K. 2011, \mnras, 415, 2920,
  \dodoi{10.1111/j.1365-2966.2011.18906.x}

\bibitem[{{Iyer} \& {Gawiser}(2017)}]{Iyer:2017}
{Iyer}, K., \& {Gawiser}, E. 2017, \apj, 838, 127,
  \dodoi{10.3847/1538-4357/aa63f0}

\bibitem[{{Iyer} {et~al.}(2019){Iyer}, {Gawiser}, {Faber}, {Ferguson},
  {Kartaltepe}, {Koekemoer}, {Pacifici}, \& {Somerville}}]{Iyer:2019}
{Iyer}, K.~G., {Gawiser}, E., {Faber}, S.~M., {et~al.} 2019, \apj, 879, 116,
  \dodoi{10.3847/1538-4357/ab2052}

\bibitem[{{Jain} {et~al.}(2024){Jain}, {Tacchella}, \& {Mosleh}}]{Jain:2024}
{Jain}, S., {Tacchella}, S., \& {Mosleh}, M. 2024, \mnras, 527, 3291,
  \dodoi{10.1093/mnras/stad3333}

\bibitem[{{Kajisawa} {et~al.}(2011){Kajisawa}, {Ichikawa}, {Tanaka}, {Yamada},
  {Akiyama}, {Suzuki}, {Tokoku}, {Katsuno Uchimoto}, {Konishi}, {Yoshikawa},
  {Nishimura}, {Omata}, {Ouchi}, {Iwata}, {Hamana}, \&
  {Onodera}}]{Kajisawa:2011}
{Kajisawa}, M., {Ichikawa}, T., {Tanaka}, I., {et~al.} 2011, \pasj, 63, 379,
  \dodoi{10.1093/pasj/63.sp2.S379}

\bibitem[{{Kaushal} {et~al.}(2024){Kaushal}, {Nersesian}, {Bezanson}, {van der
  Wel}, {Leja}, {Carnall}, {Gallazzi}, {Zibetti}, {Khullar}, {Franx}, {Muzzin},
  {de Graaff}, {Pacifici}, {Whitaker}, {Bell}, \& {Martorano}}]{Kaushal:2024}
{Kaushal}, Y., {Nersesian}, A., {Bezanson}, R., {et~al.} 2024, \apj, 961, 118,
  \dodoi{10.3847/1538-4357/ad0c4e}

\bibitem[{{Kodra} {et~al.}(2023){Kodra}, {Andrews}, {Newman}, {Finkelstein},
  {Fontana}, {Hathi}, {Salvato}, {Wiklind}, {Wuyts}, {Broussard}, {Chartab},
  {Conselice}, {Cooper}, {Dekel}, {Dickinson}, {Ferguson}, {Gawiser}, {Grogin},
  {Iyer}, {Kartaltepe}, {Kassin}, {Koekemoer}, {Koo}, {Lucas}, {Mantha},
  {McIntosh}, {Mobasher}, {Pacifici}, {P{\'e}rez-Gonz{\'a}lez}, \&
  {Santini}}]{Kodra:2023}
{Kodra}, D., {Andrews}, B.~H., {Newman}, J.~A., {et~al.} 2023, \apj, 942, 36,
  \dodoi{10.3847/1538-4357/ac9f12}

\bibitem[{{Koekemoer} {et~al.}(2011){Koekemoer}, {Faber}, {Ferguson}, {Grogin},
  {Kocevski}, {Koo}, {Lai}, {Lotz}, {Lucas}, {McGrath}, {Ogaz}, {Rajan},
  {Riess}, {Rodney}, {Strolger}, {Casertano}, {Castellano}, {Dahlen},
  {Dickinson}, {Dolch}, {Fontana}, {Giavalisco}, {Grazian}, {Guo}, {Hathi},
  {Huang}, {van der Wel}, {Yan}, {Acquaviva}, {Alexander}, {Almaini}, {Ashby},
  {Barden}, {Bell}, {Bournaud}, {Brown}, {Caputi}, {Cassata}, {Challis},
  {Chary}, {Cheung}, {Cirasuolo}, {Conselice}, {Roshan Cooray}, {Croton},
  {Daddi}, {Dav{\'e}}, {de Mello}, {de Ravel}, {Dekel}, {Donley}, {Dunlop},
  {Dutton}, {Elbaz}, {Fazio}, {Filippenko}, {Finkelstein}, {Frazer}, {Gardner},
  {Garnavich}, {Gawiser}, {Gruetzbauch}, {Hartley}, {H{\"a}ussler},
  {Herrington}, {Hopkins}, {Huang}, {Jha}, {Johnson}, {Kartaltepe},
  {Khostovan}, {Kirshner}, {Lani}, {Lee}, {Li}, {Madau}, {McCarthy},
  {McIntosh}, {McLure}, {McPartland}, {Mobasher}, {Moreira}, {Mortlock},
  {Moustakas}, {Mozena}, {Nandra}, {Newman}, {Nielsen}, {Niemi}, {Noeske},
  {Papovich}, {Pentericci}, {Pope}, {Primack}, {Ravindranath}, {Reddy},
  {Renzini}, {Rix}, {Robaina}, {Rosario}, {Rosati}, {Salimbeni}, {Scarlata},
  {Siana}, {Simard}, {Smidt}, {Snyder}, {Somerville}, {Spinrad}, {Straughn},
  {Telford}, {Teplitz}, {Trump}, {Vargas}, {Villforth}, {Wagner}, {Wandro},
  {Wechsler}, {Weiner}, {Wiklind}, {Wild}, {Wilson}, {Wuyts}, \&
  {Yun}}]{Koekemoer:2011}
{Koekemoer}, A.~M., {Faber}, S.~M., {Ferguson}, H.~C., {et~al.} 2011, \apjs,
  197, 36, \dodoi{10.1088/0067-0049/197/2/36}

\bibitem[{{Kriek} {et~al.}(2015){Kriek}, {Shapley}, {Reddy}, {Siana}, {Coil},
  {Mobasher}, {Freeman}, {de Groot}, {Price}, {Sanders}, {Shivaei}, {Brammer},
  {Momcheva}, {Skelton}, {van Dokkum}, {Whitaker}, {Aird}, {Azadi}, {Kassis},
  {Bullock}, {Conroy}, {Dav{\'e}}, {Kere{\v{s}}}, \& {Krumholz}}]{Kriek:2015}
{Kriek}, M., {Shapley}, A.~E., {Reddy}, N.~A., {et~al.} 2015, \apjs, 218, 15,
  \dodoi{10.1088/0067-0049/218/2/15}

\bibitem[{{Krogager} {et~al.}(2014){Krogager}, {Zirm}, {Toft}, {Man}, \&
  {Brammer}}]{Krogager:2014}
{Krogager}, J.~K., {Zirm}, A.~W., {Toft}, S., {Man}, A., \& {Brammer}, G. 2014,
  \apj, 797, 17, \dodoi{10.1088/0004-637X/797/1/17}

\bibitem[{{Kurczynski} {et~al.}(2016){Kurczynski}, {Gawiser}, {Acquaviva},
  {Bell}, {Dekel}, {de Mello}, {Ferguson}, {Gardner}, {Grogin}, {Guo},
  {Hopkins}, {Koekemoer}, {Koo}, {Lee}, {Mobasher}, {Primack}, {Rafelski},
  {Soto}, \& {Teplitz}}]{Kurczynski:2016}
{Kurczynski}, P., {Gawiser}, E., {Acquaviva}, V., {et~al.} 2016, \apjl, 820,
  L1, \dodoi{10.3847/2041-8205/820/1/L1}

\bibitem[{{Kurk} {et~al.}(2013){Kurk}, {Cimatti}, {Daddi}, {Mignoli},
  {Pozzetti}, {Dickinson}, {Bolzonella}, {Zamorani}, {Cassata}, {Rodighiero},
  {Franceschini}, {Renzini}, {Rosati}, {Halliday}, \& {Berta}}]{Kurk:2013}
{Kurk}, J., {Cimatti}, A., {Daddi}, E., {et~al.} 2013, \aap, 549, A63,
  \dodoi{10.1051/0004-6361/201117847}

\bibitem[{{Laidler} {et~al.}(2007){Laidler}, {Papovich}, {Grogin}, {Idzi},
  {Dickinson}, {Ferguson}, {Hilbert}, {Clubb}, \&
  {Ravindranath}}]{Laidler:2007}
{Laidler}, V.~G., {Papovich}, C., {Grogin}, N.~A., {et~al.} 2007, \pasp, 119,
  1325, \dodoi{10.1086/523898}

\bibitem[{{Le F{\`e}vre} {et~al.}(2013){Le F{\`e}vre}, {Cassata}, {Cucciati},
  {Garilli}, {Ilbert}, {Le Brun}, {Maccagni}, {Moreau}, {Scodeggio}, {Tresse},
  {Zamorani}, {Adami}, {Arnouts}, {Bardelli}, {Bolzonella}, {Bondi},
  {Bongiorno}, {Bottini}, {Cappi}, {Charlot}, {Ciliegi}, {Contini}, {de la
  Torre}, {Foucaud}, {Franzetti}, {Gavignaud}, {Guzzo}, {Iovino}, {Lemaux},
  {L{\'o}pez-Sanjuan}, {McCracken}, {Marano}, {Marinoni}, {Mazure}, {Mellier},
  {Merighi}, {Merluzzi}, {Paltani}, {Pell{\`o}}, {Pollo}, {Pozzetti},
  {Scaramella}, {Tasca}, {Vergani}, {Vettolani}, {Zanichelli}, \&
  {Zucca}}]{LeFevre:2013}
{Le F{\`e}vre}, O., {Cassata}, P., {Cucciati}, O., {et~al.} 2013, \aap, 559,
  A14, \dodoi{10.1051/0004-6361/201322179}

\bibitem[{{Le F{\`e}vre} {et~al.}(2015){Le F{\`e}vre}, {Tasca}, {Cassata},
  {Garilli}, {Le Brun}, {Maccagni}, {Pentericci}, {Thomas}, {Vanzella},
  {Zamorani}, {Zucca}, {Amorin}, {Bardelli}, {Capak}, {Cassar{\`a}},
  {Castellano}, {Cimatti}, {Cuby}, {Cucciati}, {de la Torre}, {Durkalec},
  {Fontana}, {Giavalisco}, {Grazian}, {Hathi}, {Ilbert}, {Lemaux}, {Moreau},
  {Paltani}, {Ribeiro}, {Salvato}, {Schaerer}, {Scodeggio}, {Sommariva},
  {Talia}, {Taniguchi}, {Tresse}, {Vergani}, {Wang}, {Charlot}, {Contini},
  {Fotopoulou}, {L{\'o}pez-Sanjuan}, {Mellier}, \& {Scoville}}]{LeFevre:2015}
{Le F{\`e}vre}, O., {Tasca}, L.~A.~M., {Cassata}, P., {et~al.} 2015, \aap, 576,
  A79, \dodoi{10.1051/0004-6361/201423829}

\bibitem[{{Leja} {et~al.}(2019){Leja}, {Carnall}, {Johnson}, {Conroy}, \&
  {Speagle}}]{Leja:2019}
{Leja}, J., {Carnall}, A.~C., {Johnson}, B.~D., {Conroy}, C., \& {Speagle},
  J.~S. 2019, \apj, 876, 3, \dodoi{10.3847/1538-4357/ab133c}

\bibitem[{{Lilly} {et~al.}(2007){Lilly}, {Le F{\`e}vre}, {Renzini}, {Zamorani},
  {Scodeggio}, {Contini}, {Carollo}, {Hasinger}, {Kneib}, {Iovino}, {Le Brun},
  {Maier}, {Mainieri}, {Mignoli}, {Silverman}, {Tasca}, {Bolzonella},
  {Bongiorno}, {Bottini}, {Capak}, {Caputi}, {Cimatti}, {Cucciati}, {Daddi},
  {Feldmann}, {Franzetti}, {Garilli}, {Guzzo}, {Ilbert}, {Kampczyk}, {Kovac},
  {Lamareille}, {Leauthaud}, {Le Borgne}, {McCracken}, {Marinoni}, {Pello},
  {Ricciardelli}, {Scarlata}, {Vergani}, {Sanders}, {Schinnerer}, {Scoville},
  {Taniguchi}, {Arnouts}, {Aussel}, {Bardelli}, {Brusa}, {Cappi}, {Ciliegi},
  {Finoguenov}, {Foucaud}, {Franceschini}, {Halliday}, {Impey}, {Knobel},
  {Koekemoer}, {Kurk}, {Maccagni}, {Maddox}, {Marano}, {Marconi}, {Meneux},
  {Mobasher}, {Moreau}, {Peacock}, {Porciani}, {Pozzetti}, {Scaramella},
  {Schiminovich}, {Shopbell}, {Smail}, {Thompson}, {Tresse}, {Vettolani},
  {Zanichelli}, \& {Zucca}}]{Lilly:2007}
{Lilly}, S.~J., {Le F{\`e}vre}, O., {Renzini}, A., {et~al.} 2007, \apjs, 172,
  70, \dodoi{10.1086/516589}

\bibitem[{{Lilly} {et~al.}(2009){Lilly}, {Le Brun}, {Maier}, {Mainieri},
  {Mignoli}, {Scodeggio}, {Zamorani}, {Carollo}, {Contini}, {Kneib}, {Le
  F{\`e}vre}, {Renzini}, {Bardelli}, {Bolzonella}, {Bongiorno}, {Caputi},
  {Coppa}, {Cucciati}, {de la Torre}, {de Ravel}, {Franzetti}, {Garilli},
  {Iovino}, {Kampczyk}, {Kovac}, {Knobel}, {Lamareille}, {Le Borgne}, {Pello},
  {Peng}, {P{\'e}rez-Montero}, {Ricciardelli}, {Silverman}, {Tanaka}, {Tasca},
  {Tresse}, {Vergani}, {Zucca}, {Ilbert}, {Salvato}, {Oesch}, {Abbas},
  {Bottini}, {Capak}, {Cappi}, {Cassata}, {Cimatti}, {Elvis}, {Fumana},
  {Guzzo}, {Hasinger}, {Koekemoer}, {Leauthaud}, {Maccagni}, {Marinoni},
  {McCracken}, {Memeo}, {Meneux}, {Porciani}, {Pozzetti}, {Sanders},
  {Scaramella}, {Scarlata}, {Scoville}, {Shopbell}, \&
  {Taniguchi}}]{Lilly:2009}
{Lilly}, S.~J., {Le Brun}, V., {Maier}, C., {et~al.} 2009, \apjs, 184, 218,
  \dodoi{10.1088/0067-0049/184/2/218}

\bibitem[{{Lower} {et~al.}(2020){Lower}, {Narayanan}, {Leja}, {Johnson},
  {Conroy}, \& {Dav{\'e}}}]{Lower:2020}
{Lower}, S., {Narayanan}, D., {Leja}, J., {et~al.} 2020, \apj, 904, 33,
  \dodoi{10.3847/1538-4357/abbfa7}

\bibitem[{{Madau}(1995)}]{Madau:1995}
{Madau}, P. 1995, \apj, 441, 18, \dodoi{10.1086/175332}

\bibitem[{{Madau} \& {Dickinson}(2014)}]{Madau:2014}
{Madau}, P., \& {Dickinson}, M. 2014, \araa, 52, 415,
  \dodoi{10.1146/annurev-astro-081811-125615}

\bibitem[{{Maraston}(2005)}]{Maraston:2005}
{Maraston}, C. 2005, \mnras, 362, 799, \dodoi{10.1111/j.1365-2966.2005.09270.x}

\bibitem[{{Marchesini} {et~al.}(2009){Marchesini}, {van Dokkum}, {F{\"o}rster
  Schreiber}, {Franx}, {Labb{\'e}}, \& {Wuyts}}]{Marchesini:2009}
{Marchesini}, D., {van Dokkum}, P.~G., {F{\"o}rster Schreiber}, N.~M., {et~al.}
  2009, \apj, 701, 1765, \dodoi{10.1088/0004-637X/701/2/1765}

\bibitem[{{Masters} {et~al.}(2019){Masters}, {Stern}, {Cohen}, {Capak},
  {Stanford}, {Hernitschek}, {Galametz}, {Davidzon}, {Rhodes}, {Sanders},
  {Mobasher}, {Castander}, {Pruett}, \& {Fotopoulou}}]{Masters:2019}
{Masters}, D.~C., {Stern}, D.~K., {Cohen}, J.~G., {et~al.} 2019, \apj, 877, 81,
  \dodoi{10.3847/1538-4357/ab184d}

\bibitem[{{McCracken} {et~al.}(2012){McCracken}, {Milvang-Jensen}, {Dunlop},
  {Franx}, {Fynbo}, {Le F{\`e}vre}, {Holt}, {Caputi}, {Goranova}, {Buitrago},
  {Emerson}, {Freudling}, {Hudelot}, {L{\'o}pez-Sanjuan}, {Magnard}, {Mellier},
  {M{\o}ller}, {Nilsson}, {Sutherland}, {Tasca}, \& {Zabl}}]{McCracken:2012}
{McCracken}, H.~J., {Milvang-Jensen}, B., {Dunlop}, J., {et~al.} 2012, \aap,
  544, A156, \dodoi{10.1051/0004-6361/201219507}

\bibitem[{{McLure} {et~al.}(2018){McLure}, {Pentericci}, {Cimatti}, {Dunlop},
  {Elbaz}, {Fontana}, {Nandra}, {Amorin}, {Bolzonella}, {Bongiorno}, {Carnall},
  {Castellano}, {Cirasuolo}, {Cucciati}, {Cullen}, {De Barros}, {Finkelstein},
  {Fontanot}, {Franzetti}, {Fumana}, {Gargiulo}, {Garilli}, {Guaita},
  {Hartley}, {Iovino}, {Jarvis}, {Juneau}, {Karman}, {Maccagni}, {Marchi},
  {M{\'a}rmol-Queralt{\'o}}, {Pompei}, {Pozzetti}, {Scodeggio}, {Sommariva},
  {Talia}, {Almaini}, {Balestra}, {Bardelli}, {Bell}, {Bourne}, {Bowler},
  {Brusa}, {Buitrago}, {Caputi}, {Cassata}, {Charlot}, {Citro}, {Cresci},
  {Cristiani}, {Curtis-Lake}, {Dickinson}, {Fazio}, {Ferguson}, {Fiore},
  {Franco}, {Fynbo}, {Galametz}, {Georgakakis}, {Giavalisco}, {Grazian},
  {Hathi}, {Jung}, {Kim}, {Koekemoer}, {Khusanova}, {Le F{\`e}vre}, {Lotz},
  {Mannucci}, {Maltby}, {Matsuoka}, {McLeod}, {Mendez-Hernandez},
  {Mendez-Abreu}, {Mignoli}, {Moresco}, {Mortlock}, {Nonino}, {Pannella},
  {Papovich}, {Popesso}, {Rosario}, {Salvato}, {Santini}, {Schaerer},
  {Schreiber}, {Stark}, {Tasca}, {Thomas}, {Treu}, {Vanzella}, {Wild},
  {Williams}, {Zamorani}, \& {Zucca}}]{McLure:2018}
{McLure}, R.~J., {Pentericci}, L., {Cimatti}, A., {et~al.} 2018, \mnras, 479,
  25, \dodoi{10.1093/mnras/sty1213}

\bibitem[{{Mehta} {et~al.}(2017){Mehta}, {Scarlata}, {Rafelski}, {Gburek},
  {Teplitz}, {Alavi}, {Boylan-Kolchin}, {Finkelstein}, {Gardner}, {Grogin},
  {Koekemoer}, {Kurczynski}, {Siana}, {Codoreanu}, {de Mello}, {Lee}, \&
  {Soto}}]{Mehta:2017}
{Mehta}, V., {Scarlata}, C., {Rafelski}, M., {et~al.} 2017, \apj, 838, 29,
  \dodoi{10.3847/1538-4357/aa6259}

\bibitem[{{Mehta} {et~al.}(2018){Mehta}, {Scarlata}, {Capak}, {Davidzon},
  {Faisst}, {Hsieh}, {Ilbert}, {Jarvis}, {Laigle}, {Phillips}, {Silverman},
  {Strauss}, {Tanaka}, {Bowler}, {Coupon}, {Foucaud}, {Hemmati}, {Masters},
  {McCracken}, {Mobasher}, {Ouchi}, {Shibuya}, \& {Wang}}]{Mehta:2018}
{Mehta}, V., {Scarlata}, C., {Capak}, P., {et~al.} 2018, \apjs, 235, 36,
  \dodoi{10.3847/1538-4365/aab60c}

\bibitem[{{Mehta} {et~al.}(2023){Mehta}, {Teplitz}, {Scarlata}, {Wang},
  {Alavi}, {Colbert}, {Rafelski}, {Grogin}, {Koekemoer}, {Prichard},
  {Windhorst}, {Barber}, {Conselice}, {Dai}, {Gardner}, {Gawiser}, {Guo},
  {Hathi}, {Arrabal Haro}, {Hayes}, {Iyer}, {Jansen}, {Ji}, {Kurczynski},
  {Kuschel}, {Lucas}, {Mantha}, {O'Connell}, {Ravindranath}, {Robertson},
  {Rutkowski}, {Siana}, \& {Yung}}]{Mehta:2023}
{Mehta}, V., {Teplitz}, H.~I., {Scarlata}, C., {et~al.} 2023, \apj, 952, 133,
  \dodoi{10.3847/1538-4357/acd9cf}

\bibitem[{{Meiksin}(2006)}]{Meiksin:2006}
{Meiksin}, A. 2006, \mnras, 365, 807, \dodoi{10.1111/j.1365-2966.2005.09756.x}

\bibitem[{{M{\'e}rida} {et~al.}(2023){M{\'e}rida}, {P{\'e}rez-Gonz{\'a}lez},
  {S{\'a}nchez-Bl{\'a}zquez}, {Garc{\'\i}a-Argum{\'a}nez}, {Annunziatella},
  {Costantin}, {Lumbreras-Calle}, {Alcalde-Pampliega}, {Barro},
  {Espino-Briones}, \& {Koekemoer}}]{Merida:2023}
{M{\'e}rida}, R.~M., {P{\'e}rez-Gonz{\'a}lez}, P.~G.,
  {S{\'a}nchez-Bl{\'a}zquez}, P., {et~al.} 2023, \apj, 950, 125,
  \dodoi{10.3847/1538-4357/acc7a3}

\bibitem[{{Mignoli} {et~al.}(2005){Mignoli}, {Cimatti}, {Zamorani}, {Pozzetti},
  {Daddi}, {Renzini}, {Broadhurst}, {Cristiani}, {D'Odorico}, {Fontana},
  {Giallongo}, {Gilmozzi}, {Menci}, \& {Saracco}}]{Mignoli:2005}
{Mignoli}, M., {Cimatti}, A., {Zamorani}, G., {et~al.} 2005, \aap, 437, 883,
  \dodoi{10.1051/0004-6361:20042434}

\bibitem[{{Momcheva} {et~al.}(2016){Momcheva}, {Brammer}, {van Dokkum},
  {Skelton}, {Whitaker}, {Nelson}, {Fumagalli}, {Maseda}, {Leja}, {Franx},
  {Rix}, {Bezanson}, {Da Cunha}, {Dickey}, {F{\"o}rster Schreiber},
  {Illingworth}, {Kriek}, {Labb{\'e}}, {Ulf Lange}, {Lundgren}, {Magee},
  {Marchesini}, {Oesch}, {Pacifici}, {Patel}, {Price}, {Tal}, {Wake}, {van der
  Wel}, \& {Wuyts}}]{Momcheva:2016}
{Momcheva}, I.~G., {Brammer}, G.~B., {van Dokkum}, P.~G., {et~al.} 2016, \apjs,
  225, 27, \dodoi{10.3847/0067-0049/225/2/27}

\bibitem[{{Morishita} {et~al.}(2017){Morishita}, {Abramson}, {Treu}, {Vulcani},
  {Schmidt}, {Dressler}, {Poggianti}, {Malkan}, {Wang}, {Huang}, {Trenti},
  {Brada{\v{c}}}, \& {Hoag}}]{Morishita:2017}
{Morishita}, T., {Abramson}, L.~E., {Treu}, T., {et~al.} 2017, \apj, 835, 254,
  \dodoi{10.3847/1538-4357/835/2/254}

\bibitem[{{Moutard} {et~al.}(2020){Moutard}, {Sawicki}, {Arnouts}, {Golob},
  {Coupon}, {Ilbert}, {Yang}, \& {Gwyn}}]{Moutard:2020}
{Moutard}, T., {Sawicki}, M., {Arnouts}, S., {et~al.} 2020, \mnras, 494, 1894,
  \dodoi{10.1093/mnras/staa706}

\bibitem[{{Muzzin} {et~al.}(2013){Muzzin}, {Marchesini}, {Stefanon}, {Franx},
  {McCracken}, {Milvang-Jensen}, {Dunlop}, {Fynbo}, {Brammer}, {Labb{\'e}}, \&
  {van Dokkum}}]{Muzzin:2013}
{Muzzin}, A., {Marchesini}, D., {Stefanon}, M., {et~al.} 2013, \apj, 777, 18,
  \dodoi{10.1088/0004-637X/777/1/18}

\bibitem[{{Nayyeri} {et~al.}(2017){Nayyeri}, {Hemmati}, {Mobasher}, {Ferguson},
  {Cooray}, {Barro}, {Faber}, {Dickinson}, {Koekemoer}, {Peth}, {Salvato},
  {Ashby}, {Darvish}, {Donley}, {Durbin}, {Finkelstein}, {Fontana}, {Grogin},
  {Gruetzbauch}, {Huang}, {Khostovan}, {Kocevski}, {Kodra}, {Lee}, {Newman},
  {Pacifici}, {Pforr}, {Stefanon}, {Wiklind}, {Willner}, {Wuyts}, {Castellano},
  {Conselice}, {Dolch}, {Dunlop}, {Galametz}, {Hathi}, {Lucas}, \&
  {Yan}}]{Nayyeri:2017}
{Nayyeri}, H., {Hemmati}, S., {Mobasher}, B., {et~al.} 2017, \apjs, 228, 7,
  \dodoi{10.3847/1538-4365/228/1/7}

\bibitem[{{Nedkova} {et~al.}(2021){Nedkova}, {H{\"a}u{\ss}ler}, {Marchesini},
  {Dimauro}, {Brammer}, {Eigenthaler}, {Feinstein}, {Ferguson},
  {Huertas-Company}, {Johnston}, {Kado-Fong}, {Kartaltepe}, {Labb{\'e}},
  {Lange-Vagle}, {Martis}, {McGrath}, {Muzzin}, {Oesch}, {Ordenes-Brice{\~n}o},
  {Puzia}, {Shipley}, {Simmons}, {Skelton}, {Stefanon}, {van der Wel}, \&
  {Whitaker}}]{Nedkova:2021}
{Nedkova}, K.~V., {H{\"a}u{\ss}ler}, B., {Marchesini}, D., {et~al.} 2021,
  \mnras, 506, 928, \dodoi{10.1093/mnras/stab1744}

\bibitem[{{Nedkova} {et~al.}(2024){Nedkova}, {Rafelski}, {Teplitz}, {Mehta},
  {DeGroot}, {Ravindranath}, {Alavi}, {Beckett}, {Grogin}, {H{\"a}u{\ss}ler},
  {Koekemoer}, {Oyarz{\'u}n}, {Prichard}, {Revalski}, {Snyder}, {Sunnquist},
  {Wang}, {Windhorst}, {Chartab}, {Conselice}, {Guo}, {Hathi}, {Hayes}, {Ji},
  {Kim}, {Lucas}, {Mobasher}, {O'Connell}, {Sattari}, {Smith}, {Taamoli},
  {Yung}, \& {the UVCANDELS Team}}]{Nedkova:2024}
{Nedkova}, K.~V., {Rafelski}, M., {Teplitz}, H.~I., {et~al.} 2024, arXiv
  e-prints, arXiv:2405.10908, \dodoi{10.48550/arXiv.2405.10908}

\bibitem[{{Newman} {et~al.}(2013){Newman}, {Cooper}, {Davis}, {Faber}, {Coil},
  {Guhathakurta}, {Koo}, {Phillips}, {Conroy}, {Dutton}, {Finkbeiner}, {Gerke},
  {Rosario}, {Weiner}, {Willmer}, {Yan}, {Harker}, {Kassin}, {Konidaris},
  {Lai}, {Madgwick}, {Noeske}, {Wirth}, {Connolly}, {Kaiser}, {Kirby},
  {Lemaux}, {Lin}, {Lotz}, {Luppino}, {Marinoni}, {Matthews}, {Metevier}, \&
  {Schiavon}}]{Newman:2013}
{Newman}, J.~A., {Cooper}, M.~C., {Davis}, M., {et~al.} 2013, \apjs, 208, 5,
  \dodoi{10.1088/0067-0049/208/1/5}

\bibitem[{{Noll} {et~al.}(2009){Noll}, {Burgarella}, {Giovannoli}, {Buat},
  {Marcillac}, \& {Mu{\~n}oz-Mateos}}]{Noll:2009}
{Noll}, S., {Burgarella}, D., {Giovannoli}, E., {et~al.} 2009, \aap, 507, 1793,
  \dodoi{10.1051/0004-6361/200912497}

\bibitem[{{Nonino} {et~al.}(2009){Nonino}, {Dickinson}, {Rosati}, {Grazian},
  {Reddy}, {Cristiani}, {Giavalisco}, {Kuntschner}, {Vanzella}, {Daddi},
  {Fosbury}, \& {Cesarsky}}]{Nonino:2009}
{Nonino}, M., {Dickinson}, M., {Rosati}, P., {et~al.} 2009, \apjs, 183, 244,
  \dodoi{10.1088/0067-0049/183/2/244}

\bibitem[{{Oke} \& {Gunn}(1983)}]{Oke:1983}
{Oke}, J.~B., \& {Gunn}, J.~E. 1983, \apj, 266, 713, \dodoi{10.1086/160817}

\bibitem[{{Pacifici} {et~al.}(2023){Pacifici}, {Iyer}, {Mobasher}, {da Cunha},
  {Acquaviva}, {Burgarella}, {Calistro Rivera}, {Carnall}, {Chang}, {Chartab},
  {Cooke}, {Fairhurst}, {Kartaltepe}, {Leja}, {Ma{\l}ek}, {Salmon}, {Torelli},
  {Vidal-Garc{\'\i}a}, {Boquien}, {Brammer}, {Brown}, {Capak}, {Chevallard},
  {Circosta}, {Croton}, {Davidzon}, {Dickinson}, {Duncan}, {Faber}, {Ferguson},
  {Fontana}, {Guo}, {Haeussler}, {Hemmati}, {Jafariyazani}, {Kassin}, {Larson},
  {Lee}, {Mantha}, {Marchi}, {Nayyeri}, {Newman}, {Pandya}, {Pforr}, {Reddy},
  {Sanders}, {Shah}, {Shahidi}, {Stevans}, {Triani}, {Tyler}, {Vanderhoof}, {de
  la Vega}, {Wang}, \& {Weston}}]{Pacifici:2023}
{Pacifici}, C., {Iyer}, K.~G., {Mobasher}, B., {et~al.} 2023, \apj, 944, 141,
  \dodoi{10.3847/1538-4357/acacff}

\bibitem[{{Pasquali} {et~al.}(2006){Pasquali}, {Ferreras}, {Panagia}, {Daddi},
  {Malhotra}, {Rhoads}, {Pirzkal}, {Windhorst}, {Koekemoer}, {Moustakas}, {Xu},
  \& {Gronwall}}]{Pasquali:2006}
{Pasquali}, A., {Ferreras}, I., {Panagia}, N., {et~al.} 2006, \apj, 636, 115,
  \dodoi{10.1086/497290}

\bibitem[{{Pentericci} {et~al.}(2018){Pentericci}, {McLure}, {Garilli},
  {Cucciati}, {Franzetti}, {Iovino}, {Amorin}, {Bolzonella}, {Bongiorno},
  {Carnall}, {Castellano}, {Cimatti}, {Cirasuolo}, {Cullen}, {De Barros},
  {Dunlop}, {Elbaz}, {Finkelstein}, {Fontana}, {Fontanot}, {Fumana},
  {Gargiulo}, {Guaita}, {Hartley}, {Jarvis}, {Juneau}, {Karman}, {Maccagni},
  {Marchi}, {Marmol-Queralto}, {Nandra}, {Pompei}, {Pozzetti}, {Scodeggio},
  {Sommariva}, {Talia}, {Almaini}, {Balestra}, {Bardelli}, {Bell}, {Bourne},
  {Bowler}, {Brusa}, {Buitrago}, {Caputi}, {Cassata}, {Charlot}, {Citro},
  {Cresci}, {Cristiani}, {Curtis-Lake}, {Dickinson}, {Fazio}, {Ferguson},
  {Fiore}, {Franco}, {Fynbo}, {Galametz}, {Georgakakis}, {Giavalisco},
  {Grazian}, {Hathi}, {Jung}, {Kim}, {Koekemoer}, {Khusanova}, {Le F{\`e}vre},
  {Lotz}, {Mannucci}, {Maltby}, {Matsuoka}, {McLeod}, {Mendez-Hernandez},
  {Mendez-Abreu}, {Mignoli}, {Moresco}, {Mortlock}, {Nonino}, {Pannella},
  {Papovich}, {Popesso}, {Rosario}, {Salvato}, {Santini}, {Schaerer},
  {Schreiber}, {Stark}, {Tasca}, {Thomas}, {Treu}, {Vanzella}, {Wild},
  {Williams}, {Zamorani}, \& {Zucca}}]{Pentericci:2018}
{Pentericci}, L., {McLure}, R.~J., {Garilli}, B., {et~al.} 2018, \aap, 616,
  A174, \dodoi{10.1051/0004-6361/201833047}

\bibitem[{{Picouet} {et~al.}(2023){Picouet}, {Arnouts}, {Le Floc'h}, {Moutard},
  {Kraljic}, {Ilbert}, {Sawicki}, {Desprez}, {Laigle}, {Schiminovich}, {de la
  Torre}, {Gwyn}, {McCracken}, {Dubois}, {Dav{\'e}}, {Toft}, {Weaver},
  {Shuntov}, \& {Kauffmann}}]{Picouet:2023}
{Picouet}, V., {Arnouts}, S., {Le Floc'h}, E., {et~al.} 2023, \aap, 675, A164,
  \dodoi{10.1051/0004-6361/202245756}

\bibitem[{{Planck Collaboration} {et~al.}(2016){Planck Collaboration}, {Ade},
  {Aghanim}, {Arnaud}, {Ashdown}, {Aumont}, {Baccigalupi}, {Banday},
  {Barreiro}, {Bartlett}, {Bartolo}, {Battaner}, {Battye}, {Benabed},
  {Beno{\^\i}t}, {Benoit-L{\'e}vy}, {Bernard}, {Bersanelli}, {Bielewicz},
  {Bock}, {Bonaldi}, {Bonavera}, {Bond}, {Borrill}, {Bouchet}, {Boulanger},
  {Bucher}, {Burigana}, {Butler}, {Calabrese}, {Cardoso}, {Catalano},
  {Challinor}, {Chamballu}, {Chary}, {Chiang}, {Chluba}, {Christensen},
  {Church}, {Clements}, {Colombi}, {Colombo}, {Combet}, {Coulais}, {Crill},
  {Curto}, {Cuttaia}, {Danese}, {Davies}, {Davis}, {de Bernardis}, {de Rosa},
  {de Zotti}, {Delabrouille}, {D{\'e}sert}, {Di Valentino}, {Dickinson},
  {Diego}, {Dolag}, {Dole}, {Donzelli}, {Dor{\'e}}, {Douspis}, {Ducout},
  {Dunkley}, {Dupac}, {Efstathiou}, {Elsner}, {En{\ss}lin}, {Eriksen},
  {Farhang}, {Fergusson}, {Finelli}, {Forni}, {Frailis}, {Fraisse},
  {Franceschi}, {Frejsel}, {Galeotta}, {Galli}, {Ganga}, {Gauthier}, {Gerbino},
  {Ghosh}, {Giard}, {Giraud-H{\'e}raud}, {Giusarma}, {Gjerl{\o}w},
  {Gonz{\'a}lez-Nuevo}, {G{\'o}rski}, {Gratton}, {Gregorio}, {Gruppuso},
  {Gudmundsson}, {Hamann}, {Hansen}, {Hanson}, {Harrison}, {Helou},
  {Henrot-Versill{\'e}}, {Hern{\'a}ndez-Monteagudo}, {Herranz}, {Hildebrandt},
  {Hivon}, {Hobson}, {Holmes}, {Hornstrup}, {Hovest}, {Huang}, {Huffenberger},
  {Hurier}, {Jaffe}, {Jaffe}, {Jones}, {Juvela}, {Keih{\"a}nen}, {Keskitalo},
  {Kisner}, {Kneissl}, {Knoche}, {Knox}, {Kunz}, {Kurki-Suonio}, {Lagache},
  {L{\"a}hteenm{\"a}ki}, {Lamarre}, {Lasenby}, {Lattanzi}, {Lawrence}, {Leahy},
  {Leonardi}, {Lesgourgues}, {Levrier}, {Lewis}, {Liguori}, {Lilje},
  {Linden-V{\o}rnle}, {L{\'o}pez-Caniego}, {Lubin}, {Mac{\'\i}as-P{\'e}rez},
  {Maggio}, {Maino}, {Mandolesi}, {Mangilli}, {Marchini}, {Maris}, {Martin},
  {Martinelli}, {Mart{\'\i}nez-Gonz{\'a}lez}, {Masi}, {Matarrese}, {McGehee},
  {Meinhold}, {Melchiorri}, {Melin}, {Mendes}, {Mennella}, {Migliaccio},
  {Millea}, {Mitra}, {Miville-Desch{\^e}nes}, {Moneti}, {Montier}, {Morgante},
  {Mortlock}, {Moss}, {Munshi}, {Murphy}, {Naselsky}, {Nati}, {Natoli},
  {Netterfield}, {N{\o}rgaard-Nielsen}, {Noviello}, {Novikov}, {Novikov},
  {Oxborrow}, {Paci}, {Pagano}, {Pajot}, {Paladini}, {Paoletti}, {Partridge},
  {Pasian}, {Patanchon}, {Pearson}, {Perdereau}, {Perotto}, {Perrotta},
  {Pettorino}, {Piacentini}, {Piat}, {Pierpaoli}, {Pietrobon}, {Plaszczynski},
  {Pointecouteau}, {Polenta}, {Popa}, {Pratt}, {Pr{\'e}zeau}, {Prunet},
  {Puget}, {Rachen}, {Reach}, {Rebolo}, {Reinecke}, {Remazeilles}, {Renault},
  {Renzi}, {Ristorcelli}, {Rocha}, {Rosset}, {Rossetti}, {Roudier},
  {Rouill{\'e} d'Orfeuil}, {Rowan-Robinson}, {Rubi{\~n}o-Mart{\'\i}n},
  {Rusholme}, {Said}, {Salvatelli}, {Salvati}, {Sandri}, {Santos},
  {Savelainen}, {Savini}, {Scott}, {Seiffert}, {Serra}, {Shellard}, {Spencer},
  {Spinelli}, {Stolyarov}, {Stompor}, {Sudiwala}, {Sunyaev}, {Sutton},
  {Suur-Uski}, {Sygnet}, {Tauber}, {Terenzi}, {Toffolatti}, {Tomasi},
  {Tristram}, {Trombetti}, {Tucci}, {Tuovinen}, {T{\"u}rler}, {Umana},
  {Valenziano}, {Valiviita}, {Van Tent}, {Vielva}, {Villa}, {Wade}, {Wandelt},
  {Wehus}, {White}, {White}, {Wilkinson}, {Yvon}, {Zacchei}, \&
  {Zonca}}]{Planck:2016}
{Planck Collaboration}, {Ade}, P.~A.~R., {Aghanim}, N., {et~al.} 2016, \aap,
  594, A13, \dodoi{10.1051/0004-6361/201525830}

\bibitem[{{Prevot} {et~al.}(1984){Prevot}, {Lequeux}, {Maurice}, {Prevot}, \&
  {Rocca-Volmerange}}]{Prevot:1984}
{Prevot}, M.~L., {Lequeux}, J., {Maurice}, E., {Prevot}, L., \&
  {Rocca-Volmerange}, B. 1984, \aap, 132, 389

\bibitem[{{Rafelski} {et~al.}(2009){Rafelski}, {Wolfe}, {Cooke}, {Chen},
  {Armandroff}, \& {Wirth}}]{Rafelski:2009}
{Rafelski}, M., {Wolfe}, A.~M., {Cooke}, J., {et~al.} 2009, \apj, 703, 2033,
  \dodoi{10.1088/0004-637X/703/2/2033}

\bibitem[{{Rafelski} {et~al.}(2015){Rafelski}, {Teplitz}, {Gardner}, {Coe},
  {Bond}, {Koekemoer}, {Grogin}, {Kurczynski}, {McGrath}, {Bourque}, {Atek},
  {Brown}, {Colbert}, {Codoreanu}, {Ferguson}, {Finkelstein}, {Gawiser},
  {Giavalisco}, {Gronwall}, {Hanish}, {Lee}, {Mehta}, {de Mello},
  {Ravindranath}, {Ryan}, {Scarlata}, {Siana}, {Soto}, \&
  {Voyer}}]{Rafelski:2015}
{Rafelski}, M., {Teplitz}, H.~I., {Gardner}, J.~P., {et~al.} 2015, \aj, 150,
  31, \dodoi{10.1088/0004-6256/150/1/31}

\bibitem[{{Ravikumar} {et~al.}(2007){Ravikumar}, {Puech}, {Flores}, {Proust},
  {Hammer}, {Lehnert}, {Rawat}, {Amram}, {Balkowski}, {Burgarella}, {Cassata},
  {Cesarsky}, {Cimatti}, {Combes}, {Daddi}, {Dannerbauer}, {di Serego
  Alighieri}, {Elbaz}, {Guiderdoni}, {Kembhavi}, {Liang}, {Pozzetti},
  {Vergani}, {Vernet}, {Wozniak}, \& {Zheng}}]{Ravikumar:2007}
{Ravikumar}, C.~D., {Puech}, M., {Flores}, H., {et~al.} 2007, \aap, 465, 1099,
  \dodoi{10.1051/0004-6361:20065358}

\bibitem[{{Retzlaff} {et~al.}(2010){Retzlaff}, {Rosati}, {Dickinson},
  {Vandame}, {Rit{\'e}}, {Nonino}, {Cesarsky}, \& {GOODS Team}}]{Retzlaff:2010}
{Retzlaff}, J., {Rosati}, P., {Dickinson}, M., {et~al.} 2010, \aap, 511, A50,
  \dodoi{10.1051/0004-6361/200912940}

\bibitem[{{Riess} {et~al.}(2007){Riess}, {Strolger}, {Casertano}, {Ferguson},
  {Mobasher}, {Gold}, {Challis}, {Filippenko}, {Jha}, {Li}, {Tonry}, {Foley},
  {Kirshner}, {Dickinson}, {MacDonald}, {Eisenstein}, {Livio}, {Younger}, {Xu},
  {Dahl{\'e}n}, \& {Stern}}]{Riess:2007}
{Riess}, A.~G., {Strolger}, L.-G., {Casertano}, S., {et~al.} 2007, \apj, 659,
  98, \dodoi{10.1086/510378}

\bibitem[{{Roche} {et~al.}(2006){Roche}, {Dunlop}, {Caputi}, {McLure},
  {Willott}, \& {Crampton}}]{Roche:2006}
{Roche}, N.~D., {Dunlop}, J., {Caputi}, K.~I., {et~al.} 2006, \mnras, 370, 74,
  \dodoi{10.1111/j.1365-2966.2006.10439.x}

\bibitem[{{Sanders} {et~al.}(2007){Sanders}, {Salvato}, {Aussel}, {Ilbert},
  {Scoville}, {Surace}, {Frayer}, {Sheth}, {Helou}, {Brooke}, {Bhattacharya},
  {Yan}, {Kartaltepe}, {Barnes}, {Blain}, {Calzetti}, {Capak}, {Carilli},
  {Carollo}, {Comastri}, {Daddi}, {Ellis}, {Elvis}, {Fall}, {Franceschini},
  {Giavalisco}, {Hasinger}, {Impey}, {Koekemoer}, {Le F{\`e}vre}, {Lilly},
  {Liu}, {McCracken}, {Mobasher}, {Renzini}, {Rich}, {Schinnerer}, {Shopbell},
  {Taniguchi}, {Thompson}, {Urry}, \& {Williams}}]{Sanders:2007}
{Sanders}, D.~B., {Salvato}, M., {Aussel}, H., {et~al.} 2007, \apjs, 172, 86,
  \dodoi{10.1086/517885}

\bibitem[{{Santini} {et~al.}(2009){Santini}, {Fontana}, {Grazian}, {Salimbeni},
  {Fiore}, {Fontanot}, {Boutsia}, {Castellano}, {Cristiani}, {de Santis},
  {Gallozzi}, {Giallongo}, {Menci}, {Nonino}, {Paris}, {Pentericci}, \&
  {Vanzella}}]{Santini:2009}
{Santini}, P., {Fontana}, A., {Grazian}, A., {et~al.} 2009, \aap, 504, 751,
  \dodoi{10.1051/0004-6361/200811434}

\bibitem[{{Schlegel} {et~al.}(1998){Schlegel}, {Finkbeiner}, \&
  {Davis}}]{Schlegel:1998}
{Schlegel}, D.~J., {Finkbeiner}, D.~P., \& {Davis}, M. 1998, \apj, 500, 525,
  \dodoi{10.1086/305772}

\bibitem[{{Scoville} {et~al.}(2007){Scoville}, {Aussel}, {Brusa}, {Capak},
  {Carollo}, {Elvis}, {Giavalisco}, {Guzzo}, {Hasinger}, {Impey}, {Kneib},
  {LeFevre}, {Lilly}, {Mobasher}, {Renzini}, {Rich}, {Sanders}, {Schinnerer},
  {Schminovich}, {Shopbell}, {Taniguchi}, \& {Tyson}}]{Scoville:2007}
{Scoville}, N., {Aussel}, H., {Brusa}, M., {et~al.} 2007, \apjs, 172, 1,
  \dodoi{10.1086/516585}

\bibitem[{{Shen} {et~al.}(2003){Shen}, {Mo}, {White}, {Blanton}, {Kauffmann},
  {Voges}, {Brinkmann}, \& {Csabai}}]{Shen:2003}
{Shen}, S., {Mo}, H.~J., {White}, S. D.~M., {et~al.} 2003, \mnras, 343, 978,
  \dodoi{10.1046/j.1365-8711.2003.06740.x}

\bibitem[{{Silverman} {et~al.}(2015){Silverman}, {Kashino}, {Sanders},
  {Kartaltepe}, {Arimoto}, {Renzini}, {Rodighiero}, {Daddi}, {Zahid}, {Nagao},
  {Kewley}, {Lilly}, {Sugiyama}, {Baronchelli}, {Capak}, {Carollo}, {Chu},
  {Hasinger}, {Ilbert}, {Juneau}, {Kajisawa}, {Koekemoer}, {Kovac}, {Le
  F{\`e}vre}, {Masters}, {McCracken}, {Onodera}, {Schulze}, {Scoville},
  {Strazzullo}, \& {Taniguchi}}]{Silverman:2015}
{Silverman}, J.~D., {Kashino}, D., {Sanders}, D., {et~al.} 2015, \apjs, 220,
  12, \dodoi{10.1088/0067-0049/220/1/12}

\bibitem[{{Skelton} {et~al.}(2014){Skelton}, {Whitaker}, {Momcheva}, {Brammer},
  {van Dokkum}, {Labb{\'e}}, {Franx}, {van der Wel}, {Bezanson}, {Da Cunha},
  {Fumagalli}, {F{\"o}rster Schreiber}, {Kriek}, {Leja}, {Lundgren}, {Magee},
  {Marchesini}, {Maseda}, {Nelson}, {Oesch}, {Pacifici}, {Patel}, {Price},
  {Rix}, {Tal}, {Wake}, \& {Wuyts}}]{Skelton:2014}
{Skelton}, R.~E., {Whitaker}, K.~E., {Momcheva}, I.~G., {et~al.} 2014, \apjs,
  214, 24, \dodoi{10.1088/0067-0049/214/2/24}

\bibitem[{{Speagle} {et~al.}(2014){Speagle}, {Steinhardt}, {Capak}, \&
  {Silverman}}]{Speagle:2014}
{Speagle}, J.~S., {Steinhardt}, C.~L., {Capak}, P.~L., \& {Silverman}, J.~D.
  2014, \apjs, 214, 15, \dodoi{10.1088/0067-0049/214/2/15}

\bibitem[{{Stanford} {et~al.}(2021){Stanford}, {Masters}, {Darvish}, {Stern},
  {Cohen}, {Capak}, {Hernitschek}, {Davidzon}, {Rhodes}, {Sanders}, {Mobasher},
  {Castander}, {Paltani}, {Aghanim}, {Amara}, {Auricchio}, {Balestra},
  {Bender}, {Bodendorf}, {Bonino}, {Branchini}, {Brinchmann}, {Capobianco},
  {Carbone}, {Carretero}, {Casas}, {Castellano}, {Cavuoti}, {Cimatti},
  {Cledassou}, {Conselice}, {Corcione}, {Costille}, {Cropper}, {Degaudenzi},
  {Douspis}, {Dubath}, {Dusini}, {Fosalba}, {Frailis}, {Franceschi},
  {Franzetti}, {Fumana}, {Garilli}, {Giocoli}, {Grupp}, {Haugan}, {Hoekstra},
  {Holmes}, {Hormuth}, {Hudelot}, {Jahnke}, {Kiessling}, {Kilbinger},
  {Kitching}, {Kubik}, {K{\"u}mmel}, {Kunz}, {Kurki-Suonio}, {Laureijs},
  {Ligori}, {Lilje}, {Lloro}, {Maiorano}, {Marggraf}, {Markovic}, {Massey},
  {Meneghetti}, {Meylan}, {Moscardini}, {Niemi}, {Padilla}, {Pasian},
  {Pedersen}, {Pettorino}, {Pires}, {Poncet}, {Popa}, {Pozzetti}, {Raison},
  {Roncarelli}, {Rossetti}, {Saglia}, {Scaramella}, {Schneider}, {Secroun},
  {Seidel}, {Serrano}, {Sirignano}, {Sirri}, {Taylor}, {Teplitz}, {Tereno},
  {Toledo-Moreo}, {Valentijn}, {Valenziano}, {Verdoes Kleijn}, {Wang},
  {Zamorani}, {Zoubian}, {Brescia}, {Congedo}, {Conversi}, {Copin}, {Kermiche},
  {Kohley}, {Medinaceli}, {Mei}, {Moresco}, {Morin}, {Munari}, {Polenta},
  {Sureau}, {Tallada Cresp{\'\i}}, {Vassallo}, {Zacchei}, {Andreon}, {Aussel},
  {Baccigalupi}, {Balaguera-Antol{\'\i}nez}, {Baldi}, {Bardelli}, {Biviano},
  {Borsato}, {Bozzo}, {Burigana}, {Cabanac}, {Camera}, {Cappi}, {Carvalho},
  {Casas}, {Castignani}, {Colodro-Conde}, {Coupon}, {Courtois}, {Cuby}, {Da
  Silva}, {de la Torre}, {Di Ferdinando}, {Duncan}, {Dupac}, {Fabricius},
  {Farina}, {Farrens}, {Ferreira}, {Finelli}, {Flose-Reimberg}, {Fotopoulou},
  {Galeotta}, {Ganga}, {Gillard}, {Gozaliasl}, {Graci{\'a}-Carpio}, {Keihanen},
  {Kirkpatrick}, {Lindholm}, {Mainetti}, {Maino}, {Martinet}, {Marulli},
  {Maturi}, {Maurogordato}, {Metcalf}, {Nakajima}, {Neissner}, {Nightingale},
  {Nucita}, {Patrizii}, {Potter}, {Renzi}, {Riccio}, {Romelli}, {S{\'a}nchez},
  {Sapone}, {Schirmer}, {Schultheis}, {Scottez}, {Stanco}, {Tenti}, {Teyssier},
  {Torradeflot}, {Valiviita}, {Viel}, {Whittaker}, {Zucca}, \& {Euclid
  Collaboration}}]{Stanford:2021}
{Stanford}, S.~A., {Masters}, D., {Darvish}, B., {et~al.} 2021, \apjs, 256, 9,
  \dodoi{10.3847/1538-4365/ac0833}

\bibitem[{{Stefanon} {et~al.}(2017){Stefanon}, {Yan}, {Mobasher}, {Barro},
  {Donley}, {Fontana}, {Hemmati}, {Koekemoer}, {Lee}, {Lee}, {Nayyeri}, {Peth},
  {Pforr}, {Salvato}, {Wiklind}, {Wuyts}, {Ashby}, {Castellano}, {Conselice},
  {Cooper}, {Cooray}, {Dolch}, {Ferguson}, {Galametz}, {Giavalisco}, {Guo},
  {Willner}, {Dickinson}, {Faber}, {Fazio}, {Gardner}, {Gawiser}, {Grazian},
  {Grogin}, {Kocevski}, {Koo}, {Lee}, {Lucas}, {McGrath}, {Nandra}, {Newman},
  \& {van der Wel}}]{Stefanon:2017}
{Stefanon}, M., {Yan}, H., {Mobasher}, B., {et~al.} 2017, \apjs, 229, 32,
  \dodoi{10.3847/1538-4365/aa66cb}

\bibitem[{{Straughn} {et~al.}(2009){Straughn}, {Pirzkal}, {Meurer}, {Cohen},
  {Windhorst}, {Malhotra}, {Rhoads}, {Gardner}, {Hathi}, {Jansen}, {Grogin},
  {Panagia}, {di Serego Alighieri}, {Gronwall}, {Walsh}, {Pasquali}, \&
  {Xu}}]{Straughn:2009}
{Straughn}, A.~N., {Pirzkal}, N., {Meurer}, G.~R., {et~al.} 2009, \aj, 138,
  1022, \dodoi{10.1088/0004-6256/138/4/1022}

\bibitem[{{Strolger} {et~al.}(2004){Strolger}, {Riess}, {Dahlen}, {Livio},
  {Panagia}, {Challis}, {Tonry}, {Filippenko}, {Chornock}, {Ferguson},
  {Koekemoer}, {Mobasher}, {Dickinson}, {Giavalisco}, {Casertano}, {Hook},
  {Blondin}, {Leibundgut}, {Nonino}, {Rosati}, {Spinrad}, {Steidel}, {Stern},
  {Garnavich}, {Matheson}, {Grogin}, {Hornschemeier}, {Kretchmer}, {Laidler},
  {Lee}, {Lucas}, {de Mello}, {Moustakas}, {Ravindranath}, {Richardson}, \&
  {Taylor}}]{Strolger:2004}
{Strolger}, L.-G., {Riess}, A.~G., {Dahlen}, T., {et~al.} 2004, \apj, 613, 200,
  \dodoi{10.1086/422901}

\bibitem[{{Sun} {et~al.}(2023{\natexlab{a}}){Sun}, {Faucher-Gigu{\`e}re},
  {Hayward}, {Shen}, {Wetzel}, \& {Cochrane}}]{Sun:2023}
{Sun}, G., {Faucher-Gigu{\`e}re}, C.-A., {Hayward}, C.~C., {et~al.}
  2023{\natexlab{a}}, \apjl, 955, L35, \dodoi{10.3847/2041-8213/acf85a}

\bibitem[{{Sun} {et~al.}(2023{\natexlab{b}}){Sun}, {Wang}, {Teplitz}, {Mehta},
  {Alavi}, {Rafelski}, {Windhorst}, {Scarlata}, {Gardner}, {Smith},
  {Sunnquist}, {Prichard}, {Cheng}, {Grogin}, {Hathi}, {Hayes}, {Koekemoer},
  {Mobasher}, {Nedkova}, {O'Connell}, {Robertson}, {Taamoli}, {Yung}, {Arrabal
  Haro}, {Brammer}, {Colbert}, {Conselice}, {Gawiser}, {Guo}, {Jansen}, {Ji},
  {Lucas}, {Rutkowski}, {Siana}, {Vanzella}, {Ashcraft}, {Bagley},
  {Baronchelli}, {Barro}, {Blanche}, {Broussard}, {Carleton}, {Chartab},
  {Codoreanu}, {Cohen}, {Dai}, {Darvish}, {Dav{\'e}}, {DeGroot}, {De Mello},
  {Dickinson}, {Emami}, {Ferguson}, {Ferreira}, {Finkelstein}, {Finkelstein},
  {Gburek}, {Giavalisco}, {Grazian}, {Gronwall}, {Hemmati}, {Howell}, {Iyer},
  {Kaviraj}, {Kurczynski}, {Lazar}, {MacKenty}, {Mantha}, {Martin}, {Martin},
  {McCabe}, {Olsen}, {Otteson}, {Ravindranath}, {Redshaw}, {Sattari}, {Soto},
  {Zabelle}, \& {the UVCANDELS team}}]{Sun:2024}
{Sun}, L., {Wang}, X., {Teplitz}, H.~I., {et~al.} 2023{\natexlab{b}}, arXiv
  e-prints, arXiv:2311.15664, \dodoi{10.48550/arXiv.2311.15664}

\bibitem[{{Taniguchi} {et~al.}(2007){Taniguchi}, {Scoville}, {Murayama},
  {Sanders}, {Mobasher}, {Aussel}, {Capak}, {Ajiki}, {Miyazaki}, {Komiyama},
  {Shioya}, {Nagao}, {Sasaki}, {Koda}, {Carilli}, {Giavalisco}, {Guzzo},
  {Hasinger}, {Impey}, {LeFevre}, {Lilly}, {Renzini}, {Rich}, {Schinnerer},
  {Shopbell}, {Kaifu}, {Karoji}, {Arimoto}, {Okamura}, \&
  {Ohta}}]{Taniguchi:2007}
{Taniguchi}, Y., {Scoville}, N., {Murayama}, T., {et~al.} 2007, \apjs, 172, 9,
  \dodoi{10.1086/516596}

\bibitem[{{Taniguchi} {et~al.}(2015){Taniguchi}, {Kajisawa}, {Kobayashi},
  {Shioya}, {Nagao}, {Capak}, {Aussel}, {Ichikawa}, {Murayama}, {Scoville},
  {Ilbert}, {Salvato}, {Sanders}, {Mobasher}, {Miyazaki}, {Komiyama}, {Le
  F{\`e}vre}, {Tasca}, {Lilly}, {Carollo}, {Renzini}, {Rich}, {Schinnerer},
  {Kaifu}, {Karoji}, {Arimoto}, {Okamura}, {Ohta}, {Shimasaku}, \&
  {Hayashino}}]{Taniguchi:2015}
{Taniguchi}, Y., {Kajisawa}, M., {Kobayashi}, M. A.~R., {et~al.} 2015, \pasj,
  67, 104, \dodoi{10.1093/pasj/psv106}

\bibitem[{{Teplitz} {et~al.}(2013){Teplitz}, {Rafelski}, {Kurczynski}, {Bond},
  {Grogin}, {Koekemoer}, {Atek}, {Brown}, {Coe}, {Colbert}, {Ferguson},
  {Finkelstein}, {Gardner}, {Gawiser}, {Giavalisco}, {Gronwall}, {Hanish},
  {Lee}, {de Mello}, {Ravindranath}, {Ryan}, {Siana}, {Scarlata}, {Soto},
  {Voyer}, \& {Wolfe}}]{Teplitz:2013}
{Teplitz}, H.~I., {Rafelski}, M., {Kurczynski}, P., {et~al.} 2013, \aj, 146,
  159, \dodoi{10.1088/0004-6256/146/6/159}

\bibitem[{{Treister} {et~al.}(2009){Treister}, {Virani}, {Gawiser}, {Urry},
  {Lira}, {Francke}, {Blanc}, {Cardamone}, {Damen}, {Taylor}, \&
  {Schawinski}}]{Treister:2009}
{Treister}, E., {Virani}, S., {Gawiser}, E., {et~al.} 2009, \apj, 693, 1713,
  \dodoi{10.1088/0004-637X/693/2/1713}

\bibitem[{{Trump} {et~al.}(2009){Trump}, {Impey}, {Elvis}, {McCarthy},
  {Huchra}, {Brusa}, {Salvato}, {Capak}, {Cappelluti}, {Civano}, {Comastri},
  {Gabor}, {Hao}, {Hasinger}, {Jahnke}, {Kelly}, {Lilly}, {Schinnerer},
  {Scoville}, \& {Smol{\v{c}}i{\'c}}}]{Trump:2009}
{Trump}, J.~R., {Impey}, C.~D., {Elvis}, M., {et~al.} 2009, \apj, 696, 1195,
  \dodoi{10.1088/0004-637X/696/2/1195}

\bibitem[{{Trump} {et~al.}(2011){Trump}, {Weiner}, {Scarlata}, {Kocevski},
  {Bell}, {McGrath}, {Koo}, {Faber}, {Laird}, {Mozena}, {Rangel}, {Yan},
  {Yesuf}, {Atek}, {Dickinson}, {Donley}, {Dunlop}, {Ferguson}, {Finkelstein},
  {Grogin}, {Hathi}, {Juneau}, {Kartaltepe}, {Koekemoer}, {Nandra}, {Newman},
  {Rodney}, {Straughn}, \& {Teplitz}}]{Trump:2011}
{Trump}, J.~R., {Weiner}, B.~J., {Scarlata}, C., {et~al.} 2011, \apj, 743, 144,
  \dodoi{10.1088/0004-637X/743/2/144}

\bibitem[{{Trump} {et~al.}(2013){Trump}, {Konidaris}, {Barro}, {Koo},
  {Kocevski}, {Juneau}, {Weiner}, {Faber}, {McLean}, {Yan},
  {P{\'e}rez-Gonz{\'a}lez}, \& {Villar}}]{Trump:2013}
{Trump}, J.~R., {Konidaris}, N.~P., {Barro}, G., {et~al.} 2013, \apjl, 763, L6,
  \dodoi{10.1088/2041-8205/763/1/L6}

\bibitem[{{Trump} {et~al.}(2015){Trump}, {Sun}, {Zeimann}, {Luck}, {Bridge},
  {Grier}, {Hagen}, {Juneau}, {Montero-Dorta}, {Rosario}, {Brandt},
  {Ciardullo}, \& {Schneider}}]{Trump:2015}
{Trump}, J.~R., {Sun}, M., {Zeimann}, G.~R., {et~al.} 2015, \apj, 811, 26,
  \dodoi{10.1088/0004-637X/811/1/26}

\bibitem[{{van der Wel} {et~al.}(2005){van der Wel}, {Franx}, {van Dokkum},
  {Rix}, {Illingworth}, \& {Rosati}}]{vanderWel:2005}
{van der Wel}, A., {Franx}, M., {van Dokkum}, P.~G., {et~al.} 2005, \apj, 631,
  145, \dodoi{10.1086/430464}

\bibitem[{{van der Wel} {et~al.}(2016){van der Wel}, {Noeske}, {Bezanson},
  {Pacifici}, {Gallazzi}, {Franx}, {Mu{\~n}oz-Mateos}, {Bell}, {Brammer},
  {Charlot}, {Chauk{\'e}}, {Labb{\'e}}, {Maseda}, {Muzzin}, {Rix}, {Sobral},
  {van de Sande}, {van Dokkum}, {Wild}, \& {Wolf}}]{vanderWel:2016}
{van der Wel}, A., {Noeske}, K., {Bezanson}, R., {et~al.} 2016, \apjs, 223, 29,
  \dodoi{10.3847/0067-0049/223/2/29}

\bibitem[{{Vanzella} {et~al.}(2008){Vanzella}, {Cristiani}, {Dickinson},
  {Giavalisco}, {Kuntschner}, {Haase}, {Nonino}, {Rosati}, {Cesarsky},
  {Ferguson}, {Fosbury}, {Grazian}, {Moustakas}, {Rettura}, {Popesso},
  {Renzini}, {Stern}, \& {GOODS Team}}]{Vanzella:2008}
{Vanzella}, E., {Cristiani}, S., {Dickinson}, M., {et~al.} 2008, \aap, 478, 83,
  \dodoi{10.1051/0004-6361:20078332}

\bibitem[{{Vanzella} {et~al.}(2009){Vanzella}, {Giavalisco}, {Dickinson},
  {Cristiani}, {Nonino}, {Kuntschner}, {Popesso}, {Rosati}, {Renzini}, {Stern},
  {Cesarsky}, {Ferguson}, \& {Fosbury}}]{Vanzella:2009}
{Vanzella}, E., {Giavalisco}, M., {Dickinson}, M., {et~al.} 2009, \apj, 695,
  1163, \dodoi{10.1088/0004-637X/695/2/1163}

\bibitem[{Virtanen {et~al.}(2020)Virtanen, Gommers, Oliphant, Haberland, Reddy,
  Cournapeau, Burovski, Peterson, Weckesser, Bright, {van der Walt}, Brett,
  Wilson, Millman, Mayorov, Nelson, Jones, Kern, Larson, Carey, Polat, Feng,
  Moore, {VanderPlas}, Laxalde, Perktold, Cimrman, Henriksen, Quintero, Harris,
  Archibald, Ribeiro, Pedregosa, {van Mulbregt}, \& {SciPy 1.0
  Contributors}}]{Scipy:2020}
Virtanen, P., Gommers, R., Oliphant, T.~E., {et~al.} 2020, Nature Methods, 17,
  261, \dodoi{10.1038/s41592-019-0686-2}

\bibitem[{{Wang} {et~al.}(2024){Wang}, {Teplitz}, {Sun}, {Rafelski}, {Grogin},
  {Prichard}, {Sunnquist}, {Alavi}, {Windhorst}, {Koekemoer}, {Ashcraft},
  {Bagley}, {Baronchelli}, {Barro}, {Blanche}, {Brammer}, {Broussard},
  {Carleton}, {Chartab}, {Cheng}, {Codoreanu}, {Cohen}, {Colbert}, {Conselice},
  {Dai}, {Darvish}, {Dav{\'e}}, {DeGroot}, {De Mello}, {Dickinson}, {Emami},
  {Ferguson}, {Ferreira}, {Finkelstein}, {Finkelstein}, {Gardner}, {Gawiser},
  {Gburek}, {Giavalisco}, {Grazian}, {Gronwall}, {Guo}, {Arrabal Haro},
  {Hathi}, {Hayes}, {Hemmati}, {Howell}, {Iyer}, {Jansen}, {Ji}, {Kaviraj},
  {Kurczynski}, {Lazar}, {Lucas}, {MacKenty}, {Mehta}, {Mantha}, {Martin},
  {Martin}, {McCabe}, {Mobasher}, {Nedkova}, {O'Connell}, {Olsen}, {Otteson},
  {Ravindranath}, {Redshaw}, {Robertson}, {Rutkowski}, {Sattari}, {Scarlata},
  {Siana}, {Smith}, {Soto}, {Vanzella}, {Yung}, \& {Zabelle}}]{Wang:2024}
{Wang}, X., {Teplitz}, H.~I., {Sun}, L., {et~al.} 2024, Research Notes of the
  American Astronomical Society, 8, 26, \dodoi{10.3847/2515-5172/ad1f6f}

\bibitem[{{Weaver} {et~al.}(2023){Weaver}, {Davidzon}, {Toft}, {Ilbert},
  {McCracken}, {Gould}, {Jespersen}, {Steinhardt}, {Lagos}, {Capak}, {Casey},
  {Chartab}, {Faisst}, {Hayward}, {Kartaltepe}, {Kauffmann}, {Koekemoer},
  {Kokorev}, {Laigle}, {Liu}, {Long}, {Magdis}, {McPartland}, {Milvang-Jensen},
  {Mobasher}, {Moneti}, {Peng}, {Sanders}, {Shuntov}, {Sneppen}, {Valentino},
  {Zalesky}, \& {Zamorani}}]{Weaver:2023}
{Weaver}, J.~R., {Davidzon}, I., {Toft}, S., {et~al.} 2023, \aap, 677, A184,
  \dodoi{10.1051/0004-6361/202245581}

\bibitem[{{Weibel} {et~al.}(2024){Weibel}, {Oesch}, {Barrufet}, {Gottumukkala},
  {Ellis}, {Santini}, {Weaver}, {Allen}, {Bouwens}, {Bowler}, {Brammer},
  {Carnall}, {Cullen}, {Dayal}, {Donnan}, {Dunlop}, {Giavalisco}, {Grogin},
  {Illingworth}, {Koekemoer}, {Labbe}, {Marchesini}, {McLeod}, {McLure},
  {Naidu}, {Shuntov}, {Stefanon}, {Toft}, \& {Xiao}}]{Weibel:2024}
{Weibel}, A., {Oesch}, P.~A., {Barrufet}, L., {et~al.} 2024, arXiv e-prints,
  arXiv:2403.08872, \dodoi{10.48550/arXiv.2403.08872}

\bibitem[{{Weiner}(2009)}]{Weiner:2009}
{Weiner}, B. 2009, {Star formation, extinction and metallicity at
  0.7$<$z$<$1.5: H-alpha fluxes and sizes from a grism survey of GOODS-N}, HST
  Proposal ID 11600. Cycle 17

\bibitem[{{Weisz} {et~al.}(2012){Weisz}, {Johnson}, {Johnson}, {Skillman},
  {Lee}, {Kennicutt}, {Calzetti}, {van Zee}, {Bothwell}, {Dalcanton}, {Dale},
  \& {Williams}}]{Weisz:2012}
{Weisz}, D.~R., {Johnson}, B.~D., {Johnson}, L.~C., {et~al.} 2012, \apj, 744,
  44, \dodoi{10.1088/0004-637X/744/1/44}

\bibitem[{{Whitaker} {et~al.}(2011){Whitaker}, {Labb{\'e}}, {van Dokkum},
  {Brammer}, {Kriek}, {Marchesini}, {Quadri}, {Franx}, {Muzzin}, {Williams},
  {Bezanson}, {Illingworth}, {Lee}, {Lundgren}, {Nelson}, {Rudnick}, {Tal}, \&
  {Wake}}]{Whitaker:2011}
{Whitaker}, K.~E., {Labb{\'e}}, I., {van Dokkum}, P.~G., {et~al.} 2011, \apj,
  735, 86, \dodoi{10.1088/0004-637X/735/2/86}

\bibitem[{{Whitaker} {et~al.}(2014){Whitaker}, {Franx}, {Leja}, {van Dokkum},
  {Henry}, {Skelton}, {Fumagalli}, {Momcheva}, {Brammer}, {Labb{\'e}},
  {Nelson}, \& {Rigby}}]{Whitaker:2014}
{Whitaker}, K.~E., {Franx}, M., {Leja}, J., {et~al.} 2014, \apj, 795, 104,
  \dodoi{10.1088/0004-637X/795/2/104}

\bibitem[{{Windhorst} {et~al.}(2011){Windhorst}, {Cohen}, {Hathi}, {McCarthy},
  {Ryan}, {Yan}, {Baldry}, {Driver}, {Frogel}, {Hill}, {Kelvin}, {Koekemoer},
  {Mechtley}, {O'Connell}, {Robotham}, {Rutkowski}, {Seibert}, {Straughn},
  {Tuffs}, {Balick}, {Bond}, {Bushouse}, {Calzetti}, {Crockett}, {Disney},
  {Dopita}, {Hall}, {Holtzman}, {Kaviraj}, {Kimble}, {MacKenty}, {Mutchler},
  {Paresce}, {Saha}, {Silk}, {Trauger}, {Walker}, {Whitmore}, \&
  {Young}}]{Windhorst:2011}
{Windhorst}, R.~A., {Cohen}, S.~H., {Hathi}, N.~P., {et~al.} 2011, \apjs, 193,
  27, \dodoi{10.1088/0067-0049/193/2/27}

\bibitem[{{Wirth} {et~al.}(2015){Wirth}, {Trump}, {Barro}, {Guo}, {Koo}, {Liu},
  {Kassis}, {Lyke}, {Rizzi}, {Campbell}, {Goodrich}, \& {Faber}}]{Wirth:2015}
{Wirth}, G.~D., {Trump}, J.~R., {Barro}, G., {et~al.} 2015, \aj, 150, 153,
  \dodoi{10.1088/0004-6256/150/5/153}

\bibitem[{{Wolf} {et~al.}(2004){Wolf}, {Meisenheimer}, {Kleinheinrich},
  {Borch}, {Dye}, {Gray}, {Wisotzki}, {Bell}, {Rix}, {Cimatti}, {Hasinger}, \&
  {Szokoly}}]{Wolf:2004}
{Wolf}, C., {Meisenheimer}, K., {Kleinheinrich}, M., {et~al.} 2004, \aap, 421,
  913, \dodoi{10.1051/0004-6361:20040525}

\bibitem[{{Wuyts} {et~al.}(2008){Wuyts}, {Labb{\'e}}, {F{\"o}rster Schreiber},
  {Franx}, {Rudnick}, {Brammer}, \& {van Dokkum}}]{Wuyts:2008}
{Wuyts}, S., {Labb{\'e}}, I., {F{\"o}rster Schreiber}, N.~M., {et~al.} 2008,
  \apj, 682, 985, \dodoi{10.1086/588749}

\bibitem[{{Wuyts} {et~al.}(2009){Wuyts}, {van Dokkum}, {Franx}, {F{\"o}rster
  Schreiber}, {Illingworth}, {Labb{\'e}}, \& {Rudnick}}]{Wuyts:2009}
{Wuyts}, S., {van Dokkum}, P.~G., {Franx}, M., {et~al.} 2009, \apj, 706, 885,
  \dodoi{10.1088/0004-637X/706/1/885}

\bibitem[{{Xue} {et~al.}(2016){Xue}, {Luo}, {Brandt}, {Alexander}, {Bauer},
  {Lehmer}, \& {Yang}}]{Xue:2016}
{Xue}, Y.~Q., {Luo}, B., {Brandt}, W.~N., {et~al.} 2016, \apjs, 224, 15,
  \dodoi{10.3847/0067-0049/224/2/15}

\bibitem[{{Yoshikawa} {et~al.}(2010){Yoshikawa}, {Akiyama}, {Kajisawa},
  {Alexander}, {Ohta}, {Suzuki}, {Tokoku}, {Uchimoto}, {Konishi}, {Yamada},
  {Tanaka}, {Omata}, {Nishimura}, {Koekemoer}, {Brandt}, \&
  {Ichikawa}}]{Yoshikawa:2010}
{Yoshikawa}, T., {Akiyama}, M., {Kajisawa}, M., {et~al.} 2010, \apj, 718, 112,
  \dodoi{10.1088/0004-637X/718/1/112}

\end{thebibliography}

\end{document}